\documentclass[floatfix,aps,prd,10pt,preprintnumbers,nofootinbib,superscriptaddress,showpacs,letterpaper,twocolumn]{revtex4-1}
\usepackage{amsmath,amssymb,amsbsy,booktabs,array}
\usepackage[pdftex]{graphicx,color}
\usepackage[sort&compress]{natbib}
\usepackage{rotating}
\usepackage[colorlinks=true, linkcolor=blue, filecolor=blue, urlcolor=blue, citecolor=blue, plainpages=false]{hyperref}

\newcommand{\mres}{m_{\rm res}}
\newcommand\riken{RIKEN-BNL Research Center, Brookhaven National
Laboratory, Upton, NY 11973, USA}
\newcommand\bnlaf{Physics Department, Brookhaven
National Laboratory, Upton, NY 11973, USA}

\newcommand\Higgsaf{Higgs Centre for Theoretical Physics, School of Physics \& Astronomy, The University of Edinburgh, EH9 3FD, UK}
\newcommand\Juelichaf{Forschungszentrum J\"ulich, Institute for Advanced Simulation, J\"ulich Supercomputing Centre, 52425 J\"ulich, Germany }

\newcommand\soton{School of Physics and Astronomy, University of
Southampton, Southampton SO17 1BJ, UK}
\newcommand\fermilabaf{Theoretical Physics Department, Fermi National Accelerator Laboratory, Batavia, IL 60510, USA}
\newcommand\buaff{Center for Computational Science, Boston University, Boston, MA 02215, USA}

\def\bar{\overline}

\def\spose#1{\hbox to 0pt{#1\hss}}
\def\ltapprox{\mathrel{\spose{\lower 3pt\hbox{$\mathchar"218$}}
\raise 2.0pt\hbox{$\mathchar"13C$}}}
\def\gtapprox{\mathrel{\spose{\lower 3pt\hbox{$\mathchar"218$}}
\raise 2.0pt\hbox{$\mathchar"13E$}}}
\def\inapprox{\mathrel{\spose{\lower 3pt\hbox{$\mathchar"218$}}
\raise 2.0pt\hbox{$\mathchar"232$}}}

\newcommand\Ocal{\mathcal{O}}

\newcommand\fpara{f_\parallel}
\newcommand\fperp{f_\perp}

\begin{document}
\preprint{ FERMILAB-PUB-15-019-T}
\bibliographystyle{apsrev}

\title{\texorpdfstring{$B \to \pi \ell \nu$ and $B_s \to K \ell \nu$ form factors and $|V_{ub}|$ from 2+1-flavor lattice QCD with domain-wall light quarks and relativistic heavy quarks}{The B -> pi l nu and Bs -> K l nu form factors and Vub from 2+1-flavor lattice QCD with domain-wall light quarks and relativistic heavy quarks}}

\author{J.~M.~Flynn}\affiliation\soton
\author{T.~Izubuchi}\affiliation\riken\affiliation\bnlaf
\author{T.~Kawanai}
\altaffiliation[Present address: ]{\Juelichaf}
\affiliation\riken\affiliation\bnlaf
\author{C.~Lehner}\affiliation\bnlaf
\author{A.~Soni}\affiliation\bnlaf
\author{R.~S.~Van~de~Water}\affiliation\fermilabaf
\author{O.~Witzel}
\altaffiliation[Present address: ]{\Higgsaf}
\affiliation\buaff
\collaboration{RBC and UKQCD Collaborations}\noaffiliation

%
%


\date{\today}

\begin{abstract}
We calculate the form factors for $B \to\pi\ell\nu$ and $B_s \to K \ell\nu$ decay in dynamical lattice Quantum Chromodynamics (QCD) using domain-wall light quarks and relativistic $b$ quarks.  We use the (2+1)-flavor gauge-field ensembles generated by the RBC and UKQCD collaborations with the domain-wall fermion action and Iwasaki gauge action.  For the $b$ quarks we use the anisotropic clover action with a relativistic heavy-quark interpretation.  We analyze data at two lattice spacings of $a \approx 0.11, 0.086$~fm with unitary pion masses as light as $M_\pi \approx 290$~MeV. We simultaneously extrapolate our numerical results to the physical light-quark masses and to the continuum and interpolate in the pion/kaon energy using SU(2) ``hard-pion'' chiral perturbation theory for heavy-light meson form factors.   We provide complete systematic error budgets for the vector and scalar form factors $f_+(q^2)$ and $f_0(q^2)$ for both $B \to\pi\ell\nu$ and $B_s \to K \ell\nu$ at three momenta that span the $q^2$ range accessible in our numerical simulations.  Next we extrapolate these results to $q^2 = 0$ using a model-independent $z$-parametrization based on analyticity and unitarity.  We present our final results for $f_+(q^2)$ and $f_0(q^2)$ as the coefficients of the series in $z$ and the matrix of correlations between them; this provides a parametrization of the form factors valid over the entire allowed kinematic range.  Our results agree with other three-flavor lattice-QCD determinations using staggered light quarks, and have comparable precision, thereby providing important independent cross-checks.  Both $B \to\pi\ell\nu$ and $B_s \to K \ell\nu$ decays enable determinations of the Cabibbo-Kobayashi-Maskawa matrix element $|V_{ub}|$.  To illustrate this, we perform a combined $z$-fit of our numerical $B\to\pi\ell\nu$ form-factor data with the experimental measurements of the branching fraction from BaBar and Belle leaving the relative normalization as a free parameter; we obtain $|V_{ub}| = 3.61(32) \times 10^{-3}$, where the error includes statistical and all systematic uncertainties.  The same approach can be applied to the decay $B_s\to K \ell\nu$ to provide an alternative determination of $|V_{ub}|$ once the process has been measured experimentally.  Finally, in anticipation of future experimental measurements, we make predictions for $B \to \pi\ell\nu$ and  $B_s\to K \ell\nu$ differential branching fractions and forward-backward asymmetries in the Standard Model.
\end{abstract}

\maketitle



\section{Introduction}
\label{Sec:Intro}

Semileptonic $B$-meson decays play an important role in the search for new physics in the quark-flavor sector.  Tree-level decays that occur via charged $W$-boson exchange are used to obtain the Cabibbo-Kobayashi-Maskawa (CKM) matrix elements $|V_{ub}|$ and $|V_{cb}|$, while flavor-changing neutral-current decays provide sensitive probes for heavy new particles that may enter virtual loops.  Decays involving $\tau$ leptons are especially sensitive to charged Higgs bosons that arise in many new-physics models (see {\it e.g.} Ref.~\cite{Celis:2014pza} and references therein).

The decays $B \to\pi\ell\nu$ and $B_s \to K \ell\nu$ probe the quark-flavor-changing transition $b \to u$.  In the Standard Model, the differential decay rate for these processes in the $B_{(s)}$-meson rest frame is given by 
\begin{widetext}
\begin{align}
	\frac{d\Gamma(B_{(s)}\to P\ell\nu)}{dq^2} & = \frac{G_F^2 |V_{ub}|^2}{24 \pi^3} \,\frac{(q^2-m_\ell^2)^2\sqrt{E_P^2-M_P^2}}{q^4M_{B_{(s)}}^2}
	\bigg[ \left(1+\frac{m_\ell^2}{2q^2}\right)M_{B_{(s)}}^2(E_P^2-M_P^2)|f_+(q^2)|^2 \nonumber\\
&+\,\frac{3m_\ell^2}{8q^2}(M_{B_{(s)}}^2-M_P^2)^2|f_0(q^2)|^2
\bigg]\,, \label{eq:B_semileptonic_rate}
\end{align}
\end{widetext}
where $P$ denotes the light pseudoscalar pion or kaon and $q \equiv (p_B - p_P)$ is the momentum transferred to the outgoing charged-lepton-neutrino pair.  The vector and scalar form factors $f_+(q^2)$ and $f_0(q^2)$ parametrize the hadronic contributions to the electroweak decay and must be calculated nonperturbatively, such as with lattice QCD.  Given an experimental measurement of the branching fraction and a theoretical calculation of the form factor(s), these decays enable a determination of the CKM matrix element $|V_{ub}|$.  (The contribution from $f_0^{B \pi}(q^2)$ in Eq.~(\ref{eq:B_semileptonic_rate}) can be neglected for light leptons $\ell=e,\mu$ given the current experimental and theoretical precision.)    To date, both the BaBar and Belle experiments have measured ${\mathcal B}(B \to\pi\ell\nu)$~\cite{delAmoSanchez:2010af,Ha:2010rf,Lees:2012vv,Sibidanov:2013rkk}, and the experimental uncertainty will continue to improve with the collection of data at Belle~II.  The decay $B_s \to K \ell\nu$ has not yet been measured, but we anticipate a result from LHCb in the next few years.

The CKM matrix element $|V_{ub}|$ places a constraint on the apex of the CKM unitarity triangle~\cite{CKMfitter,UTfit,Laiho:2009eu}.  Its value, however, is under scrutiny because of the long-standing $\sim 3\sigma$ disagreement between $|V_{ub}|$ obtained from exclusive $B \to \pi \ell\nu$ decay and $|V_{ub}|$ obtained from inclusive $B \to X_u \ell \nu$ decays, where $X_u$ denotes all charmless final states with up quarks~\cite{CKMfitter,UTfit,Antonelli:2009ws,Laiho:2009eu,Beringer:2012zz,Aoki:2013ldr}.   The value of $|V_{ub}|$ can also in principle be obtained from leptonic $B \to \tau \nu$ decay, but the current determination from this process lies in between those from exclusive and inclusive semileptonic decays, and is not as precise~\cite{Aoki:2013ldr}.  Further, $B \to \tau \nu$ is sensitive to charged-Higgs boson exchange, and therefore does not provide a clean Standard-Model determination of $|V_{ub}|$.  Thus the decay $B_s \to K \ell\nu$, once measured experimentally, will provide an important new determination of $|V_{ub}|$.

In this paper we present a new calculation of the semileptonic form factors for $B \to\pi\ell\nu$ and $B_s \to K \ell\nu$  in (2+1)-flavor lattice QCD.  Preliminary results were presented in Refs.~\cite{Kawanai:2012id,Kawanai:2013qxa}.  This is the second in a series of $B$-meson matrix-element calculations that uses the same lattice actions and ensembles, and our analysis follows a similar approach to our earlier work on $B$-meson decay constants~\cite{Christ:2014uea}.   We use the gauge-field ensembles generated by the RBC and UKQCD collaborations with the domain-wall fermion action and Iwasaki gluon action which include the effects of dynamical $u,d$, and $s$ quarks~\cite{Allton:2008pn,Aoki:2010dy}.  For the bottom quarks, we use the Columbia version of the relativistic heavy-quark (RHQ) action introduced by Christ, Li, and Lin in Ref.~\cite{Christ:2006us}, with the parameters of the action that were obtained nonperturbatively in Ref.~\cite{Aoki:2012xaa}.   We renormalize the lattice heavy-light vector current using the mostly nonperturbative method introduced in Ref.~\cite{ElKhadra:2001rv}, in which we compute the bulk of the matching factor nonperturbatively~\cite{Aoki:2010dy,Christ:2014uea}, with a small correction, that is close to unity, evaluated in lattice perturbation theory~\cite{Lehner:2012bt,CLehnerPT}.  We also improve the lattice heavy-light current through ${\mathcal O}(\alpha_s a)$.  

We analyze data on five sea-quark ensembles with unitary pions as light as $\approx$ 290~MeV and two lattice spacings of $a \approx$ 0.11 and 0.086~fm.  We simultaneously extrapolate our numerical results to the physical light-quark masses and to the continuum and interpolate in the pion/kaon energy using SU(2) ``hard-pion'' chiral perturbation theory ($\chi$PT) for heavy-light meson form factors~\cite{Becirevic:2002sc,Bijnens:2010ws}, which applies when the pion/kaon energy is large compared to its rest mass.  For $B \to\pi\ell\nu$ ($B_s \to K \ell\nu$), we directly simulate in the momentum region $q^2_{\rm max} > q^2 \gtapprox 19.0$~GeV$^2$ ($q^2_{\rm max} > q^2 \gtapprox 17.6$~GeV$^2$).   Both statistical errors and discretization errors increase at lower $q^2$, which corresponds to larger pion/kaon energies.  To extend our results beyond the momenta accessible in our simulations, we extrapolate our results to $q^2 = 0$ using a model-independent $z$-parametrization based on analyticity and unitarity~\cite{Boyd:1994tt,Bourrely:2008za}.   Our results can be combined with current and future experimental measurements of the experimentally measured $B \to\pi\ell\nu$ and $B_s \to K \ell\nu$ branching fractions to obtain the CKM matrix element $|V_{ub}|$.  

There are two earlier published (2+1)-flavor calculations of the $B \to \pi \ell \nu$ semileptonic form factor in the literature by the HPQCD and Fermilab/MILC collaborations~\cite{Dalgic:2006dt,Bailey:2008wp}; updates of these works are in progress~\cite{Bouchard:2013zda,Du:2013kea}.  In addition, HPQCD recently obtained the first results for the $B_s \to K \ell\nu$ form factor in Ref.~\cite{Bouchard:2014ypa}.  Both groups use the MILC collaboration's asqtad-improved staggered gauge-field ensembles~\cite{Bernard:2001av,Bazavov:2009bb}, so their results are somewhat correlated.  The differences between the two sets of calculations lie in the choices of light valence- and $b$-quark actions.  For the $b$ quarks, HPQCD uses the NRQCD action~\cite{Lepage:1992tx} while Fermilab/MILC uses a relativistic formulation similar to ours.  Specifically, they use the Fermilab interpretation of the isotropic clover action~\cite{ElKhadra:1996mp} with the tadpole-improved tree-level value of the clover coefficient $c_{SW}$.  The more recent HPQCD calculation uses the HISQ action for the light valence quarks to reduce taste-breaking discretization effects, while in the other work asqtad valence quarks are used.  

Our form-factor calculation with domain-wall light quarks and RHQ $b$ quarks has the advantage that discretization errors from the light quarks and gluons are simpler, such that the SU(2) heavy-light meson $\chi$PT expressions are continuum-like.   Further, as compared to the Fermilab/MILC calculation, we tune the coefficient of the clover term in the $b$-quark action nonperturbatively and improve the heavy-light vector current through ${\mathcal O}(\alpha_s a)$, whereas Fermilab/MILC only improve it through ${\mathcal O}(a)$.  Thus, for similar values of the lattice spacing, discretization errors from the heavy-quark action and current are smaller in our calculation.  Our new results for the $B \to \pi \ell \nu$ and $B_s \to K \ell \nu$ form factors therefore enable important independent determinations of the CKM matrix element $|V_{ub}|$. 

This paper is organized as follows.  Section~\ref{Sec:Calc} provides an overview of the lattice calculation.  First we define the needed matrix elements and form factors in Sec.~\ref{Sec:FFs}.  Next we present the lattice actions and parameters in Sec.~\ref{Sec:Actions}.  Then, in Sec.~\ref{sec:Renorm} we describe the renormalization and improvement of the heavy-light vector current operator.  Section~\ref{Sec:Analysis} presents the numerical analysis.  First, in Secs.~\ref{Sec:2pt} and~\ref{Sec:3pt} we fit lattice two-point and three-point correlators to extract the needed meson masses and matrix elements, respectively.  Then, in Sec.~\ref{Sec:Extrapolations} we extrapolate our numerical data to the physical light-quark masses and continuum, and interpolate in the pion/kaon energy, using SU(2) hard-pion $\chi$PT.  Section~\ref{Sec:SysErrors} provides complete error budgets for $f_+(q^2)$ and $f_0(q^2)$ at three momentum values that span the range accessible in our numerical simulations; for clarity, we discuss each source of systematic uncertainty in a separate subsection.  In Section~\ref{Sec:FormFactors} we extrapolate our form-factor data to $q^2=0$ using a model-independent $z$ parametrization.   We present our results for $f_+(q^2)$ and $f_0(q^2)$ as the coefficients of the series in $z$ and the matrix of correlations between them; this provides a model-independent parametrization of the form factors valid over the entire allowed kinematic range.  We illustrate the phenomenological utility of our form-factor results in Sec.~\ref{Sec:Pheno}.  First, in Sec.~\ref{Sec:Vub}, we perform a combined $z$-fit of our numerical $B\to\pi\ell\nu$ form-factor data with the experimental measurements of the branching fraction from BaBar and Belle to determine $|V_{ub}|$.  Next, in Sec.~\ref{Sec:BstoKPheno}, we make predictions for Standard-Model observables for the decay processes $B \to\pi\ell\nu$ and $B_s \to K \ell \nu$ with $\ell=\mu,\tau$ in anticipation of future experimental measurements.  Section~\ref{Sec:Conc} concludes with a comparison of our results with other lattice determinations, and with an outlook for the future.

\section{Lattice calculation}
\label{Sec:Calc}

Here we present the setup of our numerical lattice calculation.

\subsection{Form factors}
\label{Sec:FFs}

The $B\to \pi \ell \nu$ and $B_s \to K \ell \nu$ semileptonic form factors parametrize the hadronic matrix element of the
$b\to u$ vector current $\mathcal{V}^\mu\equiv \bar{u}\gamma^\mu b$:
\begin{align}
 \langle  P | \mathcal{V}^\mu | B_{(s)} \rangle 
& = f_+(q^2) \left( p_{B_{(s)}}^\mu + p_P^\mu - \frac{M_{B_{(s)}}^2-M_P^2}{q^2}q^\mu\right) \nonumber\\
& + f_0(q^2)\frac{M_{B_{(s)}}^2-M_P^2}{q^2}q^\mu \,,
\end{align}
where $f_+(q^2)$ and $f_0(q^2)$ are the vector and scalar form factors, respectively.
It is convenient in lattice simulations to instead calculate the form factors $f_{\parallel}(E_P)$ and $ f_{\perp}(E_P)$, which are defined by
\begin{equation}
 \langle  P | \mathcal{V}^\mu | B_{(s)} \rangle = \sqrt{2M_{B_{(s)}}}
\left[ v^\mu f_{\parallel}(E_P) + p^\mu_{\perp}f_{\perp}(E_P) \right] \,,
\end{equation}
where $E_P$ is the outgoing light pseudoscalar meson energy, $v^\mu \equiv p_{B_{(s)}}^\mu/M_{B_{(s)}}$ is the $B_{(s)}$-meson velocity, and $p^\mu_{\perp} \equiv p_P^\mu -(p_P\cdot v)v^\mu$.  In the $B_{(s)}$-meson rest frame, which we will use for our simulations, $f_\parallel$ and $f_\perp$ are proportional to the hadronic matrix elements of the temporal and spatial vector currents:
\begin{align}
f_{\parallel}(E_P) & = \frac{\langle P | \mathcal{V}^0 | B_{(s)} \rangle }{\sqrt{2M_{B_{(s)}}}} \,, \label{eq:fpara} \\
f_{\perp}(E_P) & = \frac{\langle P | \mathcal{V}^i | B_{(s)} \rangle}{\sqrt{2M_{B_{(s)}}}} \frac{1}{p^i_P} \,. \label{eq:fperp}
\end{align}
The vector and scalar form factors can be easily obtained from $f_\parallel$ and $f_\perp$ via
\begin{align}
 f_+(q^2) & = \frac{1}{\sqrt{2M_{B_{(s)}}}} \Big[ f_{\parallel}(E_P)+\big(M_{B_{(s)}}-E_P\big) f_{\perp}(E_P) \Big] \,, \\
 f_0(q^2) & = \frac{\sqrt{2M_{B_{(s)}}}}{M^2_{B_{(s)}} - M^2_P} \Big[ \big(M_{B_{(s)}} -E_P \big) f_\parallel(E_P) \nonumber\\ & + \left(E^2_P - M^2_P\right) f_\perp(E_P) \Big]  \,.
\end{align}

\subsection{Actions and parameters}
\label{Sec:Actions}

\begin{table*}[tb]
\caption{Lattice simulation parameters \cite{Allton:2008pn,Aoki:2010dy}.  The columns list the lattice volume, approximate lattice spacing, light ($m_l$) and strange ($m_h$) sea-quark masses, residual chiral symmetry breaking parameter $m_{\rm res}$, physical $u/d$- and $s$-quark mass, unitary pion mass, number of configurations analyzed and number of sources. The tildes over $a\widetilde m_{u/d}$ and $a \widetilde m_s$ denote that these values include the residual quark mass. 
} 
\vspace{3mm}
\label{tab:lattices}
\begin{ruledtabular}
\begin{tabular}{c@{\hskip 3mm}cc@{\hskip 4mm}ccc@{\hskip 4mm}cc@{\hskip 3mm}c@{\hskip 3mm}cccc}
$\left(\frac{L}{a}\right)^3 \times \left(\frac{T}{a}\right)$ &$\approx a$(fm)& $a^{-1}$ [GeV] & ~~$am_l$ & ~~$am_h$&$a\mres$ &$a\widetilde m_{u/d}$ &$ a \widetilde m_s$&  $M_\pi$[MeV]  & \# configs. & \# time sources\\[0.5mm] \hline
$24^3 \times 64$ &  0.11  & 1.729(25) & 0.005 & 0.040 &0.003152& 0.00136(4)&0.0379(11)& 329 & 1636 & 1 \\
$24^3 \times 64$ &  0.11  & 1.729(25) & 0.010 & 0.040 &0.003152& 0.00136(4)&0.0379(11)& 422 & 1419 & 1 \\ \hline
$32^3 \times 64$ &  0.086 & 2.281(28) & 0.004 & 0.030 &0.0006664&0.00102(5)&0.0280(7) & 289 & 628  & 2 \\ 
$32^3 \times 64$ &  0.086 & 2.281(28) & 0.006 & 0.030 &0.0006664&0.00102(5)&0.0280(7) & 345 & 889  & 2 \\
$32^3 \times 64$ &  0.086 & 2.281(28) & 0.008 & 0.030 &0.0006664&0.00102(5)&0.0280(7) & 394 & 544  & 2 \\ 
\end{tabular}
\end{ruledtabular}
\end{table*}

We use the $(2+1)$-flavor domain-wall fermion and Iwasaki gauge-field ensembles generated by the RBC and UKQCD collaborations~\cite{Allton:2008pn,Aoki:2010dy}.  We perform measurements at five different light sea-quark masses $m_l$ and at two lattice spacings of $a\approx 0.11$~fm ($a^{-1} \approx 1.729$~GeV) and $a\approx 0.086$~fm ($a^{-1} \approx 2.281$~GeV). The light sea-quark masses $m_l$  correspond to pion masses of $289\  {\rm MeV} \alt M_\pi \alt 422\  {\rm MeV}$.   The up and down sea-quark masses are degenerate and the strange sea-quark mass $m_h$ is tuned within 10\% of its physical value. The spatial volumes are approximately $(2.6\ {\rm fm})^3$, such that $M_\pi L \geq 4$.   We summarize the simulation parameters in Table~\ref{tab:lattices}.

In the valence sector we use for the light quarks the domain-wall action \cite{Shamir:1993zy,Furman:1994ky} and generate propagators with periodic boundary conditions in space and time and with the same domain-wall height ($M_5=1.8$) and extent of the fifth dimension ($L_s=16$) as in the sea sector. We generate both unitary light valence-quark propagators with the same mass as the light sea quarks and propagators with a mass close to the physical strange quark.  On the coarser ensembles we choose $a m_s = 0.0343$ and on the finer ensembles $a m_s = 0.0272$.

For the bottom quarks, we use the Columbia version of the relativistic heavy quark (RHQ) action~\cite{Christ:2006us} to control heavy-quark discretization errors introduced by the large lattice $b$-quark mass.  We use the anisotropic $\Ocal(a)$ improved Wilson-clover action with the following three parameters: the bare-quark mass $m_0 a$, clover coefficient $c_P$, and anisotropy parameter $\zeta$.  In this work we use the RHQ parameters tuned nonperturbatively in Ref.~\cite{Aoki:2012xaa} to reproduce the experimentally measured $B_s$-meson mass and hyperfine splitting; we list their values in Table~\ref{tab:RHQpara}.

\begin{table*}
\caption{Tuned RHQ parameters on the $24^3$ and $32^3$ ensembles~\cite{Aoki:2012xaa}.   
 The errors listed for $m_0a$, $c_P$, and $\zeta$ are from left to right:  
statistics, heavy-quark discretization errors, the lattice scale uncertainty, 
and the uncertainty in the experimental measurement of the $B_s$-meson hyperfine splitting, respectively.}
  \label{tab:RHQpara}
\vspace{3mm}
  \begin{tabular}{lr@{.}lr@{.}lr@{.}l} \hline\hline
    & \multicolumn{2}{c}{$m_0a$} &  \multicolumn{2}{c}{$c_P$} & \multicolumn{2}{c}{$\zeta$}  \\[0.5mm] \hline
    $a \approx 0.11$~fm  &  8&45(6)(13)(50)(7) & 5&8(1)(4)(4)(2)     & 3&10(7)(11)(9)(0) \\
    $a \approx 0.086$~fm &  3&99(3)(6)(18)(3)  & 3&57(7)(22)(19)(14) & 1&93(4)(7)(3)(0) \\ \hline\hline
  \end{tabular}
  \label{tab:RHQParamErr}
\end{table*}

We reduce autocorrelations between our lattices by shifting the gauge fields by a random 4-vector before creating the sources for the valence-quark propagators used in the 2-point and 3-point correlation functions. This random 4-vector shift is equivalent to placing the sources at random positions in spacetime but simplifies the subsequent analysis.  On the finer ensembles, we double the statistics by using two sources per configuration separated by half the lattice temporal extent.

\subsection{Operator renormalization and improvement}
\label{sec:Renorm}

\begin{table*}
  \centering
  \caption{Operator renormalization factors and improvement coefficients.  The flavor-conserving $Z$ factors were obtained nonperturbatively~\cite{Aoki:2010dy,Christ:2014uea}.  The $\rho$ factors and improvement coefficients $c_i^n$ were computed at one loop in mean-field improved lattice perturbation theory and are evaluated at $\alpha_s^{\overline{\rm MS}}(a^{-1})$~\cite{CLehnerPT}.}
  \label{tab:rho}
  \begin{ruledtabular}
    \begin{tabular}{lccccccccccc} 
      & $Z_V^{ll}$ &  $Z_V^{bb}$ & $\alpha_s^{\overline{\rm MS}}(a^{-1})$ & $\rho_{V_0}$  & $\rho_{V_i}$ & $c_t^{3}$ & $c_t^{4}$ & $c_s^{1}$ & $c_s^{2}$ & $c_s^{3}$ & $c_s^{4}$ \\[2pt]  \hline
      $a \approx 0.11$~fm   &  0.71689(51) & 10.039(25) & 0.23 & 1.02658  & 0.99723 & 0.0558 & -0.0099 & -0.00079  & 0.0018  & 0.0485 &  -0.0033 \\
      $a \approx 0.086$~fm  &  0.74469(13) & 5.256(8)   & 0.22 & 1.01661  & 0.99398 & 0.0547 & -0.0095 & -0.0012  & 0.00047 & 0.0480 & -0.0020
  \end{tabular}
\end{ruledtabular}
\end{table*}

To match the lattice amplitudes to the continuum matrix elements, we multiply by the heavy-light renormalization factor $Z_{V_\mu}^{bl}$:
\begin{equation}
\langle P | \mathcal{V}_\mu | B_{(s)}\rangle = Z_{V_\mu}^{bl} \langle P | V_\mu | B_{(s)}\rangle \,,
\end{equation}
where $\mathcal{V}_\mu$ and $V_\mu$ are the continuum and lattice current operators, respectively.  Following Ref.~\cite{ElKhadra:2001rv} we calculate the renormalization factor $Z_{V_\mu}^{bl}$ using a mostly nonperturbative method in which we express $Z_{V_\mu}^{bl}$ as the following product: 
\begin{equation}
Z_{V_\mu}^{bl} = \rho_{V_\mu}^{bl} \sqrt{Z_{V}^{bb} Z_{V}^{ll}} .  
\end{equation}
Most of the heavy-light current renormalization comes from the flavor-conserving factors $Z_V^{bb}$ and $Z_V^{ll}$.  The remaining factor $\rho^{bl}_V$ is expected to be close to unity because most of the radiative corrections, including contributions from tadpole graphs, cancel~\cite{Harada:2001fi}.

Both flavor-conserving renormalization factors $Z_{V}^{bb}$ and  $Z_{V}^{ll}$ were computed nonperturbatively in previous works. We computed $Z_V^{bb}$ for our earlier calculation of $B$-meson decay constants from the matrix element of the $b \to b$ vector current between two $B_s$ mesons~\cite{Christ:2014uea}.  We can also take advantage of the fact that for domain-wall fermions $Z_V^{ll} = Z_A^{ll}$ up to corrections of $\Ocal(am_{\rm res})$ and use the determination of $Z_A^{ll}$ from Ref.~\cite{Aoki:2010dy}.  The flavor off-diagonal renormalization factor $\rho^{bl}_V$ is calculated at ${\mathcal O}(\alpha_s)$ in mean-field improved lattice perturbation theory~\cite{Lepage:1992xa} and evaluated at the $\overline{\rm MS}$ coupling $\alpha_s^{\overline{\rm MS}}(\mu= a^{-1})$.  Our perturbative computation extends the work of Ref.~\cite{Aoki:2002iq} to bilinears with one relativistic heavy quark in the Columbia formulation and one domain-wall light quark.   For $\alpha_s^{\overline{\rm MS}}$, we use Eq.~(167) of Ref.~\cite{Aoki:2002iq}, which does not take into account sea-quark effects.  Because sea-quark effects enter at two loops, however, and the rest of the computation is performed at one loop, the error introduced by setting $N_f=0$ is of the same size as other remaining truncation errors. Further details of the perturbative calculation will be provided in a forthcoming publication~\cite{CLehnerPT}.  Table~\ref{tab:rho} shows the renormalization factors used in this work. 

We reduce discretization errors in the heavy-light vector current by improving it through $\mathcal{O}(\alpha_s a)$. We compute the matrix element of the tree-level heavy-light vector current
\begin{equation}
 V_\mu^{0}(x) = \bar{q}(x)\gamma^\mu Q(x), \label{eq_V0}
\end{equation}
plus matrix elements of these additional single-derivative operators
\begin{eqnarray}
 V_\mu^{1}(x) &=& \bar{q}(x) 2\overrightarrow{D}_\mu Q(x), \\
 V_\mu^{2}(x) &=& \bar{q}(x) 2\overleftarrow{D}_\mu Q(x), \\
 V_\mu^{3}(x) &=& \bar{q}(x) 2\gamma_\mu \gamma_i \overrightarrow{D}_i Q(x), \\
 V_\mu^{4}(x) &=& \bar{q}(x) 2\gamma_\mu \gamma_i \overleftarrow{D}_i Q(x), \label{eq_Vi}
\end{eqnarray}
where the covariant derivatives are defined by 
\begin{align}
 \overrightarrow{D}_\mu Q(x)    =  \frac{1}{2}( & U_\mu(x)Q(x+\hat{\mu}) \notag \\
 &  - U_\mu^\dagger(x-\hat{\mu}) Q(x-\hat{\mu})),  \\
\bar{q}(x)\overleftarrow{D}_\mu  =   \frac{1}{2}( & \bar{q}(x+\hat{\mu})U_\mu^\dagger(x)  \notag \\
&- \bar{q}(x-\hat{\mu})U_\mu(x-\hat{\mu}) ).
\end{align}
The temporal and spatial $\Ocal (a)$-improved vector-current operators are given by
the following sums:
\begin{eqnarray}
 V_0^{\rm imp}(x) &=& V_0^{0}(x) + c_t^{3} V_0^{3}(x) + c_t^{4} V_0^{4}(x),  \label{eq:V0_imp} \\
 V_i^{\rm imp}(x) &=& V_i^{0}(x) + c_s^{1} V_i^{1}(x) + c_s^{2} V_i^{2}(x) \notag \\
&& \ \ \ \ \ \ \ \ \ \ \ \ 
+ c_s^{3} V_i^{3}(x) + c_s^{4} V_i^{4}(x) \,. \label{eq:Vi_imp}
\end{eqnarray}
We calculate the coefficients $c_t^n$ and $c_s^n$ at one loop in mean-field improved lattice perturbation theory~\cite{CLehnerPT}; the values of the coefficients evaluated at $\alpha_s^{\overline{\rm MS}}(a^{-1})$ are shown in Table~\ref{tab:rho}.
\vfill
\section{Analysis}
\label{Sec:Analysis}

Here we present our determinations of the form factors $f_+(q^2)$ and $f_0(q^2)$ for $B \to\pi\ell\nu$ ($B_s \to K \ell\nu$) at large values of $q^2 \gtapprox 19.0~{\rm GeV}^2$ ($q^2 \gtapprox 17.6~{\rm GeV}^2$) accessible in our numerical simulations.

Our analysis proceeds in three steps:  First, in Sec.~\ref{Sec:2pt}, we fit the pion, kaon, and $B_{(s)}$-meson 2-point correlation functions to obtain the ground-state meson masses.  The results for these meson masses then enter our 3-point correlator fits in Sec.~\ref{Sec:3pt} to obtain the lattice form factors $f_\parallel(E_P)$ and $f_\perp(E_P)$ at fixed values of the pion/kaon energy $E_P$. In Sec.~\ref{Sec:Extrapolations}, we interpolate the renormalized values for $f_\parallel(E_P)$ and $f_\perp(E_P)$ in energy, and extrapolate to the physical light-quark masses and the continuum limit, using SU(2) hard-pion $\chi$PT formulated for heavy-light mesons.  To avoid possible biases due to analysis choices, we use the same fit functions in the correlator and chiral fits for both processes $B \to\pi\ell\nu$ and $B_s \to K \ell\nu$, and fitting ranges that are as close as possible.

We propagate statistical errors throughout the analysis via a single-elimination jackknife procedure. We avoid a direct dependence on the lattice  scale by carrying out our analysis in units of the $B_s$-meson mass. The $B_s$-meson mass plays a special role because we tuned parameters of the $b$-quark action to match the experimental value.  Thus we can obtain the form factors in physical units after the chiral-continuum extrapolation by multiplying by $M_{B_s}^{\rm exp.}$ to the appropriate power.  With this approach, the uncertainty on the lattice scale enters only indirectly via the values of the RHQ parameters.

We use the Chroma software library for lattice QCD  to compute our numerical data for the lattice 2-point and 3-point correlation functions~\cite{Edwards:2004sx}. 
\vfill

\begin{table*}[t]
 \caption{Time ranges used in two-point and three-point fits to determine the lattice meson masses and form factors.  For the three-point fits, we use the same range for all operators and momenta.}
 \vspace{3mm}
 \label{tab:fit_ranges}
    \begin{tabular}{c|cccc|cccc} \hline \hline
     \multicolumn{1}{c}{ } & \multicolumn{8}{c}{\ \  [$t_{\rm min}$, $t_{\rm max}$]} \\
    & \multicolumn{4}{c|}{2-point fits} & \multicolumn{4}{c}{3-point fits} \\
   & $M_\pi$ &  $M_K$   &    $M_B$   &    $M_{B_s}$  & $\fpara^{B\pi}$ & $\fperp^{B\pi}$ & $\fpara^{B_sK}$ & $\fperp^{B_sK}$\\  \hline
   $a\approx 0.11$~fm    & $[12,23]$ & $[12,23]$ & $[7,30]$  & $[10,30]$  & [6,10] & [6,10] & [6,10] & [6,10]\\
   $a\approx 0.086$~fm   & $[16,30]$ & $[16,30]$ & $[9,30]$  & $[13,30]$  & [8,13] & [8,13] & [8,13] & [8,13]\\ \hline\hline
  \end{tabular}
\end{table*}

\subsection{Two-point correlator fits}
\label{Sec:2pt}

To obtain the lattice $B_{(s)}\to P$ amplitude, we first calculate the following two-point correlation functions:
\begin{eqnarray}
    C_{P}(t,\vec{p}_P) &=&  \sum_{\vec{x}} e^{i\vec{p}_P\cdot \vec{x}}
        \langle \mathcal{O}_P^\dagger(\vec{x},t) \mathcal{O}_P(\vec{0},0) \rangle, \label{eq_pion} \\
      C_{{B_{(s)}}}(t) &=&  \sum_{\vec{x}}
      \langle \mathcal{O}_{B_{(s)}}^\dagger(\vec{x},t) \tilde{\mathcal{O}}_{B_{(s)}}(\vec{0},0) \rangle, \label{eq_Bmeson} \\
    \tilde{C}_{{B_{(s)}}}(t) &=&  \sum_{\vec{x}}
        \langle \tilde{\mathcal{O}}_{B_{(s)}}^\dagger(\vec{x},t) \tilde{\mathcal{O}}_{B_{(s)}}(\vec{0},0) \rangle, \label{eq_Bmeson2}
\end{eqnarray}
where $\mathcal{O}_P  = \bar{q} \gamma_5 q$ and $\mathcal{O}_{B_{(s)}} = \bar{Q} \gamma_5 q$ are interpolating operators for the light pseudoscalar and ${B_{(s)}}$-meson, respectively.  Both pions and kaons are simulated with a point source and point sink, whereas $b$-quark propagators are generated with a gauge-invariant Gaussian smeared source \cite{Alford:1995dm,Lichtl:2006dt} to reduce excited state contamination. We employ the same smearing parameters optimized in Ref.~\cite{Aoki:2012xaa} and denote a smeared source in Eqs.~(\ref{eq_pion})-(\ref{eq_Bmeson2}) with a tilde above the operator.

We obtain the pion or kaon energy and ${B_{(s)}}$-meson mass from the exponential decay of the correlators in  Eqs.~(\ref{eq_pion}) and (\ref{eq_Bmeson}).  The correlator in Eq.~(\ref{eq_Bmeson2}) is used to normalize the $B_{(s)}\to P$ three-point function.  We work in the ${B_{(s)}}$-meson rest frame such that only pions or kaons carry nonzero momentum. In our analysis we use data with discrete lattice pion momenta through $2\pi(1,1,1)/L$ and kaon momenta through $2\pi(2,0,0)/L$.  We average the results for all equivalent momenta, {\it i.e.} with different spatial directions but the same total $|\vec{p}_P|$.  We effectively double our statistics by folding the two-point correlators at the temporal midpoint of the lattice, thereby averaging forward- and backward-propagating states.

\begin{figure*}[p] 
  \centering
  \includegraphics[width=.47\textwidth]{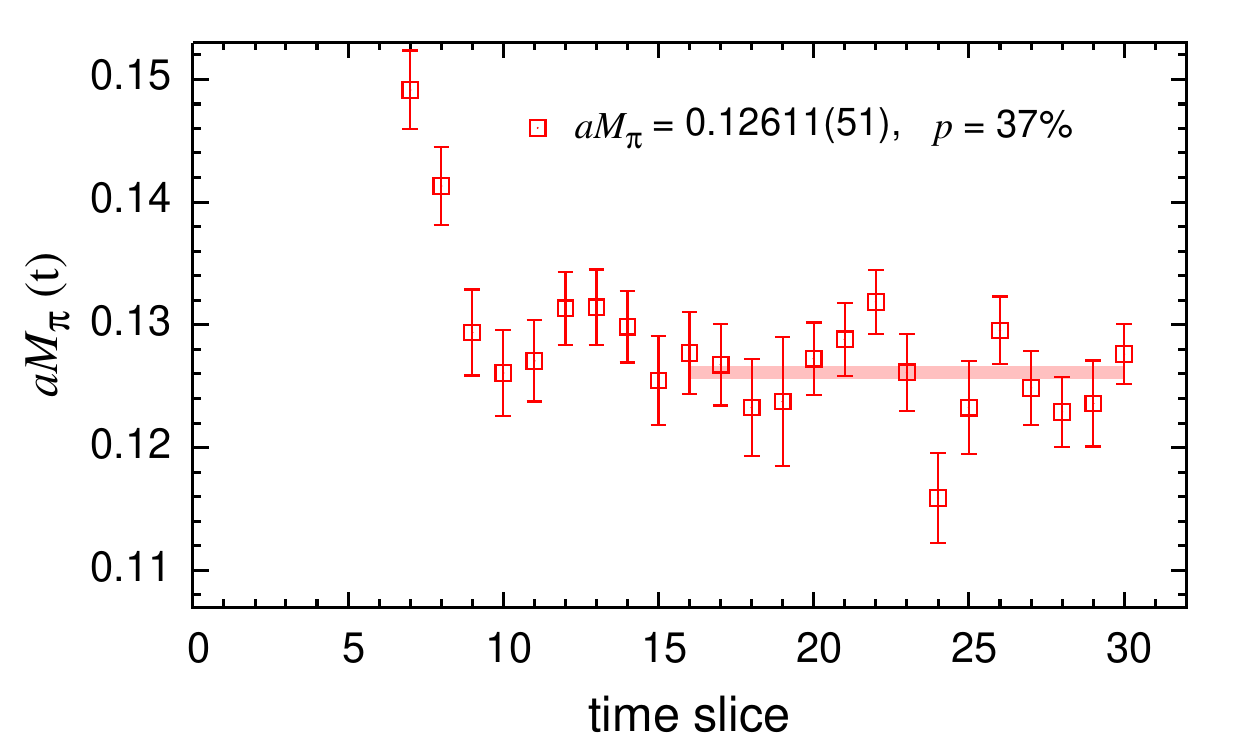}
  \includegraphics[width=.47\textwidth]{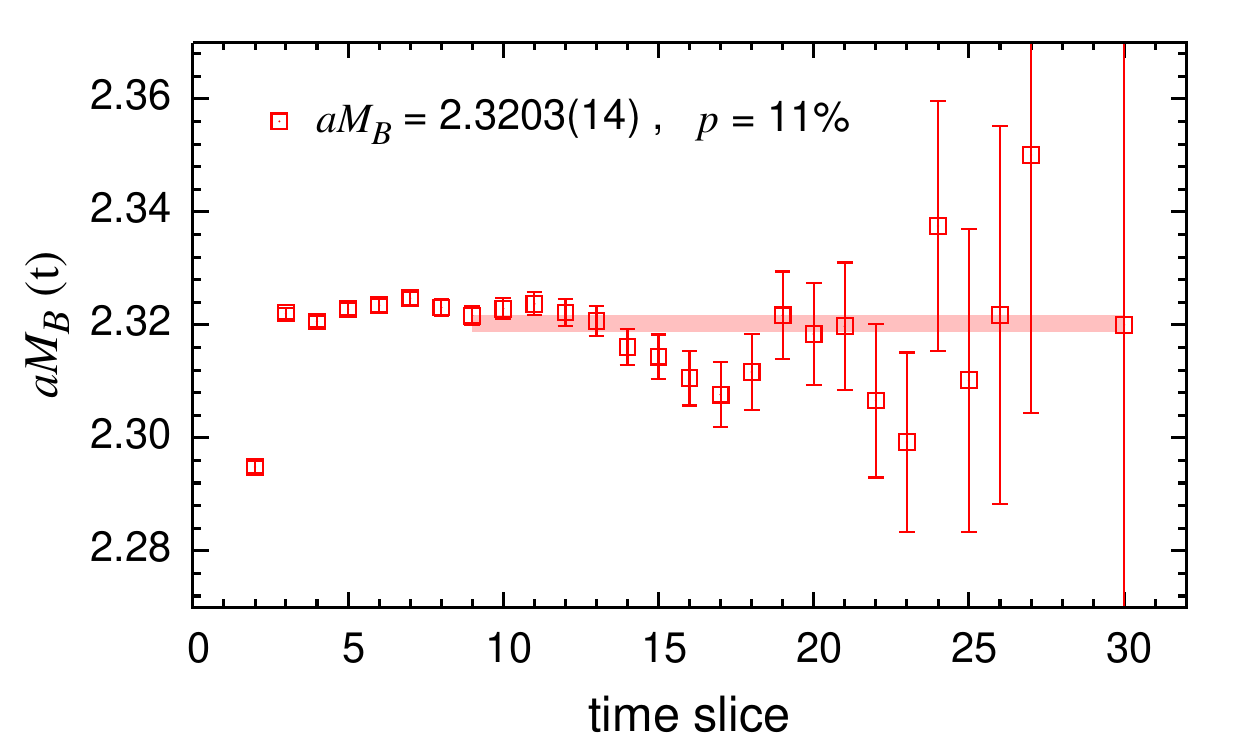}
  \includegraphics[width=.47\textwidth]{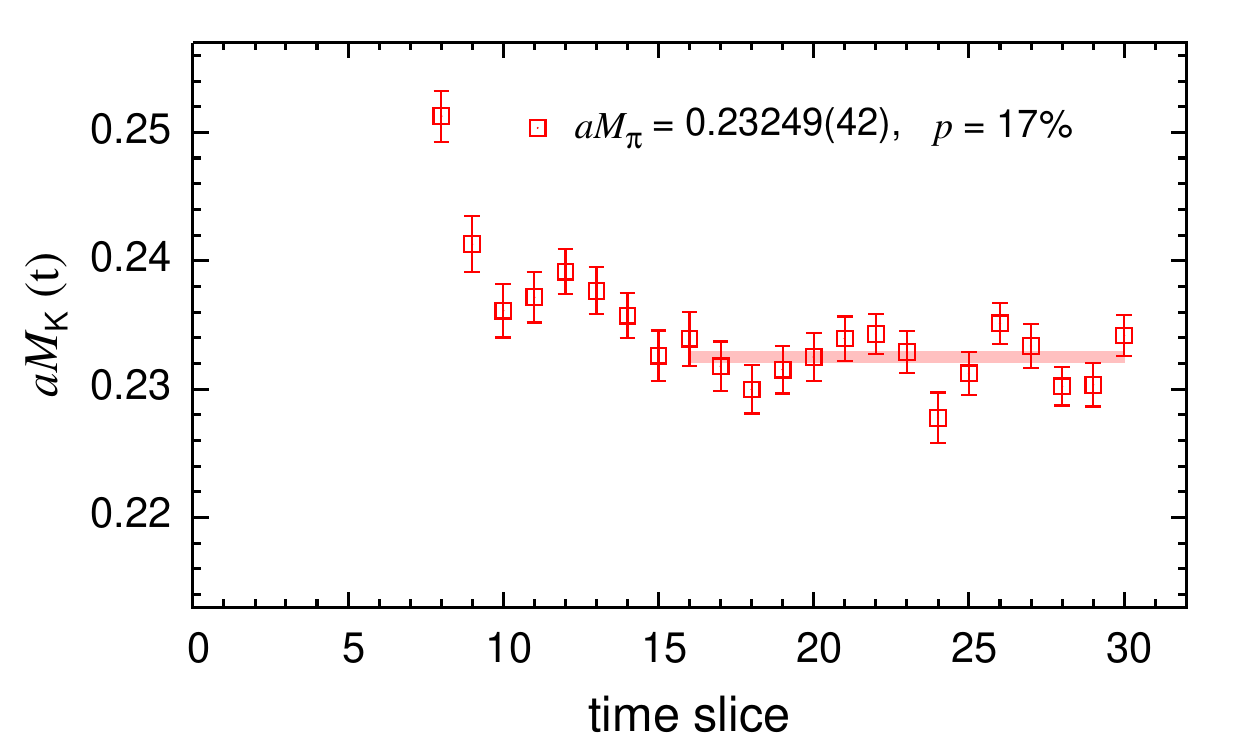}
  \includegraphics[width=.47\textwidth]{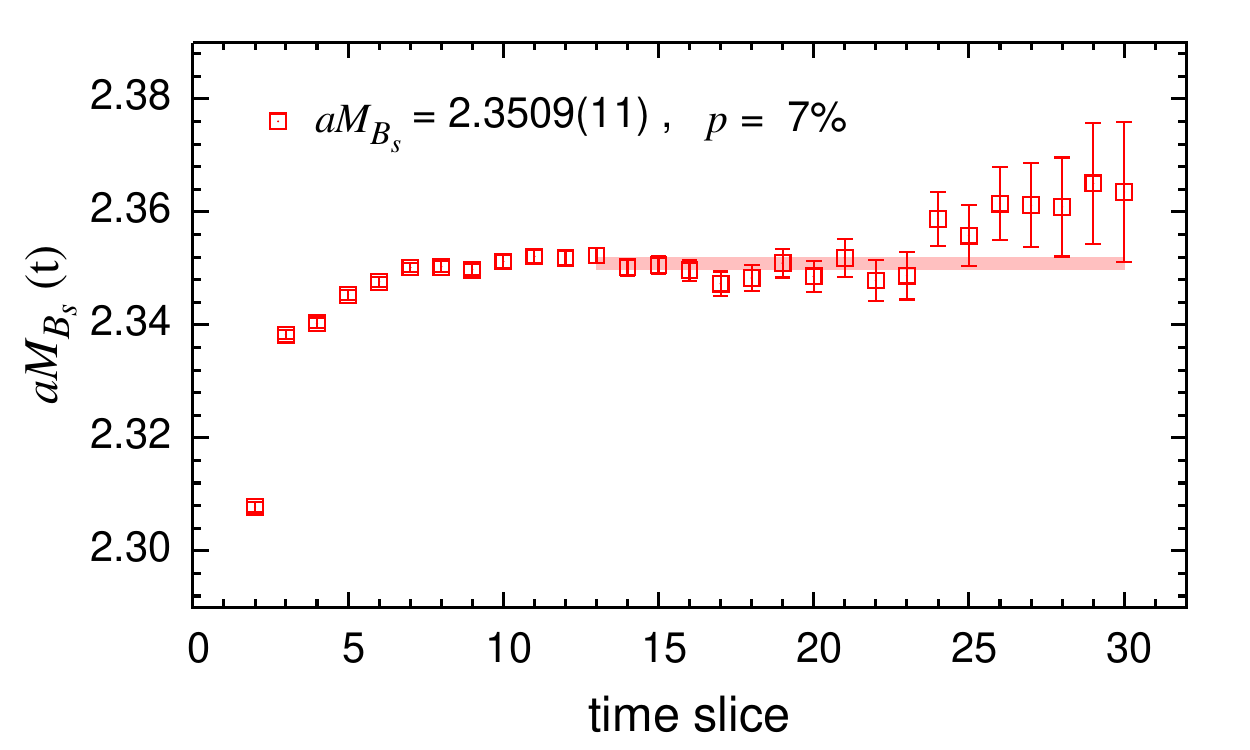}
 \caption{Effective masses of the pion~(upper left), kaon~(bottom left), $B$ meson~(upper right) and $B_s$ meson~(bottom right) on the $a\approx 0.086$~fm ensemble with $am_l = 0.004$.    Shaded bands show the correlated fit results with jackknife statistical errors over the fit ranges used.  All results are shown in lattice units.
}
   \label{fig:2pt}
\end{figure*}

At sufficiently large lattice times, the ground-state masses and energies can be determined from simple two-point correlator ratios.  We define the light pseudoscalar-meson effective energy and $B_{(s)}$-meson effective mass as
\begin{eqnarray}
 E_P(t, \vec{p}_P) &=& \cosh^{-1} \left[ \frac{C_P(t, \vec{p}_P)+C_P(t+2, \vec{p}_P)}{C_P(t+1, \vec{p}_P)} \right] , \quad
\label{eq_mP} \\
   M_{B_{(s)}}(t)  &=&  \cosh^{-1} \left[ \frac{C_{B_{(s)}}(t)+C_{B_{(s)}}(t+2)}{C_{B_{(s)}}(t+1)} \right] .
\label{eq_mB}
\end{eqnarray}
We perform correlated, constant-in-time, fits to these expressions, choosing fit ranges without visible excited-state contamination that lead to acceptable $p$ values.  Figure~\ref{fig:2pt} shows example meson-mass determinations on our fine ensemble with $am_l = 0.004$.  To minimize bias, we use the same fit range for all ensembles with the same lattice spacing (although different for light and heavy-light mesons); these fit ranges are given in Table~\ref{tab:fit_ranges}. The resulting pion/kaon and $B_{(s)}$-meson masses on all ensembles are given in Tables~\ref{tab:data_2pt_pion} and~\ref{tab:data_2pt_B}, respectively.

\begin{figure*}[p] 
  \centering
  \includegraphics[width=.47\textwidth]{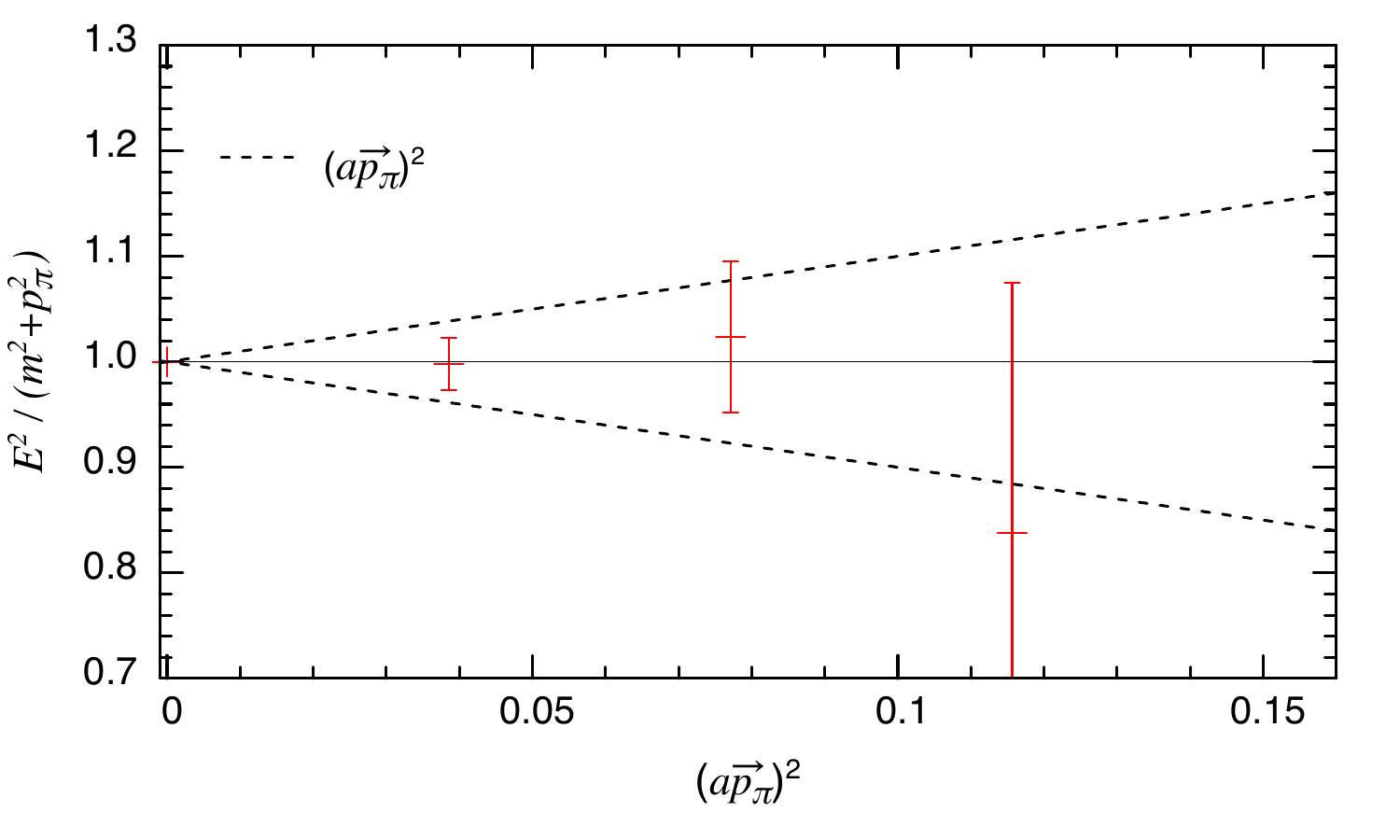}
  \includegraphics[width=.47\textwidth]{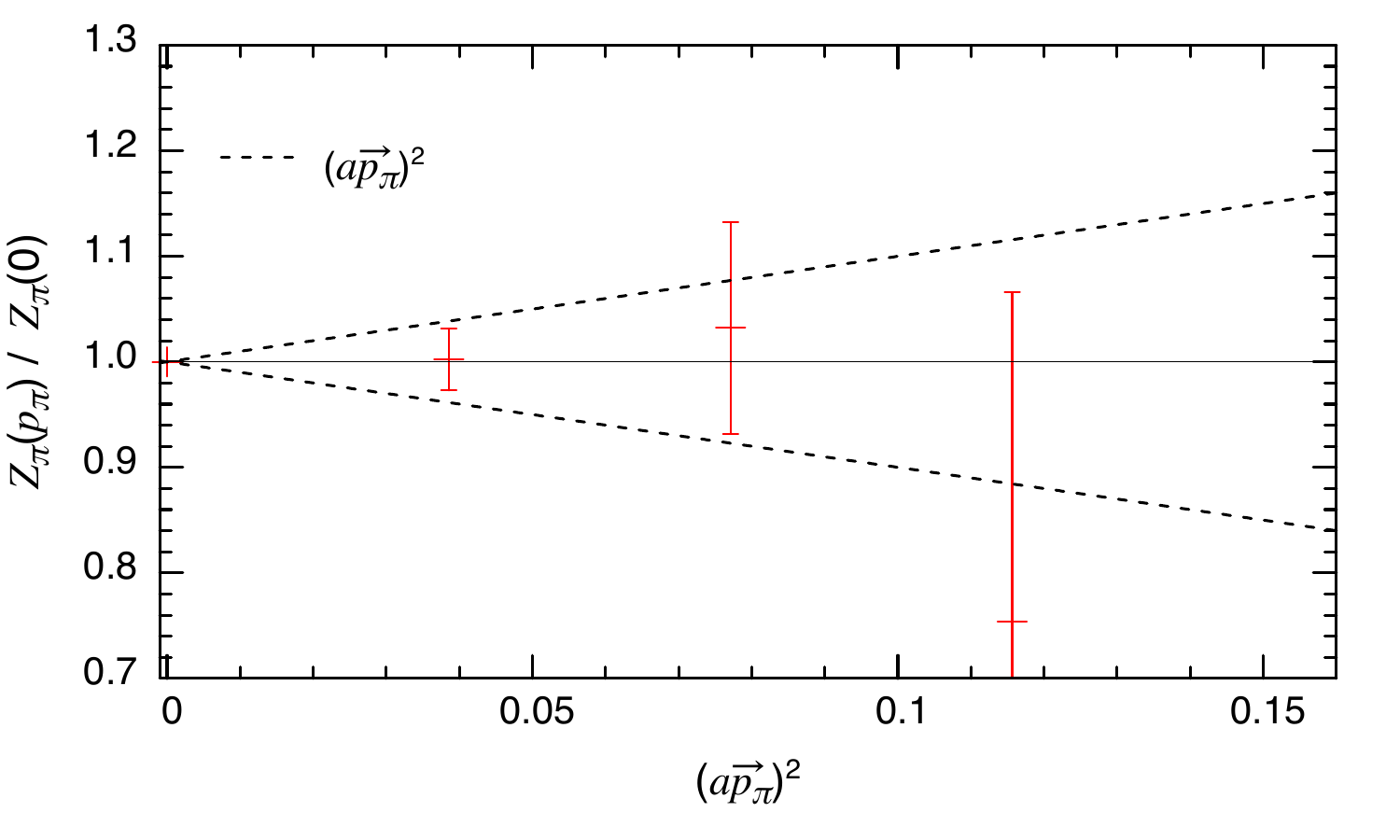}
  \includegraphics[width=.47\textwidth]{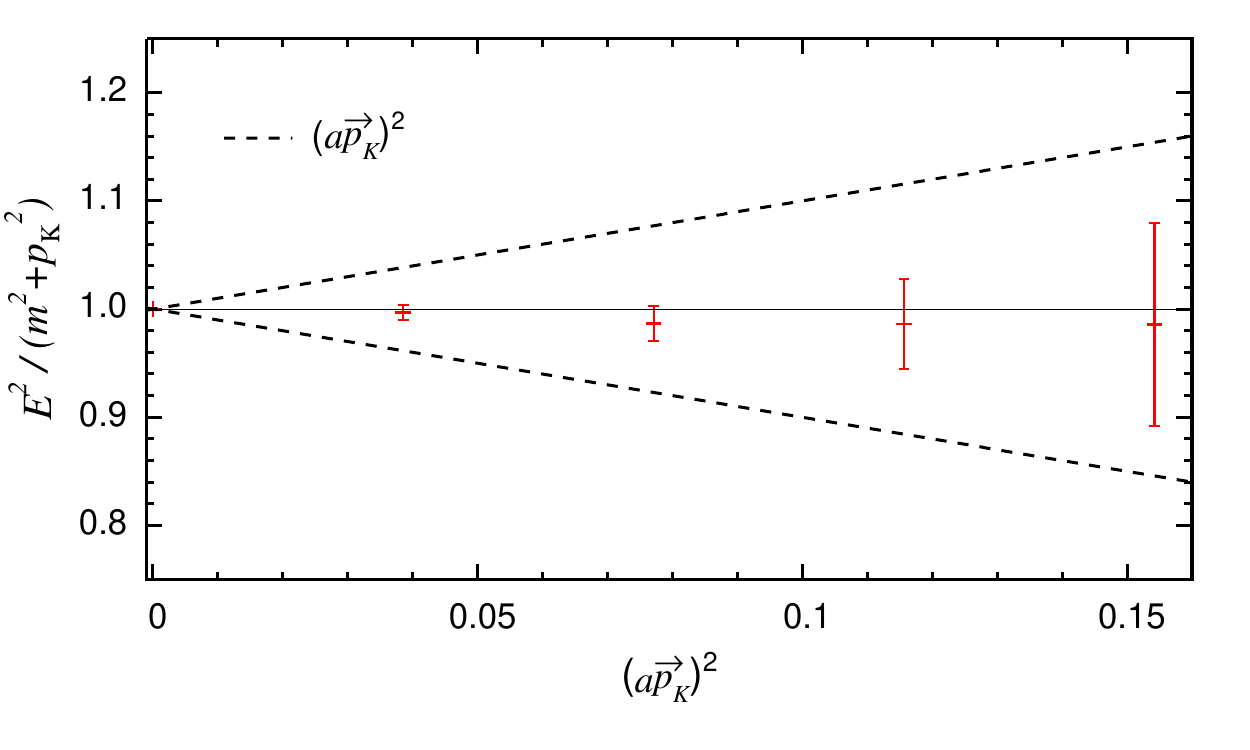}
  \includegraphics[width=.47\textwidth]{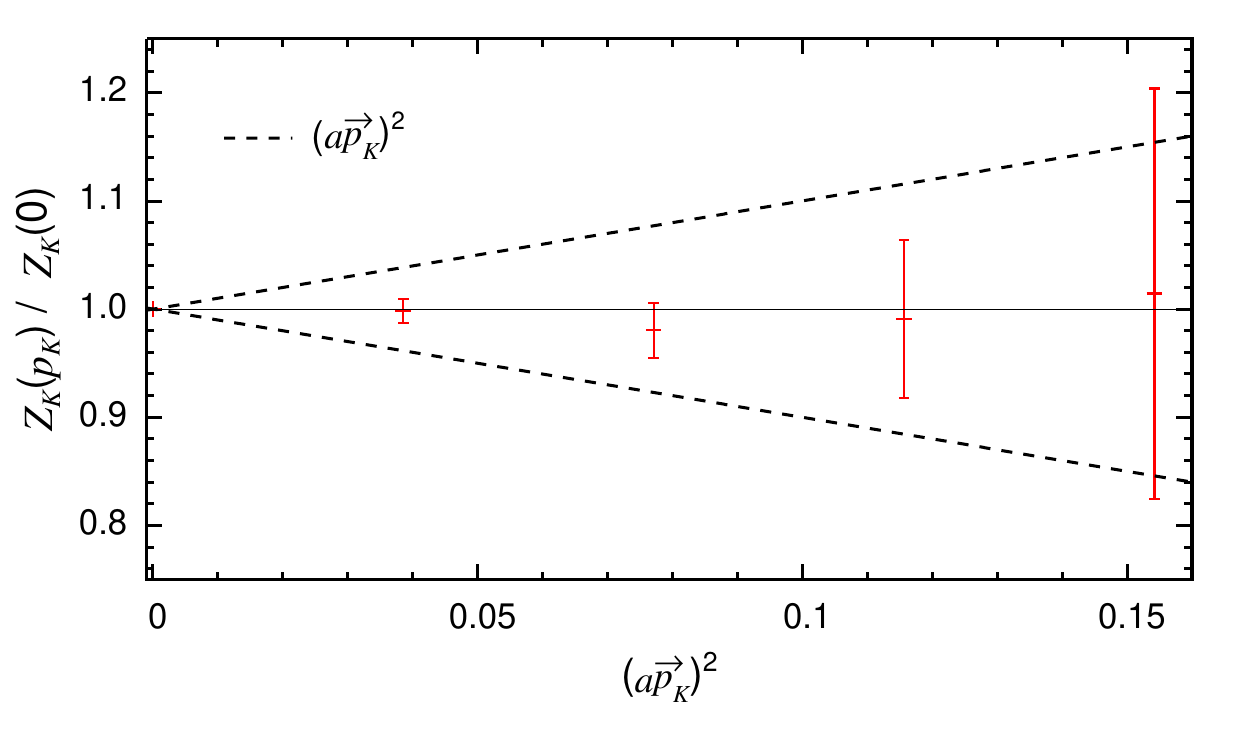}
 \caption{Comparison of pion (top) and kaon (bottom) energies~(left) and amplitudes~(right) with continuum-limit expectations on the $a\approx 0.086$~fm ensemble with $am_l = 0.004$. The dashed lines show a power-counting estimate of the leading $\Ocal((a\vec{p})^2)$ momentum-dependent discretization errors.}
   \label{fig:dispersion}
\end{figure*}

In the continuum limit, the pion and kaon energies should satisfy the dispersion relation $E_P^2 = M_P^2 + \vec{p}_P^2$ and the amplitudes of the two-point functions $Z_P = |\langle 0 | \Ocal_P |\pi\rangle | $ should be independent of the momentum $\vec{p}_P$.  We obtain the amplitudes from correlated plateau fits to 
\begin{equation}
 Z_P(t)=  \sqrt{\frac{2E_P C_P(t, \vec{p}_P)}{e^{-E_Pt}}} \label{eq:ZP}
\end{equation}
using the same fit ranges as for the masses.  Figure~\ref{fig:dispersion} compares the measured pion and kaon energies and amplitudes with continuum expectations on the $a \approx 0.086$~fm, $am_l = 0.004$ ensemble.  The measured kaon energies and amplitudes agree remarkably well with the predictions from the continuum dispersion relation,  to within 5\% even at the largest momentum $a\vec{p}_K= 2\pi(2,0,0)/L$.  Although the pion data is not precise enough to draw strong quantitative conclusions, the measured energies and amplitudes still agree with continuum expectations within the large statistical uncertainties for all momenta.  Dispersion-relation plots for the other ensembles show similar behavior.

The kaon data, for which both the energies and amplitudes are statistically well resolved, provides an accurate measure of momentum-dependent discretization effects, while the pion data provides only a rough cross-check. On all ensembles, the measured pion and kaon energies both agree within statistical errors with the predictions from the continuum dispersion relation, and the measured pion and kaon amplitudes agree with the zero-momentum result.  Thus, in our determinations of the lattice form factors $\fpara$ and $\fperp$ in the next section, we use pion and kaon energies calculated from the continuum dispersion relation (rather than the measured values) to reduce the statistical uncertainties.  Although we do not use the amplitudes obtained from Eq.~(\ref{eq:ZP}) in our subsequent form-factor determinations, the observed momentum independence of $Z_P$ provides further support for this strategy.

\subsection{Three-point correlator fits}
\label{Sec:3pt}

\begin{figure}[tb]
  \centering
  \includegraphics[width=.39\textwidth]{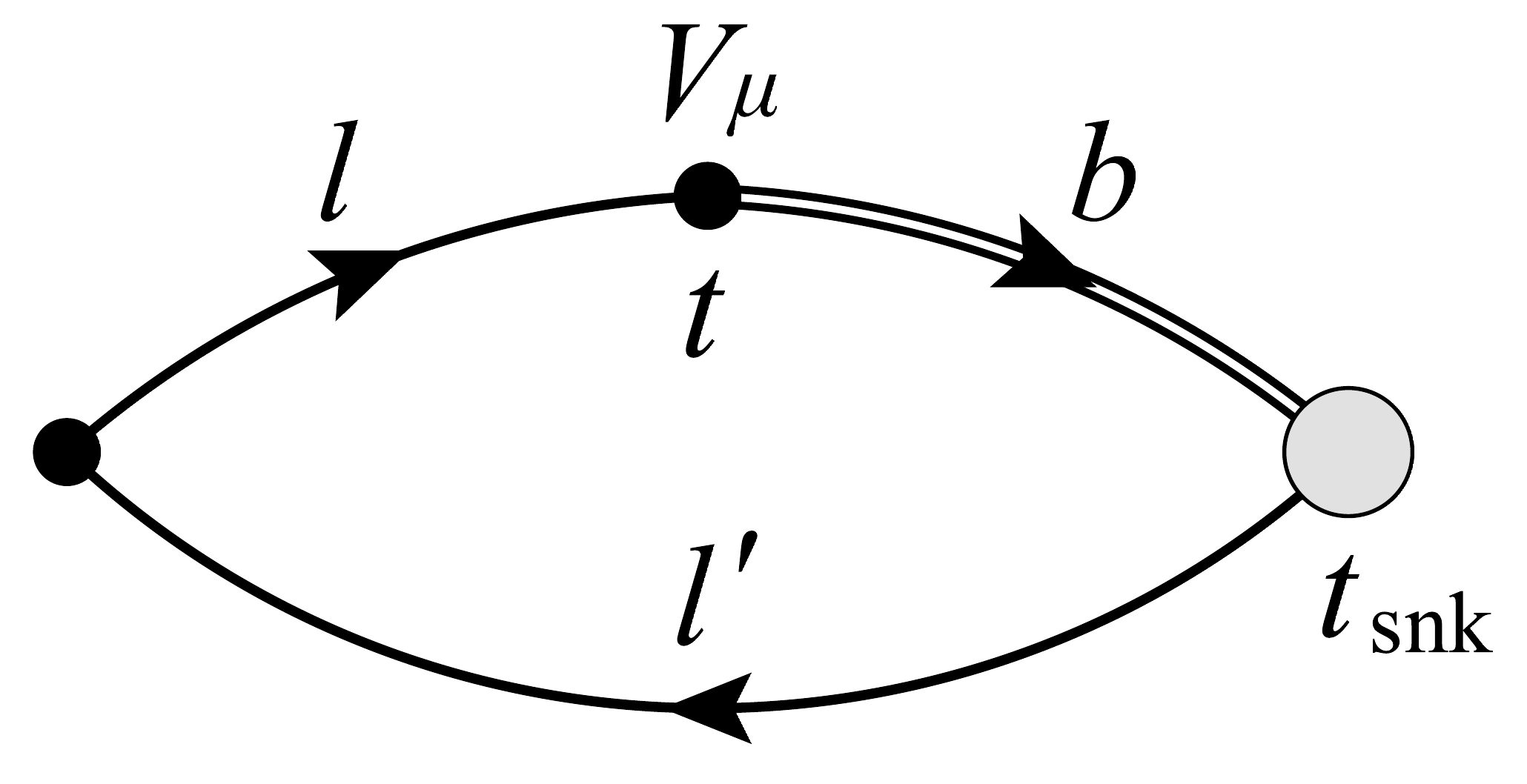}
  \caption{Three-point correlation function used to obtain the $B\to P$ form factors. The single and double lines correspond to light- and $b$-quark propagators, respectively.  For $B\to\pi\ell\nu$, the spectator-quark mass ($m_{l'}$) is the same as the light sea-quark mass, while for $B_s \to K \ell \nu$, the spectator-quark mass is close to the physical $m_s$.  The light daughter-quark mass ($m_l$) is always equal to to the light sea-quark mass.  Black and grey circles denote local and smeared operators, respectively.}
   \label{fig:diagram_3pt}
\end{figure}

To extract the desired $B_{(s)}\to P$ hadronic amplitudes, we calculate the following three point correlation functions:
\begin{multline}
 C_{3,\mu}^{\rm imp}(t,t_{\rm snk}, \vec{p}_P) \\
    =  \sum_{\vec{x},\vec{y}} e^{i\vec{p}_P\cdot \vec{y}} 
        \langle \tilde{\Ocal}_{B_{(s)}}^\dagger(\vec{x}, t_{\rm snk}) V^{\rm imp}_\mu(\vec{y},t)
	  \mathcal{O}_P(\vec{0},0) \rangle,
\end{multline}
where the improved lattice temporal and spatial lattice vector currents $V_\mu^{\rm imp}$ are defined in Eqs.~(\ref{eq:V0_imp}) and~(\ref{eq:Vi_imp}).  As shown in Fig.~\ref{fig:diagram_3pt}, we fix the location of the pion or kaon at the temporal origin and the location of the $B_{(s)}$ meson at time $t_{\rm snk}$, and vary the location of the current operator over all time slices in between.  In our calculations, the mass of the light daughter quark ($l$) is always equal to the light sea-quark mass. For $B\to\pi$ decay, the spectator-quark mass ($l^\prime$) also equals the light sea-quark mass.  For $B_s\to K$ decay, the spectator-quark mass is close to that of the physical strange quark.  We use a Gaussian-smeared sequential source for the $b$ quark in the $B_{(s)}$ meson to reduce excited-state contamination. We insert discrete nonzero momentum at the local current operator through $\vec{p}_\pi = 2\pi(1,1,1)/L$ for $B\to \pi$ and $\vec{p}_K=2\pi(2,0,0)/L$ for $B_s\to K$ (recall that the $B_{(s)}$ meson is at rest).  To improve statistics,  we compute the three-point correlators with both positive and negative source-sink separations ($\pm t_{\rm snk}$); we also average over equivalent spatial momenta.

\begin{figure}[tb]
  \centering
  \includegraphics[width=.48\textwidth]{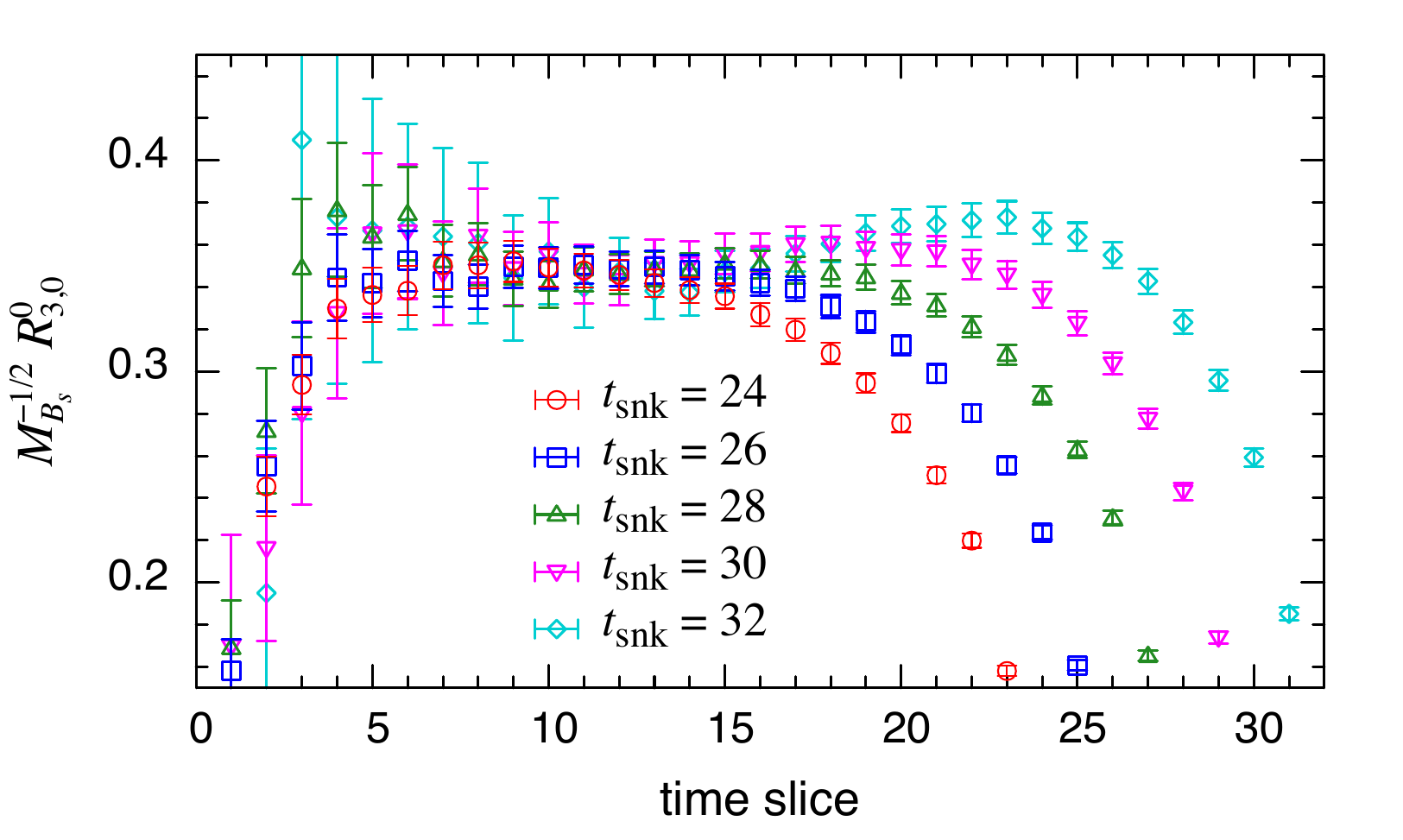}
 \caption{Unimproved three-point ratio $R_{3,0}^0$ for $\vec{p}_\pi = 0$ with several source-sink separations $t_{\rm snk}$
   on the $a \approx 0.086$~fm ensemble with $am_l = 0.004$.}
   \label{fig:3pt_separation}
\end{figure}

The lattice form factors $f_{\parallel}^{\rm lat}$ and $ f_{\perp}^{\rm lat}$ are obtained from the following ratios of correlation functions far away from both the pion/kaon source and the $B_{(s)}$-meson sink:
\begin{eqnarray}
    f^{\rm lat}_{\parallel}(\vec{p}_P) &=& \lim_{0\ll t \ll t_{\rm snk}}R_{3,0}(t,t_{\rm snk},\vec{p}_P), \\
    f^{\rm lat}_{\perp}(\vec{p}_P)     &=& \lim_{0\ll t \ll t_{\rm snk}}\frac{1}{p^i_\pi}R_{3,i}(t,t_{\rm snk},\vec{p}_P),
\end{eqnarray}
with
\begin{multline}
R_{3,\mu}(t,t_{\rm snk}, \vec{p}_P)  \\
= \frac{C_{3,\mu}^{\rm imp}(t,t_{\rm snk},\vec{p}_P)}{\sqrt{C_2^{P}(t,\vec{p}_P) \tilde{C}_2^{B_{(s)}}(t_{\rm snk}-t) }}
   \sqrt{\frac{2E_P}{e^{-E_P t}e^{-M_B(t_{\rm snk} - t)}}},
\label{Eq.ratioR}
\end{multline}
 where we use the continuum dispersion relation and measured light pseudoscalar-meson mass $M_P$ to construct the energy $E_P$.  To determine the optimal source-sink separation for $C_{3,\mu}(t,t_{\rm snk},\vec{p}_P)$, we carried out a dedicated study.  We computed the {\it unimproved}\/ ratio $R_{3,\mu}^0$ for several values of the source-sink separation on one $a\approx 0.086$~fm and one $a\approx 0.11$~fm ensemble, choosing those with the lightest sea-quark mass because they are most sensitive to excited-state contamination.  Figure \ref{fig:3pt_separation} shows the ratio $R_{3,0}^0$ for $B \to \pi$ with $\vec{p}_\pi = 0$ for several source-sink separations on the $a\approx 0.086$~fm ensemble.  All plateaus overlap within statistical uncertainties in the region far from both the source and the sink. The results for the ratios $R_{3,0}^0$ and $R_{3,i}^0$ at nonzero momenta and on the $a\approx 0.11$~fm ensemble look similar.  Because the statistical errors increase with larger source-sink separation, we chose $t_{\rm snk} = 26$ (20) on the $a \approx 0.086$~fm ($a\approx 0.11$~fm) ensembles.  This corresponds to approximately the same physical distance for the two lattice spacings.

\begin{figure*}[tb]
  \centering
  \includegraphics[width=.49\textwidth]{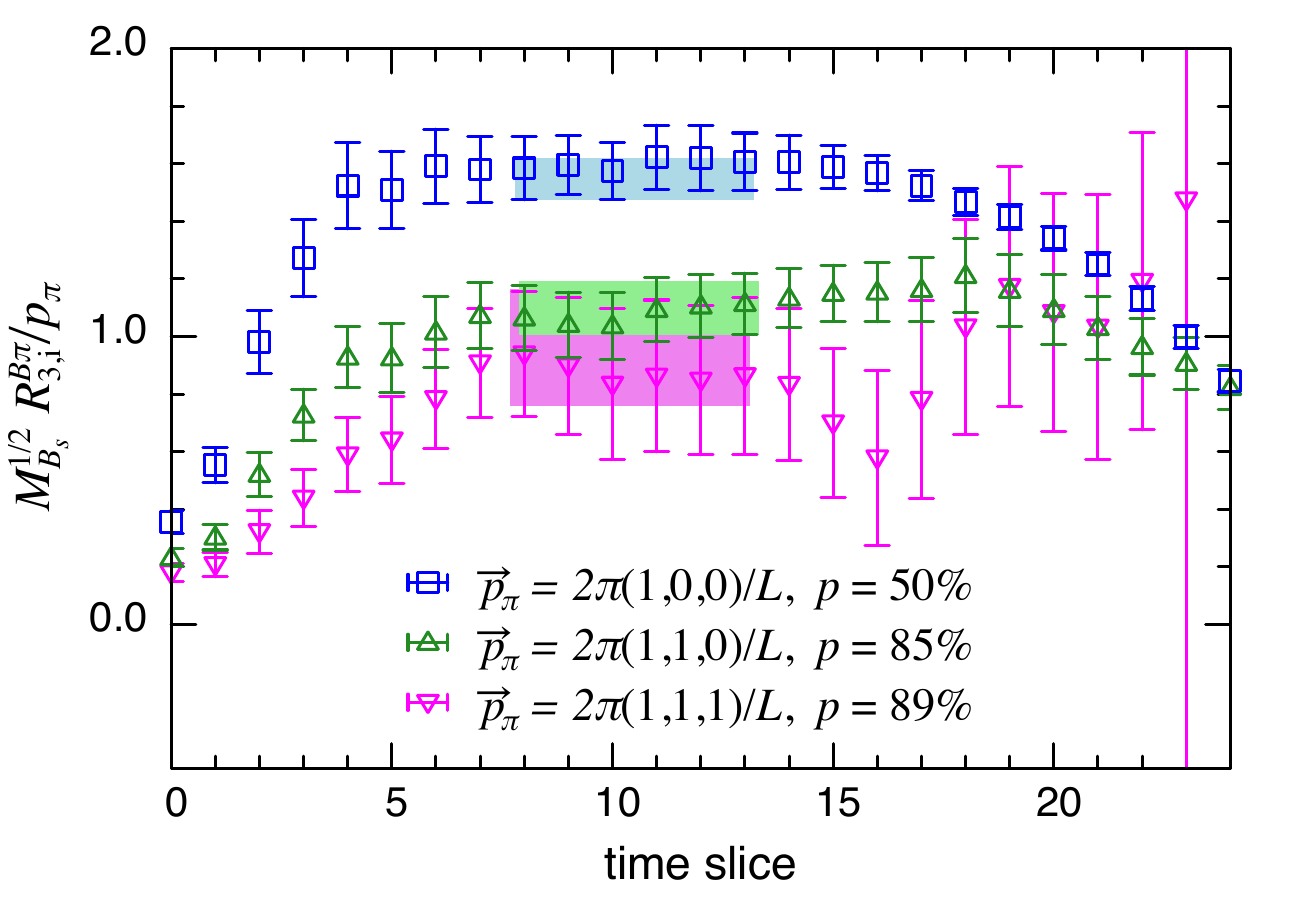}
  \includegraphics[width=.49\textwidth]{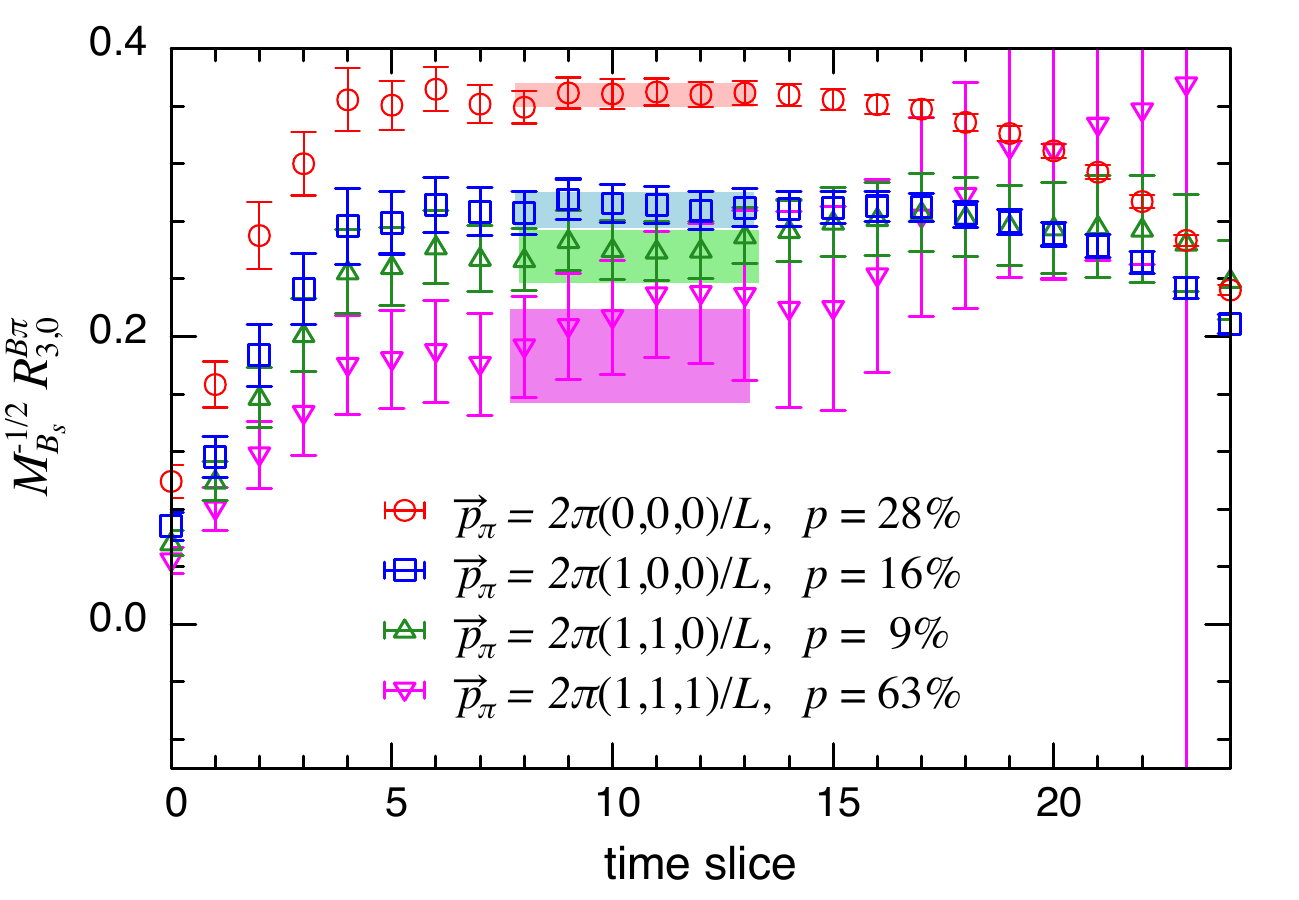}
  \includegraphics[width=.49\textwidth]{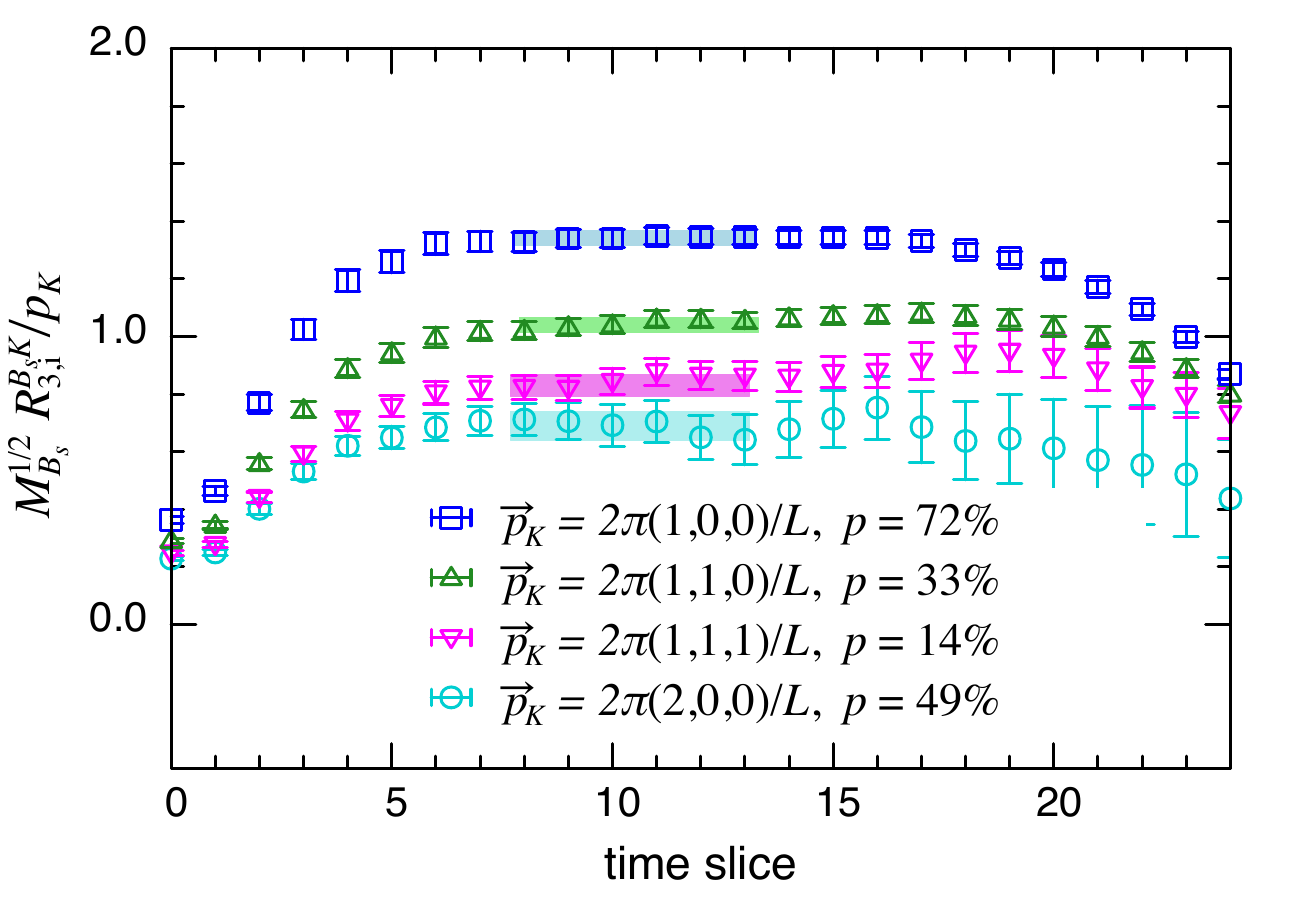}
  \includegraphics[width=.49\textwidth]{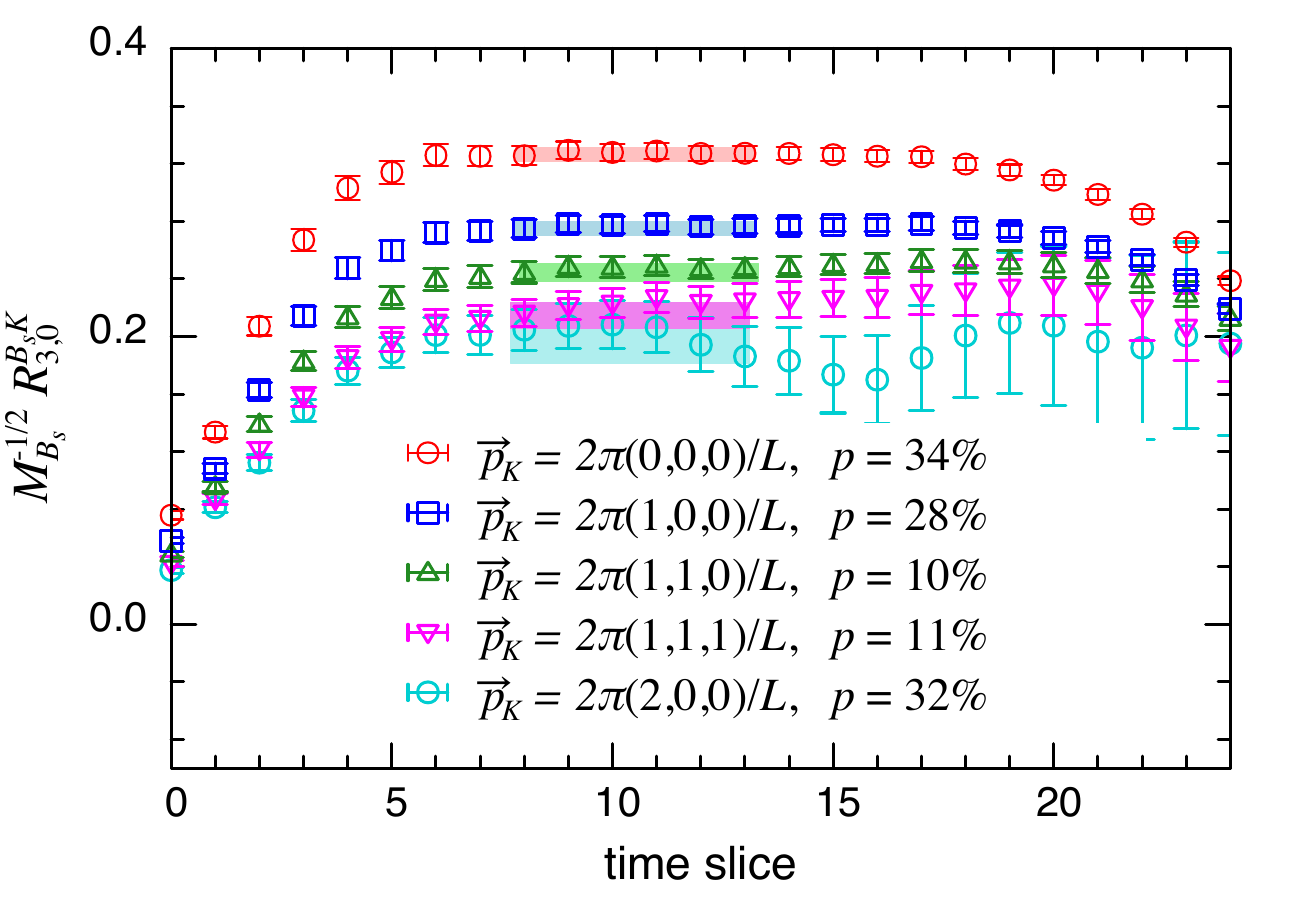}
  \caption{${\mathcal O}(\alpha_s a)$-improved ratios $R_{3,i}/p_{P}^i$ (left) and $R_{3,0}$ (right) with $t_{\rm snk} = 26$ on the $a\approx 0.086$~fm ensemble with $am_l = 0.004$. Plots for $B\to \pi l\nu$ are on the top and $B_s\to K l\nu$ are on the bottom.  Fit ranges and fit results with jackknife statistical errors are shown as horizontal bands.}
   \label{fig:3pt}
\end{figure*}

Figure~\ref{fig:3pt} shows the ${\mathcal O}(\alpha_s a)$-improved ratios $R_{3,0}$  and $R_{3,i}/p_{P}^i$ for different momenta on the $a \approx 0.086$~fm ensemble with $a m_l = 0.004$.  Results for other ensembles look similar.  We perform correlated, constant-in-time, fits to these ratios using fit ranges without visible excited-state contamination that lead to acceptable $p$ values.  To minimize bias, we use the same fit range for all momenta and ensembles with the same lattice spacing; these fit ranges are given in Table~\ref{tab:fit_ranges}.  The complete fit results for the three-point ratios are given in Tables~\ref{tab:data_R_BtoPi} and~\ref{tab:data_R_BstoK}. 

Finally, we obtain the renormalized $B_{(s)}\to P\ell\nu$ form factors $f_{\parallel}$ and $ f_{\perp}$  in the continuum after multiplying by the heavy-light renormalization factors $Z_{V_\mu}^{bl}$ given in Table~\ref{tab:rho}:
\begin{eqnarray}
     f_{\parallel}(\vec{p}_P) &=& Z_{V_0}^{bl}  f^{\rm lat}_{\parallel}(\vec{p}_P) ,\\
     f_{\perp}(\vec{p}_P)     &=& Z_{V_i}^{bl}  f^{\rm lat}_{\perp}(\vec{p}_P).
\end{eqnarray}

\subsection{Chiral-continuum extrapolation}
\label{Sec:Extrapolations}

We extrapolate the renormalized lattice form factors to the physical light-quark mass, and interpolate in the pion or kaon energy using next-to-leading order (NLO) SU(2) chiral perturbation theory for heavy-light mesons (HM$\chi$PT) in the ``hard-pion'' limit.   In the SU(2) theory, the strange-quark mass is integrated out, and only the light-quarks' degrees-of-freedom are included.  Therefore the chiral logarithms for $B \to \pi \ell\nu$ ($B_s \to K \ell \nu$) depend on the pion mass and the pion (kaon) energy.  The SU(2) low-energy constants depend upon the value of the strange-quark mass, as well as on the value of the $b$-quark mass for $B$-meson form factors.  ``Hard-pion'' $\chi$PT, which was introduced by Flynn and Sachrajda for the light-pseudoscalar-meson decay $K\to\pi\ell\nu$ in Ref.~\cite{Flynn:2008tg} and later extended to heavy-light-meson decays by Bijnens and Jemos in Ref.~\cite{Bijnens:2010ws}, applies in the kinematic regime where the pion or kaon energy is large compared to its rest mass.  Almost all of our lattice simulation data is in this hard-pion (or kaon) regime.  We can obtain the expressions for the $B \to \pi \ell\nu$ and $B_s \to K \ell \nu$ form factors in hard-pion/kaon $\chi$PT by taking the limit of the continuum expressions from Ref.~\cite{Becirevic:2002sc} as $M_\pi / E_P \to 0$, where $P = \pi, K$ denotes the final-state pseudoscalar meson.  

The NLO SU(2) $\chi$PT full-QCD expressions for the $B \to \pi \ell \nu$ and $B_s \to K \ell \nu$ form factors in the hard-pion/kaon limit are functions of the pion mass $M_\pi$, pion or kaon energy $E_P$, and lattice spacing $a$.  They have two general forms:
\begin{widetext}
\begin{eqnarray}
 f_{\rm no\  pole}^{B_{(s)}  P}(M_{\pi}, E_P, a^2) &=&  
 c_{\rm np}^{(1)}\left[1+  \left( \frac{\delta f_\parallel}{(4 \pi f)^2}
  + c_{\rm np}^{(2)} \frac{M_{\pi}^2}{\Lambda^2} 
  + c_{\rm np}^{(3)}\frac{E_P}{\Lambda}  + c_{\rm np}^{(4)}\frac{E_P^2}{\Lambda^2}
  + c_{\rm np}^{(5)}\frac{a^2}{\Lambda^2 a_{32}^4}  \right) \right]  \label{eq:fpar_ChPT} \\
 f_{\rm pole}^{B_{(s)} P}(M_{\pi}, E_P, a^2) &=&  
  \frac{1}{E_P+\Delta}
  c_{\rm p}^{(1)}\left[1+ \left( \frac{\delta f_\perp}{(4 \pi f)^2} 
  + c_{\rm p}^{(2)} \frac{M_{\pi}^2}{\Lambda^2}  
  + c_{\rm p}^{(3)}\frac{E_P}{\Lambda}  + c_{\rm p}^{(4)}\frac{E_P^2}{\Lambda^2} 
  + c_{\rm p}^{(5)}\frac{a^2}{\Lambda^2 a_{32}^4} \right)  \right] \label{eq:fperp_ChPT}  \,,
\end{eqnarray}
\end{widetext}
one with a pole at $E_P = - \Delta = M_{B^*} - M_{B_{(s)}}$ and one without.  Here the $B^*$ resonance corresponds to a state with flavor $bu$ and quantum numbers $J^P = 0^+$ for $f_\parallel$ and $1^-$ for $f_\perp$.  The experimentally measured vector-meson mass is $M_{B^*}=5.3252(4)$~GeV~\cite{Beringer:2012zz}.  The scalar $B^*$ meson has not been observed experimentally, but its value has been estimated theoretically using heavy-quark and chiral-symmetry arguments to be $M_{B^*}(0^+)=5.63(4)$~GeV~\cite{Bardeen:2003kt}, while the $0^+$-$0^-$ splitting has been estimated in (2+1)-flavor lattice QCD to be $M_{B^*}(0^+)-M_B \sim 400$~MeV~\cite{Gregory:2010gm}.  In our chiral-continuum extrapolations we include the effects of resonances below the $B\pi$ and $B_sK$ production thresholds, {\it i.e.} $q^2 < (M_{B_{(s)}} + M_P)^2$.  For $B \to\pi\ell\nu$, the $B^*$ meson lies below the $B\pi$ production threshold, so we include a pole in the fit for $f_\perp^{B \pi}$ taking $\Delta_\perp^{B  \pi} = 45.78$~MeV from experiment~\cite{Beringer:2012zz}.  The predicted value of $M_{B^*}(0^+)$ is well above $M_B + M_{\pi}$, however, so we do not include a pole in the fit of $f_\parallel^{B\pi}$.  For $B_s \to K$, both $M_{B^*}$ and $M_{B^*}(0^+)$ are below $M_{B_s} + M_K$, so we include a pole in the fits for both $f_\perp^{B_s  K}$ and $f_\parallel^{B_s  K}$, taking $\Delta_\perp^{B_s  K} = -41.6$~MeV from experiment~\cite{Beringer:2012zz} and taking $\Delta_\parallel^{B_s  K} = 263$~MeV from the model estimate in Ref.~\cite{Bardeen:2003kt}.  The precise value of $M_{B^*}(0^+)$ has little impact on the fit because the pole location is so far outside the semileptonic region, but we vary its value by a generous amount when estimating the chiral-continuum extrapolation error in Sec.~\ref{Sec:ChiPT}.

The one-loop chiral logarithms are the same for $f_\parallel$ and $f_\perp$, but differ for $B \to \pi \ell \nu$ and $B_s \to K \ell \nu$:
\begin{eqnarray}
	 \delta f^{B  \pi} & = &  -\frac{3}{4} \left( 3 g_b^2 + 1 \right) M_\pi^2 {\rm log} \left( \frac{M_\pi^2}{\Lambda^2} \right) \label{eq:chpt_na_btopi}\\
	 \delta f^{B_s  K} & = &  -\frac{3}{4} M_\pi^2 {\rm log} \left( \frac{M_\pi^2}{\Lambda^2} \right) \,, \label{eq:chpt_na_bstok}
\end{eqnarray}
where $g_b$ is the $B^\ast B \pi$ coupling constant.  At tree level, the mass of a pion composed of two domain-wall quarks is given in terms of the light-quark mass by
\begin{align}
M_\pi^2 = 2 \mu(m_l+ \mres) \,,
\label{eq:MPiSq}
\end{align}
where $\mu$ is a leading-order low-energy constant.

We include a term proportional to $a^2$ in the chiral fit functions Eqs.~(\ref{eq:fpar_ChPT}) and~(\ref{eq:fperp_ChPT}) to account for the dominant lattice-spacing dependence.  To make the $a^2$ analytic term dimensionless with an expected coefficient of ${\mathcal O}(1)$ in $\chi$PT, we normalize it using the lattice spacing on the finer  $32^3$ ensembles $a_{32}$.  Discretization errors from the domain-wall and Iwasaki actions are of ${\mathcal O}\left(a \Lambda_{\rm QCD} \right)^2$; using $\Lambda_{\rm QCD} = 500$~MeV,\footnote{Recent three- and four-flavor lattice-QCD calculations typically give values for $\Lambda_{\bar{{\rm MS}}}$ in the range of about 300--400~MeV~\cite{Maltman:2008bx,Aoki:2009tf,McNeile:2010ji,Bazavov:2012ka,Blossier:2013ioa}.  The 2013 Flavor Lattice Averaging Group (FLAG) review quotes the range $\Lambda^{(3)}_{\rm{\bar{MS}}} = 339(17)$~MeV for three active flavors~\cite{Aoki:2013ldr}.  To be conservative, we take a slightly larger value $\Lambda_{\rm QCD} = 500$~MeV for the power-counting estimates throughout this work.} we estimate these to be about 5\% on the $32^3$ ensembles.   The remaining discretization errors --  light-quark and gluon discretization errors in the heavy-light current, and heavy-quark discretization errors from both the action and current -- are expected from power counting to be much smaller.  In Secs.~\ref{Sec:HeavyQuarkDiscErrors} and~\ref{Sec:LightQuarkDiscErrors}, we estimate their sizes to be below 2\%.   We therefore expect light-quark and gluon discretization errors from the action to dominate the scaling behavior of the form factors, such that including an $a^2$ term in the fit will largely remove these contributions.  We will add the remaining subdominant discretization errors {\it a posteriori} to the systematic error budget after the chiral fit.

In addition to the pion masses and pion/kaon energies, several parameters enter the expressions in Eqs.~(\ref{eq:fpar_ChPT}) and~(\ref{eq:fperp_ChPT}).   For completeness, we compile the values of the fixed parameters in our chiral fits in Table~\ref{Tab:ChiPTconst}.  We use the lattice spacings and low-energy constant $\mu$ obtained in Ref.~\cite{Aoki:2010dy} from the RBC/UKQCD analysis of light pseudoscalar meson masses and decay constants.  We use the PDG value of $f_\pi = 130.4(2)$~MeV~\cite{Beringer:2012zz}, and take $\Lambda_\chi = 1$~GeV for the scale in the chiral logarithms.   We use the $B^\ast B \pi$ coupling constant $g_b = 0.57(8)$ obtained in our companion analysis also using the RBC/UKQCD domain-wall+Iwasaki ensembles and the RHQ action for the $b$-quarks~\cite{Flynn:2013kwa}. 

\begin{table}[tb]
\caption{Constants used in the chiral and continuum extrapolations of the $B \to \pi \ell \nu$ and $B_s \to K \ell \nu$ form factors~\cite{Aoki:2010dy,Beringer:2012zz,Flynn:2013kwa}. }
\label{Tab:ChiPTconst}
\begin{tabular}{ccc}
\hline\hline
 $a$ & $\approx 0.11$~fm & $\approx 0.086$~fm \\ \hline
$a^{-1}$ & 1.729~GeV & 2.281~GeV \\
$a\mu$ & 2.348 & 1.826\\
$f_\pi$ &\multicolumn{2}{c}{130.4 MeV}\\
$g_b$ &\multicolumn{2}{c}{0.57}\\
$\Lambda_\chi$ &\multicolumn{2}{c}{1~GeV}\\
\hline\hline
\end{tabular}
\end{table}

We perform correlated chiral-continuum fits to the data calculated on all five sea-quark ensembles listed in Table~\ref{tab:lattices} using the full-QCD NLO SU(2) hard-pion/kaon HM$\chi$PT expressions.  For $B\to\pi\ell\nu$, we include discrete lattice momenta up to $\vec{p_\pi}=2\pi(1,1,1)/L$, which corresponds to $\approx 0.78$~GeV on the coarser ensembles.  
For $B_s\to K\ell\nu$, where the statistical errors are smaller, we include momenta up to $\vec{p}_{K}=2\pi(2,0,0)/L$, or $\approx 0.91$~GeV on the coarser ensembles.  
For the pion masses in Eqs.~(\ref{eq:fpar_ChPT}) and~(\ref{eq:fperp_ChPT}), we use the tree-level expression in Eq.~(\ref{eq:MPiSq}).  We obtain the physical form factors after the chiral-continuum fit by setting the light quark mass to the physical average $u/d$-quark mass $a_{32}m_{ud} = 0.00102(5)$~\cite{Aoki:2010dy} and the lattice spacing to zero. 

\begin{figure*}[tb]
   \includegraphics[width=.49\textwidth]{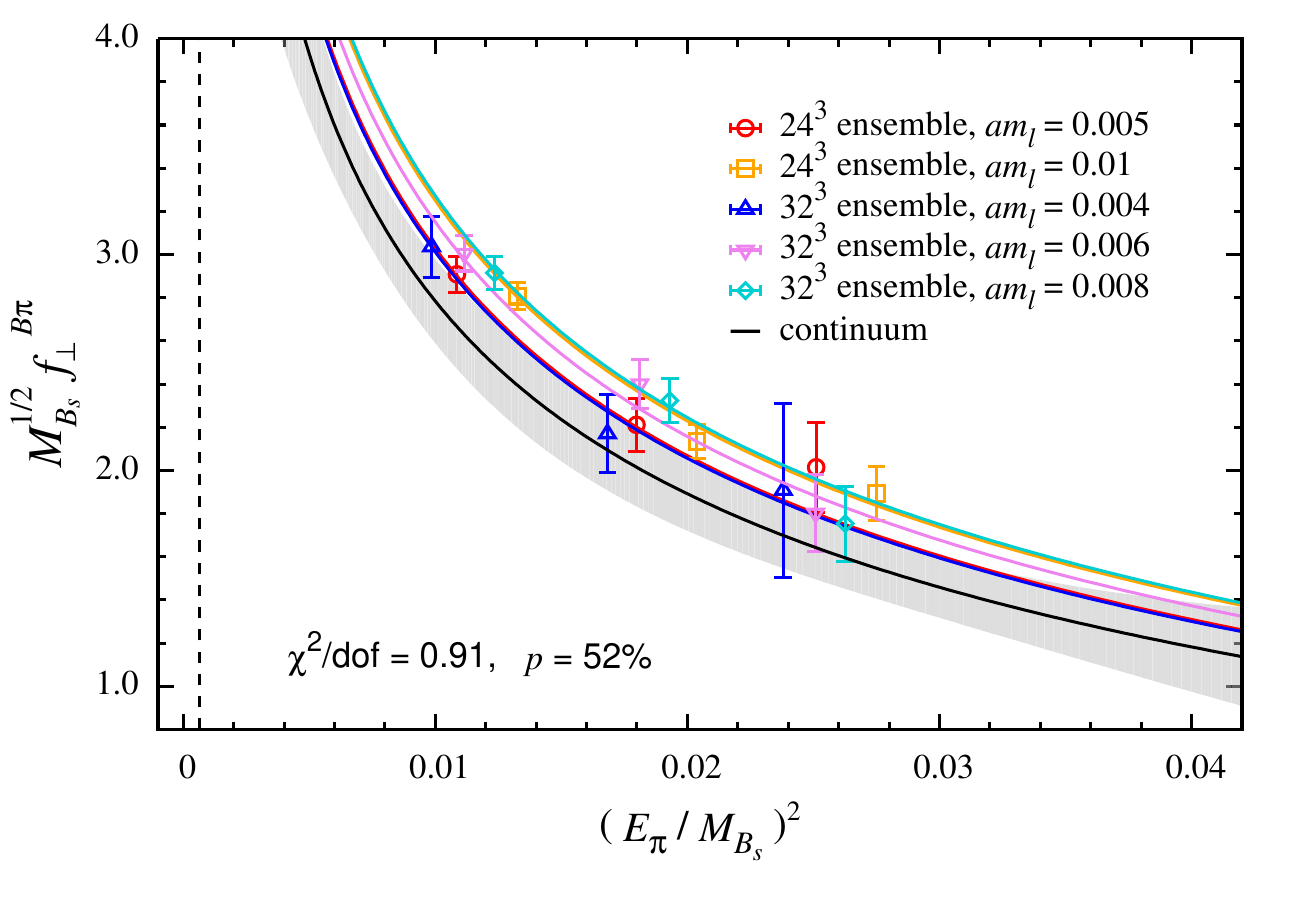}
   \includegraphics[width=.49\textwidth]{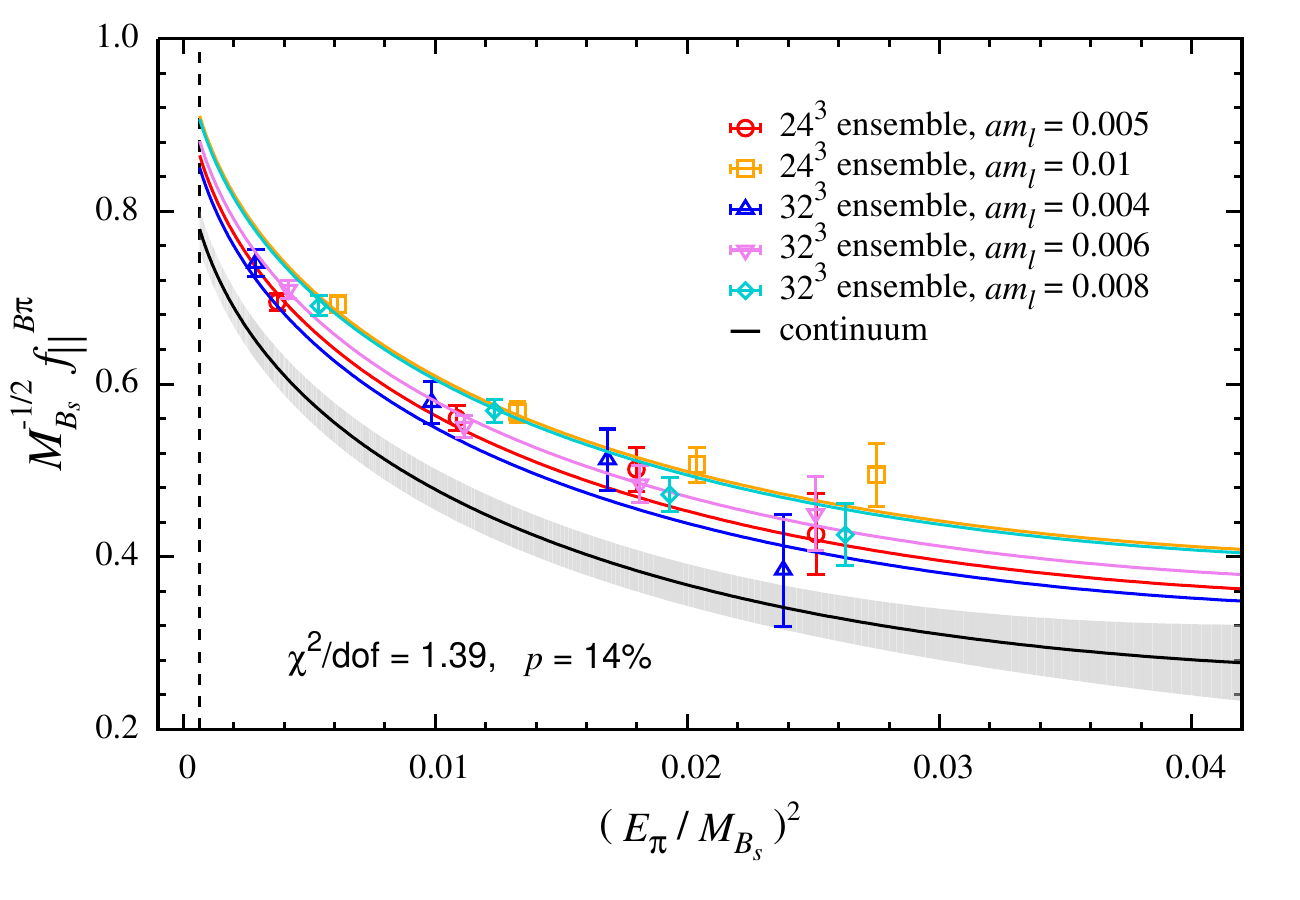}
   \includegraphics[width=.49\textwidth]{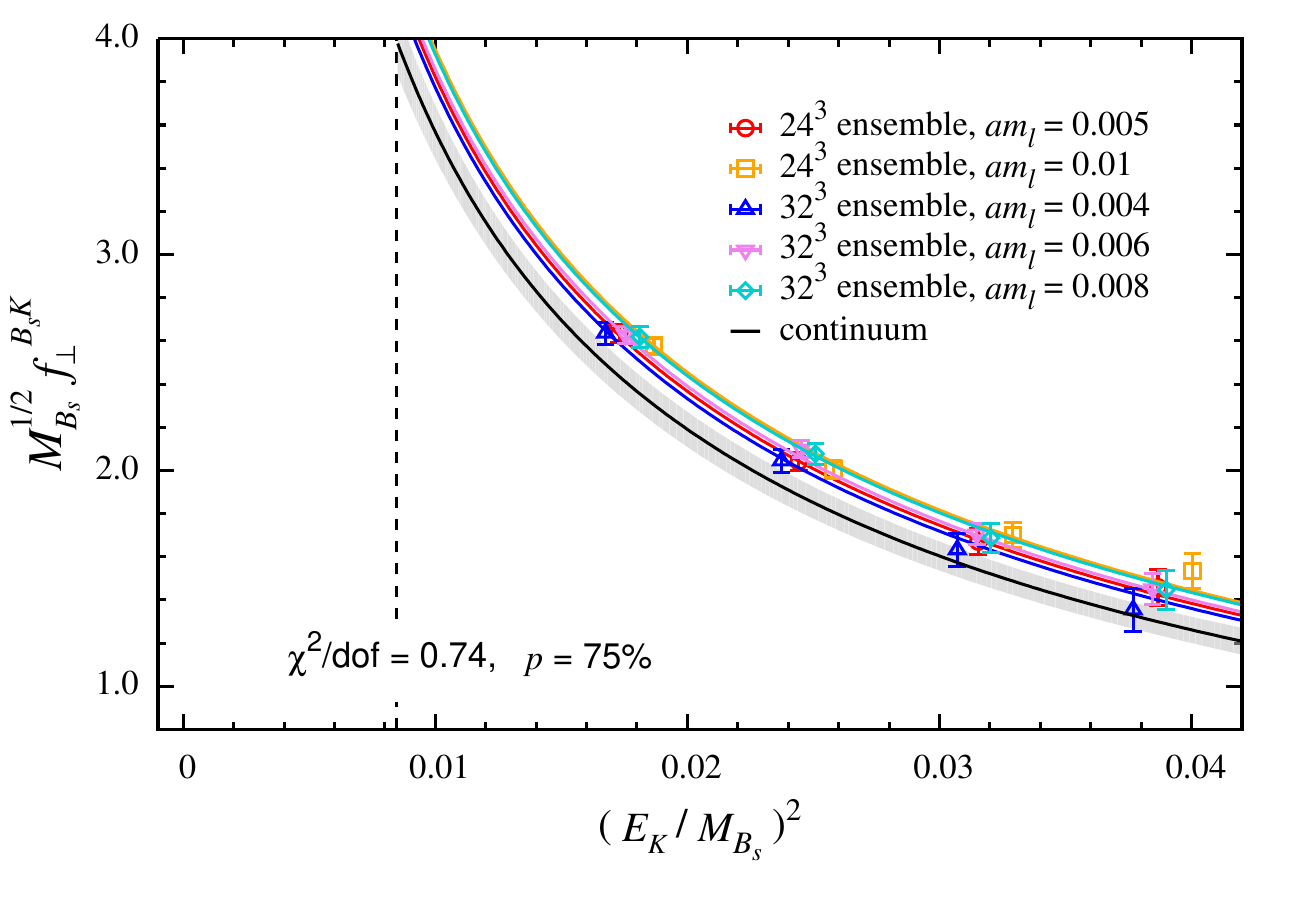}
   \includegraphics[width=.49\textwidth]{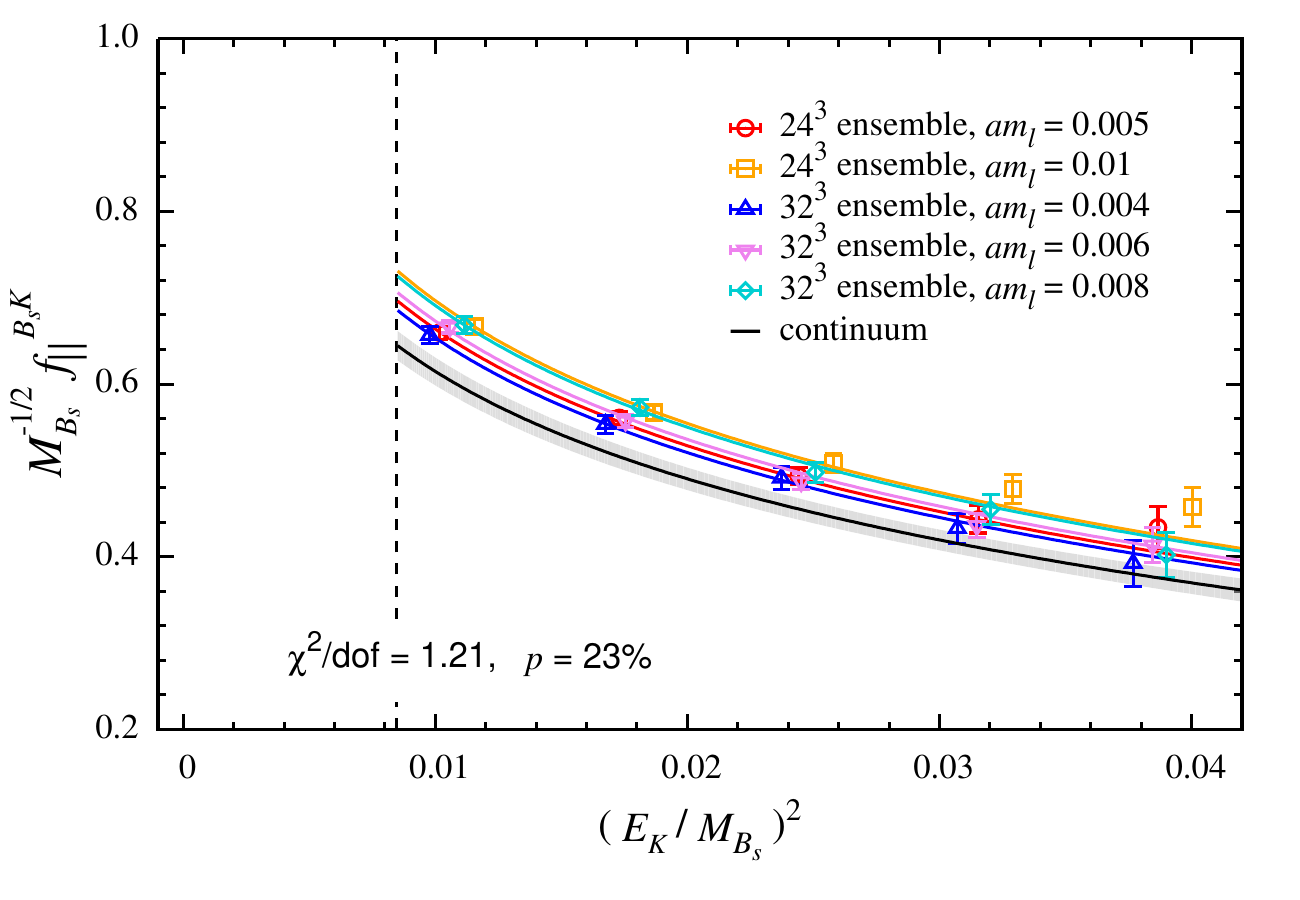}
\caption{Chiral-continuum extrapolation of the $B\to\pi\ell\nu$ (upper plots) and $B_s\to K \ell\nu$ (lower plots) form factors from correlated fits using NLO SU(2) hard-pion/kaon HM$\chi$PT.  Fits of $f_\perp$ are on the left and of $f_\parallel$ are on the right.  In each plot, the colors distinguish between data points on the five different ensembles:  circles and squares correspond to the $a \approx 0.11$~fm data and triangles and diamonds to $a\approx 0.086$~fm data.  The colored fit curves show the interpolation/extrapolation in pion/kaon energy:  the fit function is evaluated at the unphysical sea-quark masses and nonzero lattice spacings on the different ensembles, such that the curves should go through the data points of the same color.  The continuum, physical-quark-mass form factors are shown as a function of pion/kaon energy by the black lines with gray error band.  The vertical dashed line on the left-hand side of each plot shows the physical pion or kaon mass.}
\label{fig:ChiralFits}
\end{figure*}

Figure~\ref{fig:ChiralFits} shows the resulting fits, which all have good $\chi^2/{\rm dof}$ and $p$-values.  We do not observe any statistically-significant lattice-spacing dependence for any of the form factors, and cannot resolve the coefficients of the $a^2$ terms in the four fits.  Dropping the $a^2$ term altogether does not reduce the fit quality, and we consider this alternate fit as one of many possibilities when estimating the systematic uncertainty due to the chiral-continuum extrapolation in Sec.~\ref{Sec:ChiPT}.  We observe a mild sea-quark mass dependence for $f_\parallel^{B_s  K}$, and cannot resolve any sea-quark mass dependence in the other form factors.  Dropping the term proportional to $M_\pi^2$ reduces the $p$-value of the $f_\parallel^{B_s  K}$ to $\sim 5\%$, which is still acceptable, and does not impact the quality of the other fits.  Again, we consider this alternative when estimating the chiral-continuum extrapolation error.  Finally, we do not see any evidence for the onset of chiral logarithms given that our lightest pion $M_\pi \approx 290$~MeV is still quite heavy, and consider fits without the logarithms in Eqs.~(\ref{eq:fpar_ChPT}) and~(\ref{eq:fperp_ChPT}) among the alternate fits for assessing the systematic uncertainty.  

As a consistency check of our chiral-continuum extrapolation, we can use our $B\to\pi\ell\nu$ form-factor fit results to obtain a rough estimate for the $B^\ast B \pi$ coupling at lowest order in the $1/m_b$ expansion of HM$\chi$PT.  From our preferred fits of $f_\parallel$ and $f_\perp$, we find that the ratio of leading-order coefficients gives 
\begin{eqnarray}
	g_{b} \approx c^{(1)}_{\rm p} / c^{(1)}_{\rm np} = 0.35(15) \, \label{eq:gbLO}
\end{eqnarray}
where the error is statistical only (and does not include omitted higher-order corrections in the chiral and $1/m_b$ expansions).  The value for $g_b$ in Eq.~(\ref{eq:gbLO}) is consistent with our independent determination of the $B^\ast B \pi$ coupling in Ref.~\cite{Flynn:2013kwa}, and mostly independent of the input value of $g_b$ in the chiral logarithms.  

We also considered chiral-continuum extrapolations of the $B\to\pi\ell\nu$ form factors using NLO SU(3) HM$\chi$PT, in which the logarithms have explicit strange-quark mass dependence, but were unable to obtain good fits for $f_\parallel^{B\pi}$.  Fits of $f_\parallel^{B\pi}$ to NLO SU(2) ``soft-pion'' $\chi$PT, in which the logarithms have explicit dependence on the pion energy, also failed to describe the data.  All fits tried led to acceptable $p$-values for the case of $f_\perp^{B\pi}$ due to the fact that the shape is largely dictated by the $B^*$ pole term in the denominator.  Finally, we tried supplementing the NLO expressions for the $B\to\pi\ell\nu$ form factors with NNLO analytic terms.  The resulting partly NNLO fits yielded form-factor results consistent with those from our preferred fits, but with significantly larger uncertainties due to the fact our data could not resolve any of the higher-order terms.

\section{Estimation of systematic errors}
\label{Sec:SysErrors}

\begin{figure*}[tb]
  \includegraphics[width=.49\textwidth]{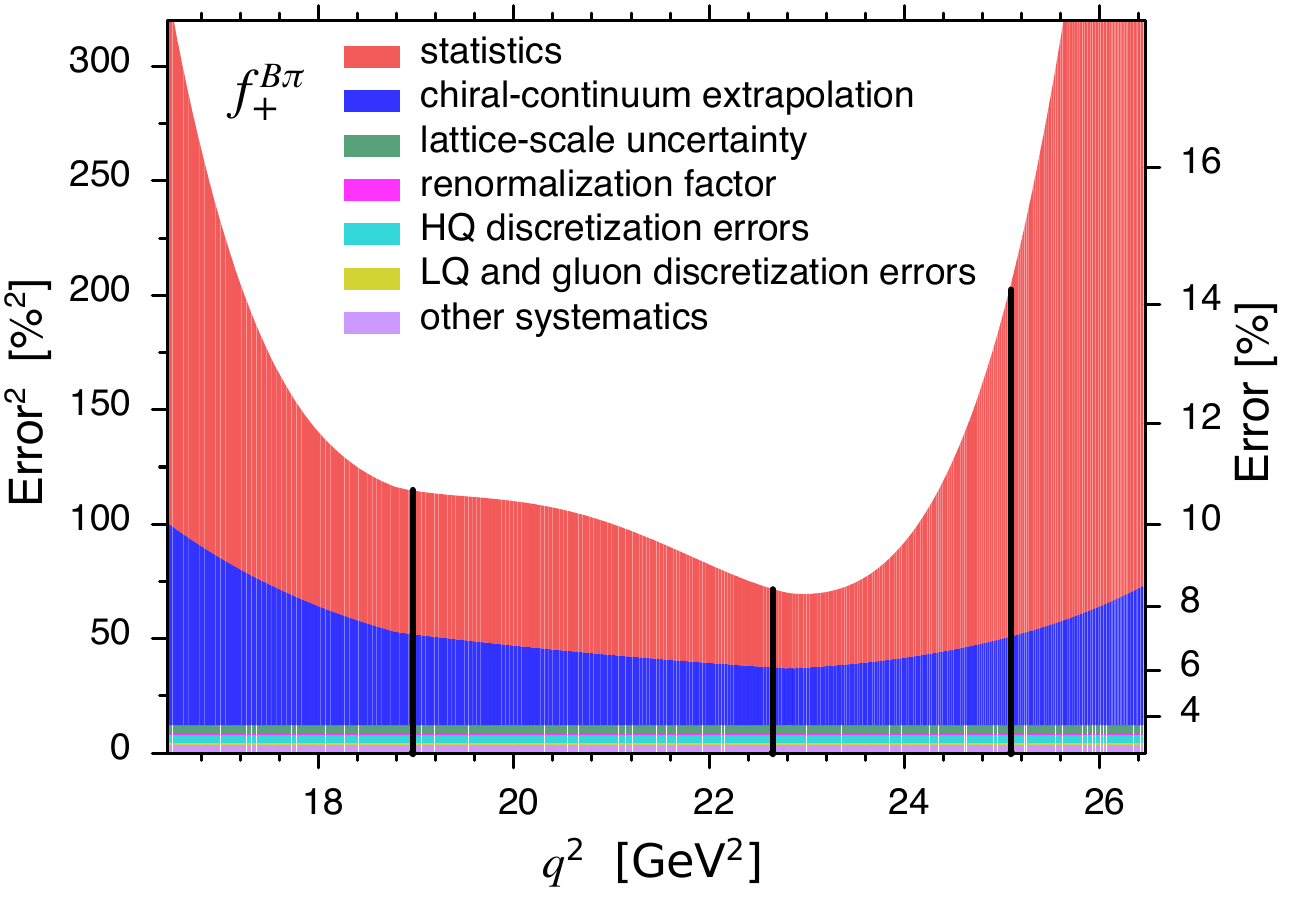}
  \includegraphics[width=.49\textwidth]{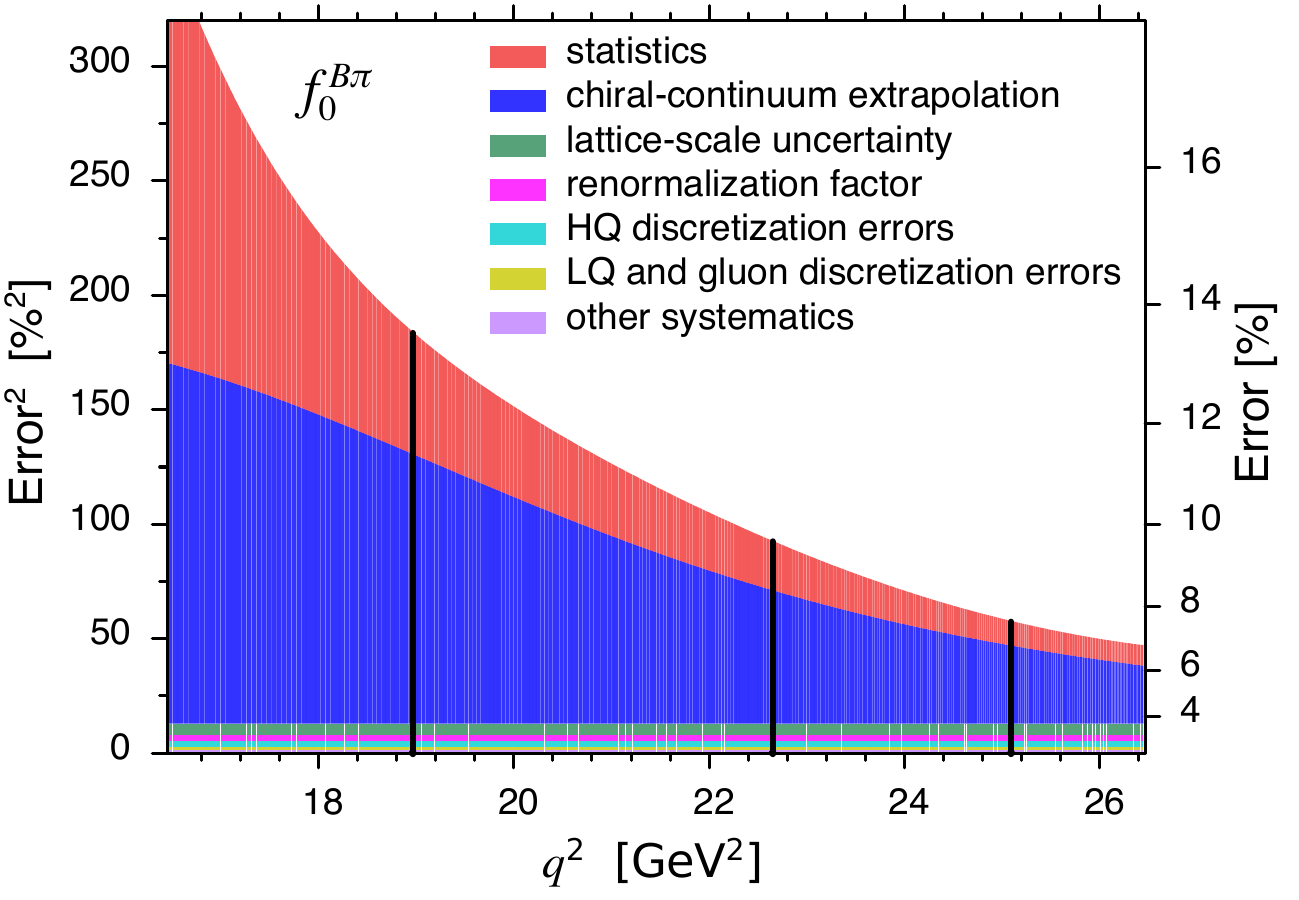}
  \includegraphics[width=.49\textwidth]{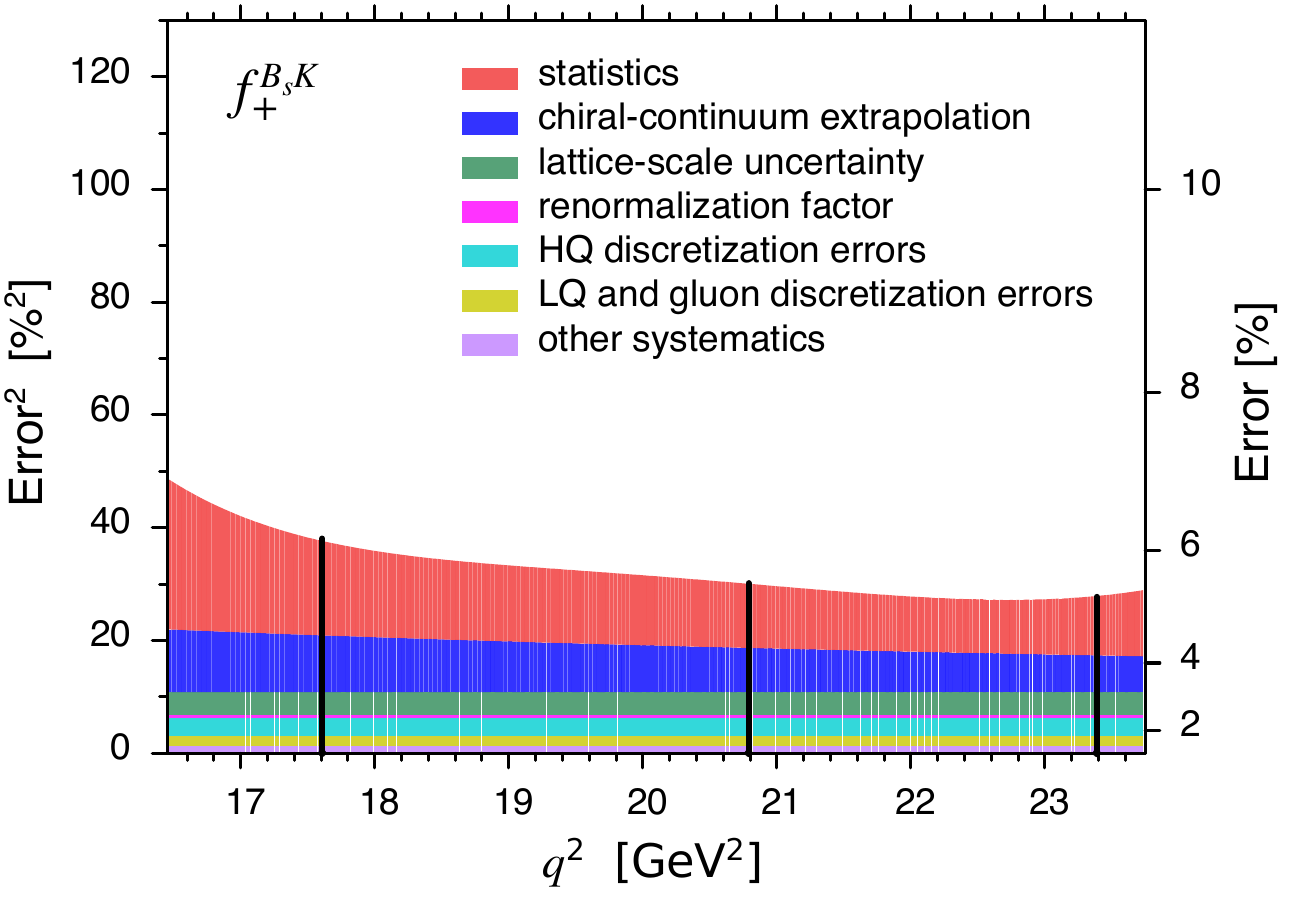}
  \includegraphics[width=.49\textwidth]{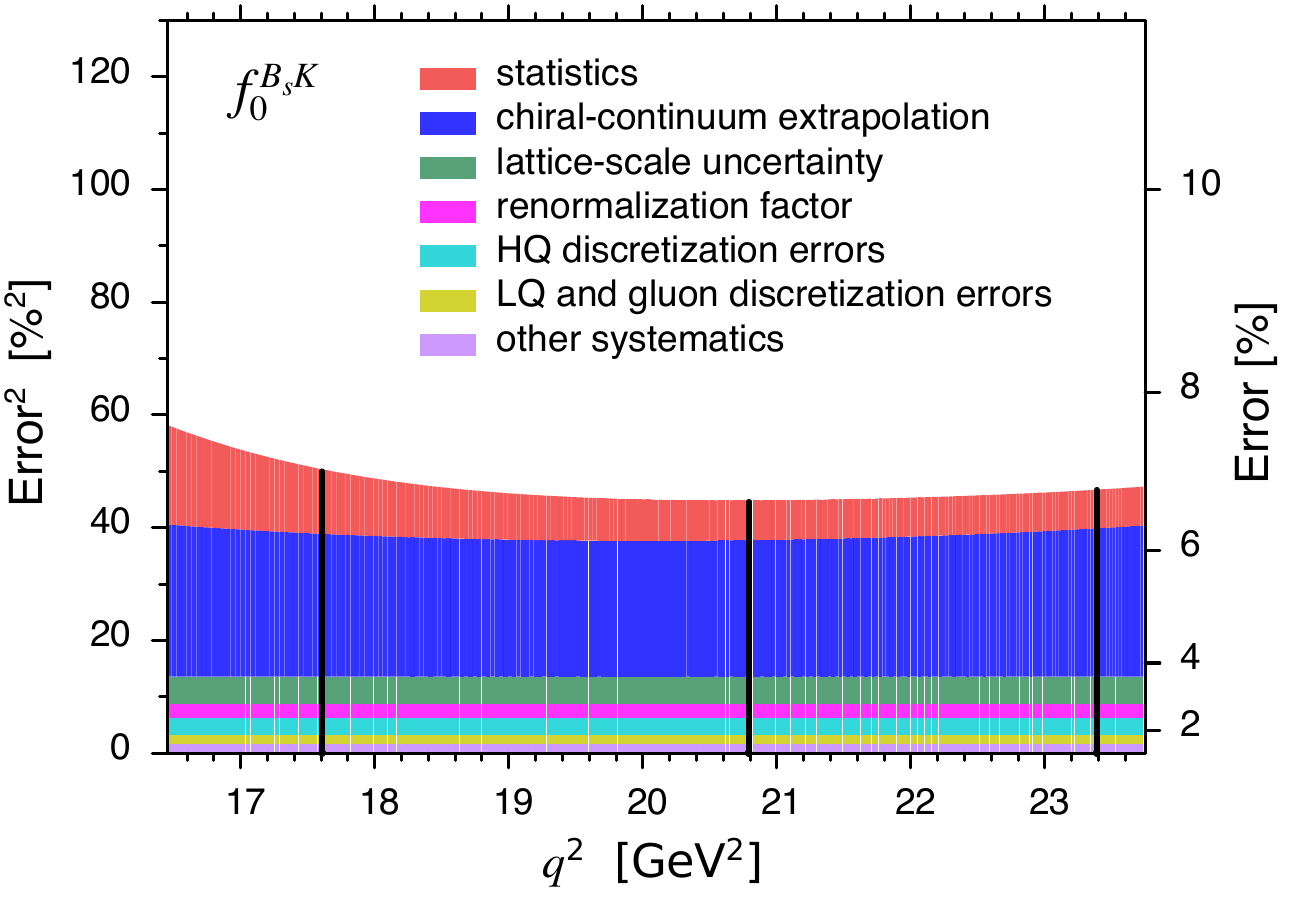}
\caption{Visualization of the error budgets for the $B\to\pi\ell\nu$ (upper plots) and $B_s\to K \ell\nu$ (lower plots) form factors.  Error budgets for $f_\perp$ are on the left and of $f_\parallel$ are on the right.   The curves from bottom-to-top show the increase in the total percentage error as we add each individual source of error in quadrature.  In each plot, the left y-axis label shows the squared error, while the right y-axis label shows the error in the form factor.  For readability, we have combined all of the sources of uncertainty that we estimate to be below $\sim 1\%$ into a single entry labeled ``other systematics.'' 
The three vertical lines in each plot show the location of the synthetic data points used in the subsequent extrapolation to $q^2 = 0$.  Detailed error budgets at these $q^2$ values are given in Table~\ref{tab:ErrorBudget}.
}
\label{fig:ErrorBudgets}
\end{figure*}

We now discuss the sources of systematic uncertainty in our determinations of the $B\to\pi\ell\nu$ and $B_s\to K \ell\nu$ form factors.  Each uncertainty is discussed in a separate subsection.  We visually summarize the error budgets for the form factors versus $q^2$ in Fig.~\ref{fig:ErrorBudgets}, and provide a detailed numerical error budget for the form factors at three representative $q^2$ values within the range of simulated lattice momenta in Table~\ref{tab:ErrorBudget}.  The form factors at these three points will be used later in Sec.~\ref{Sec:FormFactors} for the extrapolation to $q^2=0$ via the $z$ expansion.

\begin{table*}[tb]
\centering
\caption{\label{tab:ErrorBudget} Error budgets for the $B\to\pi\ell\nu$ and $B_s\to K \ell\nu$ form factors 
 at three representative $q^2$ values in the range of simulated lattice momenta.  For convenience, we also show the corresponding pion or kaon energy, $E_P$. 
Errors are given in \%.  The total error is obtained by adding the individual errors in quadrature.}
\begin{ruledtabular}
 \begin{tabular}{l|ccc|ccc|ccc|ccc} 
& \multicolumn{3}{c|}{$f_+^{B\pi}$}  & \multicolumn{3}{c|}{$f_0^{B\pi  }$} 
& \multicolumn{3}{c|}{$f_+^{B_s K}$} & \multicolumn{3}{c}{$f_0^{B_s K }$} \\ \hline
$E_P$ [GeV]			&  0.85 & 0.50  & 0.27  &  0.85 & 0.50  & 0.27 & 1.07  & 0.77  & 0.53 & 1.07  & 0.77  & 0.53  \\
$q^2$ [GeV$^2$]                          & 19.0  & 22.6  & 25.1  & 19.0  & 22.6  & 25.1 & 17.6  & 20.8  & 23.4  & 17.6  & 20.8  & 23.4\\[2pt] \hline
$f(q^2)$                                 & 1.21  & 2.27  & 4.11  & 0.46  & 0.68  & 0.92	& 0.99  & 1.64  & 2.77  & 0.48  & 0.63  & 0.81\\[2pt] \hline
Statistics                               & 7.9   & 5.9   & 12.4  & 7.3   & 4.6   & 3.3 	& 4.1   & 3.4   & 3.2   & 3.4   & 2.7   & 2.6 \\
Chiral-continuum extrapolation           & 6.3   & 5.0   & 6.2   & 10.9  & 7.6   & 5.8 	& 3.2   & 2.8   & 2.5   & 5.0   & 4.9   & 5.1 \\
Light-quark mass  $m_{ud}$               & 0.3   & 0.2   & 0.2   & 0.4   & 0.3   & 0.2 	& 0.1   & 0.1   & 0.1   & 0.1   & 0.1   & 0.1 \\
Strange-quark mass  $m_{s}$              & 0.0   & 0.0   & 0.0   & 0.0   & 0.0   & 0.0  & 0.1   & 0.1   & 0.1   & 0.0   & 0.0   & 0.0 \\
Lattice-scale uncertainty                & 2.0   & 2.0   & 2.0   & 2.2   & 2.2   & 2.2 	& 2.0   & 2.0   & 2.0   & 2.2   & 2.2   & 2.2 \\
RHQ parameter tuning                     & 0.9   & 0.9   & 0.8   & 1.0   & 1.0   & 1.0 	& 0.9   & 0.9   & 0.9   & 1.0   & 1.0   & 1.0 \\
Renormalization factor                   & 0.8   & 0.8   & 0.7   & 1.6   & 1.6   & 1.7 	& 0.9   & 0.8   & 0.8   & 1.6   & 1.6   & 1.7 \\
Finite volume                            & 0.5   & 0.4   & 0.3   & 0.7   & 0.5   & 0.4 	& 0.2   & 0.2   & 0.2   & 0.2   & 0.1   & 0.1 \\
Heavy-quark discretization errors                 & 1.8   & 1.8   & 1.8   & 1.7   & 1.7   & 1.7 	& 1.8   & 1.8   & 1.8   & 1.7   & 1.7   & 1.7 \\
Light-quark \& gluon discretization errors       & 1.1   & 1.1   & 1.1   & 1.1   & 1.1   & 1.1 	& 1.3   & 1.3   & 1.3   & 1.3   & 1.3   & 1.3 \\
Isospin breaking                 & 0.7   & 0.7   & 0.7   & 0.7   & 0.7   & 0.7 	& 0.7   & 0.7   & 0.7   & 0.7   & 0.7   & 0.7 \\[2pt] \hline
Total  (\%)                        & 10.6  & 8.4   & 14.3  & 13.6  & 9.6   & 7.6 	& 6.2   & 5.5   & 5.3   & 7.1   & 6.7   & 6.8 \\
 \end{tabular}
\end{ruledtabular}
\end{table*}

In cases where the estimation of a systematic uncertainty requires the explicit variation of simulation parameters, we use the $a\approx 0.11$~fm ensemble with $am_l = 0.005$, and take the dependence of that ensemble to be representative of all ensembles.  We choose this ensemble because it has very high statistics, and therefore allows us to most reliably measure the dependence of the form factors on on the input parameters.  We expect the behavior of the form factors on this ensemble to provide conservative bounds on the errors since it has the largest lattice spacing and heaviest kaons.

\subsection{Chiral-continuum extrapolation}
\label{Sec:ChiPT}

We estimate the systematic uncertainty due to the chiral-continuum extrapolation of the $B\to \pi$ and $B_s \to K$ form factors by varying the chiral-continuum fit Ans\"atze.
We consider the following fit alternatives:
\begin{itemize}
  \item standard HM$\chi$PT including explicit $E_P$ dependence in the chiral logarithms
  \item omitting  the term proportional to $a^2$ in Eqs.~(\ref{eq:fpar_ChPT}) and (\ref{eq:fperp_ChPT})
  \item omitting  the term proportional to $M_\pi^2$ in Eqs.~(\ref{eq:fpar_ChPT}) and (\ref{eq:fperp_ChPT})
  \item omitting  terms proportional to $a^2$ and $M_\pi^2$ in Eqs.~(\ref{eq:fpar_ChPT}) and (\ref{eq:fperp_ChPT})
  \item analytic fits omitting the chiral logarithms in Eqs.~(\ref{eq:fpar_ChPT}) and (\ref{eq:fperp_ChPT})
  \item analytic fits omitting the chiral logarithms and the term proportional to $a^2$  
  in Eqs.~(\ref{eq:fpar_ChPT}) and (\ref{eq:fperp_ChPT})
  \item varying the value of $f_\pi$ in the coefficients of the chiral logarithms from $f_0$= 112(2)~MeV~\cite{Aoki:2010dy} in the chiral limit to $f_K = 156.1(8)$~MeV~\cite{Beringer:2012zz}
  \item varying the $B^\ast B \pi$ coupling in the coefficients of the chiral logarithms $g_b=0.57(8)$ by plus/minus one standard deviation~\cite{Flynn:2013kwa}
   \item varying the scalar pole mass $M_{B^*}(0^+)=5.63$~GeV in $f_0^{B_s  K}$ by plus/minus 100~MeV
  \item omitting the data point at zero momentum
  \item omitting the data point at the highest momentum $\vec{p} = 2\pi/L (2,0,0)$ for $f_{+/0}^{B_s  K}$
  \item excluding ensembles with pion masses $M_\pi \gtrsim 400$ MeV
\end{itemize}

\begin{figure*}[tb]
  \centering
  \includegraphics[width=0.99\textwidth]{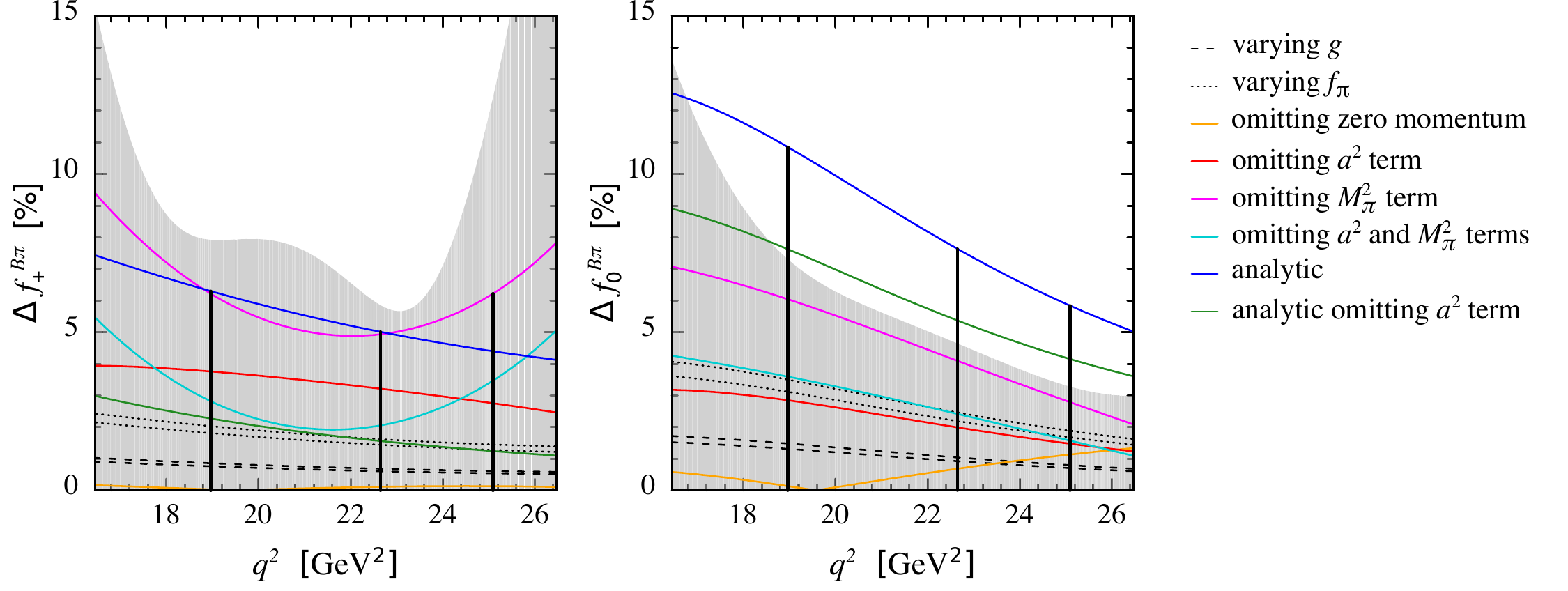}
  \includegraphics[width=0.99\textwidth]{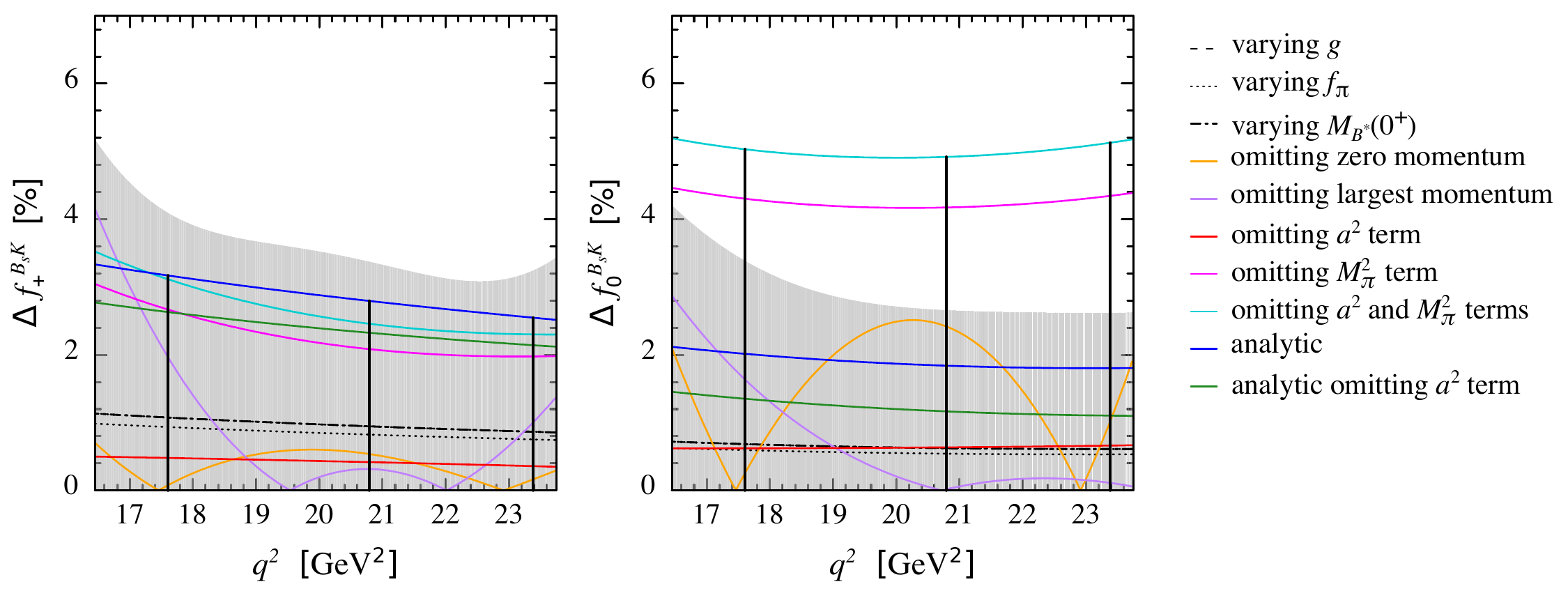}
  \caption{Relative change of the form-factor central value under the considered fit variations for $B\to\pi\ell\nu$ (upper) and $B_s\to K \ell\nu$ (lower).  In each plot, the shaded band shows the statistical uncertainty of the preferred fit. The three vertical lines show the location of the synthetic data points used in the subsequent extrapolation to $q^2=0$. }
  \label{fig:syst_ChPT}
\end{figure*}

Figure~\ref{fig:syst_ChPT} shows the relative changes of the form-factor central values under each fit variation
\begin{align}
\Delta f_i = |f_i^{\rm pref.} - f_i^{\rm alt.}| / f_i^{\rm pref.} \,,
\end{align}
where $i=\{0,+\}$. We take the largest difference between our preferred fit and any of the alternate fits as systematic uncertainty due to the chiral-continuum extrapolation.  We do not use fits with $p$-values below 1\% or those that cannot resolve the coefficients within statistical uncertainties for our error estimate.  Thus we exclude the fit omitting ensembles with pion masses $M_\pi\gtrsim 400$~MeV and the fit using standard ``soft-pion'' HM$\chi$PT. 

For each form factor, we obtain the largest difference from our preferred fit using the following variation:
\begin{align*}
 f_+^{B\pi} : \ & \text{analytic for $18.7 \ {\rm GeV}^2 \leq q^2 \leq 22.7 \ {\rm GeV}^2$} \\
             \ &  \text{and omitting the $M_\pi$ term elsewhere,}\\
f_0^{B\pi} : \ & \text{analytic,} \\
f_+^{B_sK} : \ & \text{analytic,} \\
f_0^{B_sK} : \ & \text{omitting the $a^2$ and $M_\pi$ terms,} 
\end{align*}
We therefore use these fits to obtain the $q^2$-dependent chiral-continuum extrapolation errors quoted in Table~\ref{tab:ErrorBudget} and shown in Fig.~\ref{fig:ErrorBudgets}.

\subsection{Lattice-scale uncertainty}
\label{Sec:Scale}

\begin{figure*}[tb]
  \centering
  \includegraphics[width=0.99\textwidth]{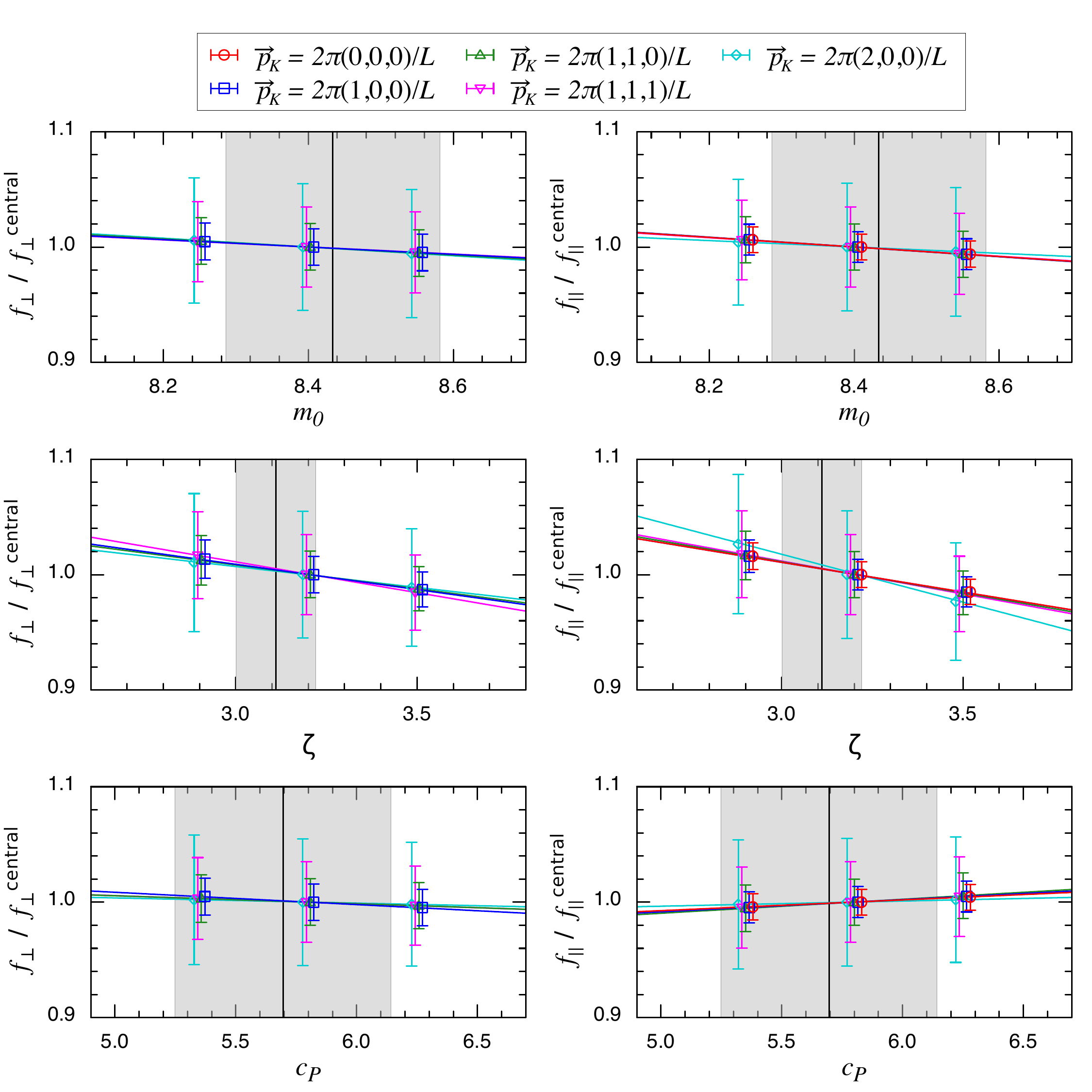}
 \caption{RHQ parameter dependence of the $B_s \to K$ form factors  $\fperp$~(left) and $\fpara$~(right) on the $24^3$ ensembles with $am_l=0.005$ using the unimproved heavy-light vector current in Eq.~(\ref{eq_V0}).  The slopes are normalized using the form factors obtained at the central set of RHQ parameters.  From left to right, the plots show the dependence on $m_0a$, $c_P$, and $\zeta$.  The colored lines show the results of a linear fit to the three data points at each momentum.   The black vertical lines indicate the tuned values of the RHQ parameters.  The shaded vertical bands indicate the systematic errors in the RHQ parameters due to the lattice-scale uncertainty.
For clarity, data points at equal RHQ parameter values are plotted with a slight horizontal offset.}
  \label{fig:RHQ_slopes}
\end{figure*}

We tuned the parameters of the $b$-quark action to reproduce the experimental value of the ${B_s}$ meson mass, and carry out our analysis in terms of dimensionless ratios over $M_{B_s}$ to remove all explicit dependence on the lattice scale.  We then obtain the form factors and momentum transfers in physical units by multiplying by the appropriate power of $M_{B_s}=5.367(4)$~GeV.  The uncertainties in the form factors due to to the experimental error on $M_{B_s}$ are negligible.

We do, however, still need to consider the implicit dependence on the lattice spacing through the parameters of the $b$-quark action. We estimate the size of this dependence, by computing the form factors $\fperp$ and $\fpara$ for seven sets of RHQ parameters. We then calculate the slopes with respect to the parameters -- $\Delta f/\Delta m_0a$, $\Delta f/\Delta c_P$ and  $\Delta f/\Delta \xi$ --  for all momenta used in the analysis. Next we multiply each slope by the uncertainty in the corresponding RHQ parameter due to the lattice spacing from Table~\ref{tab:RHQpara}, {\it e.g.} $\Delta f/\Delta m_0 \times \sigma_{m_0a}^a$.  Finally, we add the individual contributions from the three RHQ parameters in quadrature to obtain the total systematic error due the lattice spacing.

We examined the slopes with respect to the RHQ parameters for both $B \to \pi$ and $B_s \to K$, and found them to be consistent. We therefore base our estimates for the systematic uncertainty due to the lattice spacing on the slopes obtained for the $B_s \to K$ form factors because the smaller statistical errors in $B_s\to K$ enable the slopes to be resolved more precisely.  Figure~\ref{fig:RHQ_slopes} shows the slopes of the $B_s \to K \ell\nu$ form factors with respect to the $\{m_0a, c_P, \zeta \}$ on the $a\approx 0.11$~fm ensemble with $am_l = 0.005$.  For this slope estimate, we use the unimproved heavy-light vector current from Eq.~(\ref{eq_V0}).  We find the largest slopes at $\vec{p} = 2\pi(2,0,0)$ for $\fperp$ and $\vec{p} = 2\pi(1,1,0)$ for $\fpara$.   Following the procedure outlined above, we estimate lattice-spacing errors in $\fperp$ and $\fpara$ of $1.9$~\% and $2.2$~\%, respectively. In the continuum this corresponds to errors on $f_+$ ($f_0$) of $2.0$\% ($2.2$\%) which we take for both $B_s \to K$ and $B \to \pi$.

\subsection{Light- and strange-quark mass uncertainties}
\label{Sec:Quark}

Here we estimate the error in the form factors due to the uncertainty in the light-quark mass and the mistuning of the strange sea quark.  For clarity we discuss separately each place where the light- or strange-quark mass enters the analysis.

\subsubsection{\texorpdfstring{$u/d$-quark mass uncertainty}{u/d quark mass}}

We obtain the physical form factors $\fperp$ and $\fpara$ after the chiral-continuum fit by evaluating Eqs.~(\ref{eq:fpar_ChPT}) and (\ref{eq:fperp_ChPT}) at the physical average $u/d$-quark mass $a_{32}\tilde{m}_{ud}^{\rm phys} = 0.00102(5)$. We estimate the error in the form factors due to the light-quark mass uncertainty by varying $\tilde{m}_{ud}^{\rm phys}$ by plus/minus one sigma.  For $B\to\pi$ the central value shifts by $0.2-0.3$\% for $f_+^{B \pi}$ and $0.2-0.4$\% for $f_0^{B \pi}$, while for $B_s\to K$ both $f_+^{B_s K}$ and $f_0^{B_s K}$ change by $0.1$\%.

\subsubsection{Strange sea-quark mistuning}
\label{Sec:SeaQuark}

Our preferred chiral-continuum fit employs $SU(2)$ chiral perturbation theory, in which the strange quark mass is integrated out, so our fit function has no explicit dependence on $m_s$.  Further, at each lattice spacing, results for the form factors are only available at a single value of the strange sea-quark mass, so we cannot directly compute the strange sea-quark mass dependence of $\fpara$ and $\fperp$.  We therefore study the light sea-quark mass dependence and  use it to bound the strange sea-quark mass dependence.  We cannot resolve any light sea-quark mass dependence within statistical uncertainties, and expect the strange sea-quark mass dependence to be even smaller.  Thus we take the error due to mistuning the strange sea-quark mass to be negligible.

\subsubsection{Valence strange-quark mass uncertainty}

\begin{figure}[tb]
  \centering
  \includegraphics[width=0.48\textwidth]{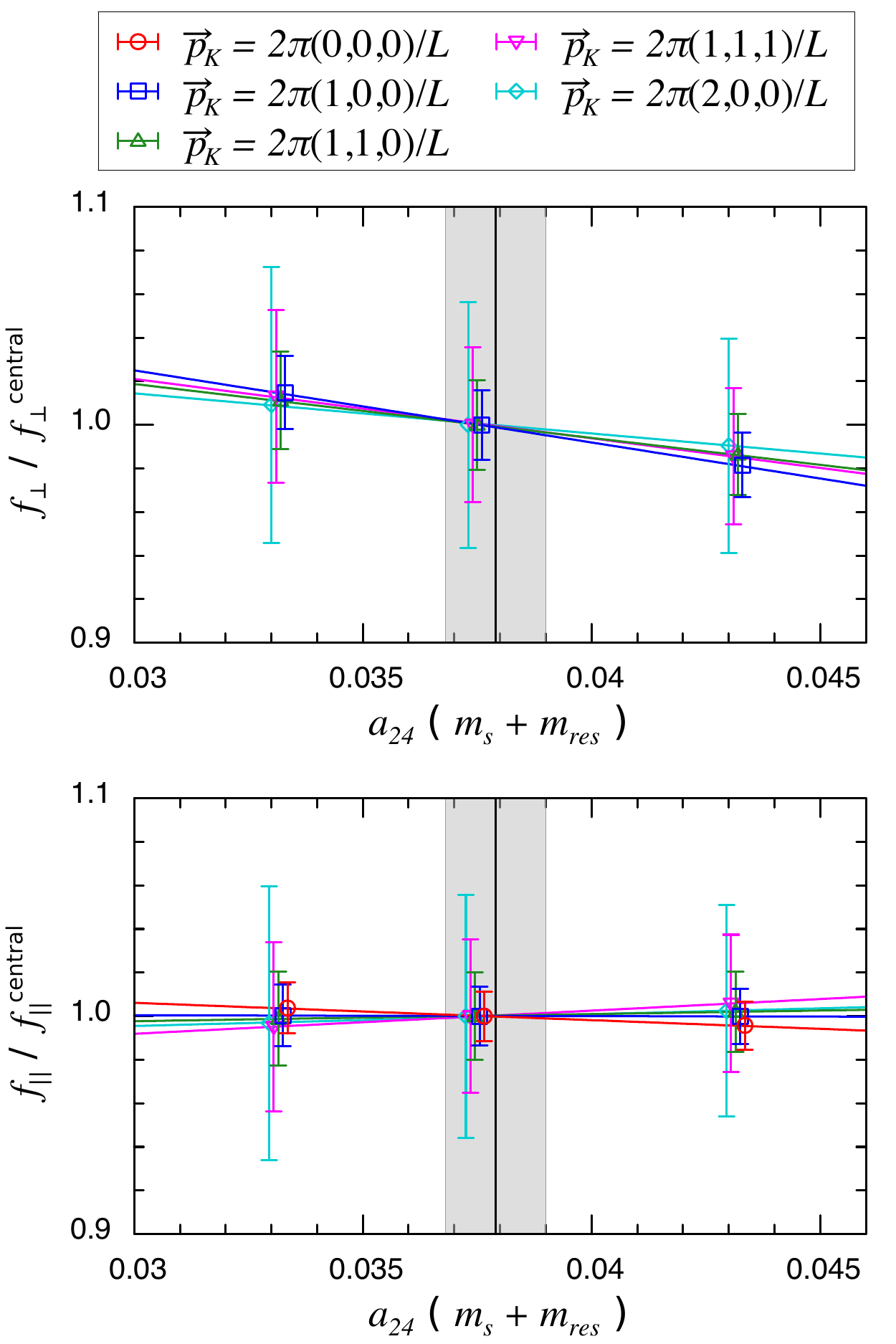}
 \caption{Valence strange-quark mass dependence of the $B_s \to K$ form factors $\fperp$~(top) and $\fpara$~(bottom) on the $a\approx 0.11$~fm ensemble with $am_l=0.005$.  The slopes are normalized by the form factors obtained with the strange-quark mass used in our production simulations.  The colored lines show the results of a linear fit to the three data points at each momentum.  The black vertical line with error band shows the total (statistical plus systematic) uncertainty in the physical strange-quark mass~\cite{Aoki:2010dy}.
 For clarity, data points at equal strange-quark masses are plotted with a slight horizontal offset.}
  \label{fig:ms_slopes}
\end{figure}

The $B_s \to K$ form factors have explicit strange valence-quark mass dependence.  The strange-quark masses employed in our simulations differ slightly from the physical, tuned values $a_{24}\tilde{m}_{s}^{phys} = 0.0379(11)$ and $a_{32}\tilde{m}_{s}^{phys} = 0.0280(7)$~\cite{Aoki:2010dy}.  To study the valence strange-quark mass dependence, we calculated the $B_s \to K$ form factors on the $a \approx 0.11$~fm, $am_l = 0.005$, ensemble with two additional spectator-quark masses of $a_{24}\tilde{m}_s=0.033$ and $0.043$.  Figure \ref{fig:ms_slopes} shows the valence-quark mass dependence of the $B_s \to K$ form factors; we observe the largest slopes for $\fpara$ at $p=(0,0,0)$ and for $\fperp$ at $p=(1,0,0)$.  Multiplication of these measured slopes by the discrepancy between the simulated and tuned strange-quark masses, $\Delta({m_s}) \equiv (\tilde{m}_s - \tilde{m}_s^{\rm phys}) = 0.004$, leads to estimates for the error due to mistuning the valence strange-quark mass of about $0.1\%$ for $f_+$ and below this for $f_0$ (which we consider as negligible).

\subsection{RHQ parameter uncertainty}

We compute the semileptonic form factors using the nonperturbatively tuned RHQ parameters obtained in Ref.~\cite{Aoki:2012xaa} and given in Table~\ref{tab:RHQpara}. The RHQ parameters have four significant sources of uncertainty:  statistics, heavy-quark discretization errors, lattice scale, and the experimental inputs.  We already discussed the uncertainty due to the lattice scale in Sec.~\ref{Sec:Scale}.  We follow the same approach for propagating the uncertainty in the RHQ parameters due to heavy-quark discretization errors and experimental inputs.  We multiply the estimated slopes of the form factors with respect to changes in $m_0$, $c_P$, and $\zeta$ (shown in Fig.~\ref{fig:RHQ_slopes}) by the uncertainties in the corresponding parameters due to heavy-quark discretization errors and experimental inputs.  Adding the contributions from the three RHQ parameters and the two uncertainty sources in quadrature, we obtain error estimates for $f_+^{B\pi}$ of 0.8--0.9\%, $f_0^{B\pi}$ of 1.0\%, $f_+^{B_sK}$ of 0.9\%, and $f_0^{B_sK}$ of 1.0\%.

We neglect the statistical uncertainties in the RHQ parameters in our final analysis, after checking that they have a negligible impact on the form factors. On the $a \approx 0.11$~fm, $am_l = 0.005$ ensemble, we computed the form factors with seven sets of RHQ parameter values. We then used the approach detailed in Refs.~\cite{Aoki:2012xaa,Christ:2014uea} to interpolate to the tuned RHQ parameters.  This procedure automatically propagates the statistical errors in the RHQ parameters via the jackknife.  Indeed we find that the statistical errors obtained from the two procedures are identical. Thus we do not need to perform the more complicated and computationally expensive procedure of interpolating to the tuned RHQ parameters in our analysis.

\subsection{Heavy-quark discretization errors}
\label{Sec:HeavyQuarkDiscErrors}

\begin{table*}[tb]
	\centering 
	\caption{Percentage errors from mismatches in the action and current for the bottom quark on the $24^3$ and $32^3$ ensembles.  For this estimate, we calculate the mismatch functions for the nonperturbatively tuned parameters of the RHQ action from Table~\ref{tab:RHQpara}.  We estimate the size of operators using HQET power counting with $\Lambda_{\rm QCD}=500$~MeV and the coupling constant $\alpha_s^{\bar{\rm MS}}(1/a) = 1/3$ on the  $24^3$ ensemble and $0.22$ on the $32^3$ ensembles.  
	  To obtain the total, we add the individual errors in quadrature, including each contribution the number of times that operator occurs.  Contribution $E$ is counted twice, and 3 is counted twice for $f_\parallel$ and four times for $f_\perp$.  The definitions of operators ``E," ``$X_1,$" ``$X_2$," ``$Y,$" and ``3" and expressions for the mismatch functions are given in Appendix~B of Ref.~~\cite{Christ:2014uea}.}  \vspace{2mm}
	\label{tab:HQDiscErrs_BtoPi}
\begin{tabular}{l@{\hskip 2mm}c@{\hskip 2mm}ccc @{\hskip 2mm} c@{\hskip 2mm}c@{\hskip 2mm}c}
\hline\hline
	& ${\mathcal O}(a^2)$ error & \multicolumn{3}{c}{${\mathcal O}(a^2)$ errors} {\hskip 1.mm} & ${\mathcal O}(\alpha_s^2 a)$ error  \\
	& from action & \multicolumn{3}{c}{from current} {\hskip 1.mm} & from current & \multicolumn{2}{c}{Total (\%)}  \\	
        & $E$ & $X_1$ & $X_2$ & $Y$ & 3  &  $f_\parallel$ & $f_\perp$ \\\hline
	$a \approx$~0.11 fm   & 0.55 & 0.67 & 1.27 & 1.34 & 1.48 & 2.97 & 3.64\\
	$a \approx $~0.086 fm & 0.42 & 0.46 & 0.85 & 0.91 & 0.55 & 1.65 & 1.82 \\
\hline\hline
\end{tabular}
\end{table*}
The RHQ action gives rise to nontrivial lattice-spacing dependence in the form factors in the region $m_0 a \sim 1$.  
To estimate the size of the resulting discretization errors, we use the same power-counting approach as 
in our companion papers on bottomonium masses and splittings~\cite{Aoki:2012xaa} and $B_{(s)}$-meson decay constants~\cite{Christ:2014uea}.  

We tune the parameters of the operators in the dimension-5 RHQ action nonperturbatively,  such that the leading heavy-quark discretization errors from the action are of ${\mathcal O}(a^2)$.  We use an ${\mathcal O}(\alpha_s a)$-improved vector current and calculate the improvement coefficient to 1-loop;
therefore the leading heavy-quark discretization errors from the current are of ${\mathcal O}(\alpha_s^2 a, a^2)$. 
Because we use the same actions and simulation parameters as in our earlier calculation of the $B_{(s)}$-meson leptonic decay constants~\cite{Christ:2014uea}, 
the numerical error estimates are almost identical in the two works.  The same operators contribute in both cases, 
but enter a different number of times for the spatial and temporal vector currents.  
Table~\ref{tab:HQDiscErrs_BtoPi} quotes the estimate of heavy-quark discretization errors 
from the five different operators in the action and current on the $24^3$ and $32^3$ ensembles, 
and we refer the reader to Sec.~V.~E. and Appendix~B of Ref.~\cite{Christ:2014uea} for details. 
We take the size of heavy-quark discretization errors in our calculation of 
the $B\to\pi\ell\nu$ and $B_s\to K \ell\nu$  semileptonic form factors 
to be the estimate on our finer $a^{-1} = 2.281$~GeV lattices, 
which is $\sim 1.7$\% for the lattice form factors $f_{\parallel}$ and $\sim 1.8$\% for $f_\perp$.  
These lead to errors in the continuum form factors $f_+$ and $f_0$  of $\sim  1.8$\% and $\sim  1.7$\%, respectively.

\subsection{Light-quark discretization errors}
\label{Sec:LightQuarkDiscErrors}

The dominant discretization errors from the light-quark and gluon sectors are of $\Ocal((a\Lambda_{\rm QCD})^2)$ from the action, and is about 5\% using $\Lambda_{\rm QCD}=500$~MeV.  We remove this $\Ocal((a\Lambda_{\rm QCD})^2)$ error by including a term proportional to $a^2$ in the chiral-continuum extrapolation (see Eqs.~(\ref{eq:fpar_ChPT}) and~(\ref{eq:fperp_ChPT})). Then the leading light-quark and gluon discretization errors in the heavy-light vector current are of
 $\Ocal(\alpha_s a \tilde{m}_q, (a\tilde{m}_q)^2, \alpha_s^2 a \Lambda_{\rm QCD}, (ap)^2)$ where $\tilde{m}_q$ denotes the bare lattice mass. The first entry of $\Ocal(\alpha_s a \tilde{m}_q)$ leads to estimated errors of $\sim 0.1\%$ in $B\to\pi$  and $\sim 0.6\%$ in $B_s \to K $ form factors on the $32^3$ ensembles.
The second is negligible ($<0.1\%$) and the third is estimated to be $\sim 1.1\%$ in both $B\to\pi$ and $B_s \to K$ form factors.  Adding these contributions in quadrature, we estimate the total uncertainty from light-quark and gluon discretization errors in the heavy-light current to be $1.1$\% in $B\to\pi$ and $1.3$\% in $B_s \to K$.

We do not observe any evidence of sizable momentum-dependent discretization errors in our data.  Figure~\ref{fig:dispersion} shows that the pion and kaon energies and amplitudes are consistent with continuum expectations, and smaller than power-counting estimates of $\Ocal((ap)^2)$. Thus we do not include a systematic error due to momentum-dependent discretization errors. 

\subsection{Renormalization factor}
\label{Sec:rho}

We renormalize the lattice form factors using a mostly nonperturbative approach in which we separate $Z_{V_\mu}^{bl}$ into three components.  We consider the uncertainties from these three multiplicative factors separately, and then add them in quadrature to obtain the total error on the form factors.

For $Z_V^{ll}$, we use the nonperturbatively determined value of the axial-current renormalization factor $Z_A$ in the chiral limit from Ref.~\cite{Aoki:2010dy}. We can neglect the statistical uncertainty in $Z_A$ (which is only $0.02$\% on the finer ensembles)  and the difference between $Z_V^{ll}$ and $Z_A$ (which is about $\Ocal(am_{\rm res}) \sim 7\times 10^{-4}$ at $a \approx 0.086$~fm).   For $Z_V^{bb}$, we use the nonperturbative determination from~\cite{Christ:2014uea}.  The statistical uncertainty in $Z_V^{bb}$ on the finer ensemble is $0.15$\%.  We conservatively estimate the perturbative truncation error in $\rho_{V}^{bl}$ to be the full size of the 1-loop correction at the finer $a \approx 0.086$~fm lattice spacing, which leads to $1.7$\% for $\rho_{V_0}$ and $0.6$\% for $\rho_{V_i}$.  These are significantly larger than what we would estimate for two-loop contributions from naive power counting.  Taking $\alpha_s \sim 0.23$ on the coarser lattice spacing and a coefficient of $1/(4\pi)^2$ from two loop-suppression factors, we would obtain an estimate of 0.03\%.  Even with a coefficient of $1/\pi^2$, we would obtain an estimate of 0.5\%, which is slightly smaller than the perturbative uncertainty that we assign to $\rho_{V_i}$ and 3 times smaller than the error we assign to $\rho_{V_0}$.  Because we use the values of $\rho_{V_\mu}$ and  $Z_V^{ll}$ in the chiral limit, we must consider the errors due to the nonzero physical up, down, and strange-quark masses. The leading quark-mass dependent errors in  $\rho_{V_\mu}$ and  $Z_V^{ll}$ are $\Ocal(\alpha_s a \tilde{m}_q)$ and $\Ocal((a\tilde{m}_q)^2)$, respectively, but these are already accounted for in our estimate of light-quark and gluon discretization errors~(see Sec.~\ref{Sec:LightQuarkDiscErrors}). Thus we do not count them again here.

Perturbative truncation errors are by far the dominant source of uncertainty in the renormalization factor, and the quadrature sum of the three error contributions is 1.7\% for $\fpara$ and 0.6\% for $\fperp$.

\subsection{Finite-volume errors}
\label{Sec:FVerr}

We compute the form factors on a finite-sized lattice. We estimate the effect of the finite spatial volume using one-loop finite-volume SU(2) hard-pion $\chi$PT, in which loop integrals are replaced by a sum over lattice sites. Only a single integral enters the NLO SU(2) hard-pion ChPT expression.  The correction to Eqs.~(\ref{eq:chpt_na_btopi}) and (\ref{eq:chpt_na_bstok}) to account for the finite spatial volume is given by a sum over modified Bessel functions~\cite{Arndt:2004bg,Aubin:2007mc}:
\begin{equation}
  M_\pi^2 {\rm log} \left( \frac{M_\pi^2}{\Lambda^2} \right) 
\rightarrow M_\pi^2 {\rm log} \left( \frac{M_\pi^2}{\Lambda^2} \right)
+\frac{4M_\pi}{L}\sum_{\vec{r}\neq 0}^{L^3} \frac{K_1(|\vec{r}|M_\pi L)}{|\vec{r}|}.
\label{eq:fv_corre}
\end{equation}
From Eq.~(\ref{eq:fv_corre}), the ensemble with the lightest quark mass receives the largest correction.  For $B\to\pi$ we find corrections to $\fperp$ ($\fpara$) of 0.3-0.4\% (0.6--0.8\%), while for $B_s\to K$ corrections to $\fperp$($\fpara$) of 0.2--0.3\% (0.4--0.5\%). These result in the following errors on the continuum form factors: 0.3--0.5\% for $f_+^{B\pi}$,  0.4--0.7\% for $f_0^{B\pi}$, 0.2\% for $f_+^{B_sK}$, and 0.1--0.2\% for $f_0^{B_sK}$.

\subsection{Isospin breaking}
\label{Sec:Isospin}

Our $B\to\pi\ell\nu$ and $B_s\to K \ell\nu$ form factors are calculated in the isospin limit.
The form factors of the charged and neutral $B$~($B_s$)-mesons, however,  differ due to 
both the masses and the charges of the constituent light $u$ and $d$ quarks.
The leading quark-mass contribution to the isospin breaking from the valence-quark masses is 
of $\Ocal((m_d-m_u)/\Lambda_{\rm QCD}) \sim 0.5$\%, which is obtained 
using the determination of the quark masses $(m_d-m_u)= 2.35(8)(24)$~MeV from FLAG~\cite{Aoki:2013ldr} 
and $\Lambda_{\rm QCD}=500$~MeV.
The difference between the $u$- and $d$-quark masses in the sea sector 
should have a negligible effect on the $B\to\pi\ell\nu$~($B_s\to K \ell\nu$) form factor
because the sea quarks couple to the valence quarks through $I = 0$ gluon exchange,
and they give only the uncertainty of $\Ocal(((m_d-m_u)/\Lambda_{\rm QCD})^2) \sim 0.003$\%.
The electromagnetic contribution to the isospin breaking is 
expected to be $\Ocal(\alpha_s) \sim 1/137 \sim 0.7\%$ which is the typical size of 1-loop QED corrections.
We therefore take 0.7\% as the uncertainties due to the isospin breaking and electromagnetism effects.


\subsection{Correlation matrices}
\label{Sec:correlations}

In the next section we fit synthetic lattice data generated at three values of $q^2$ to the $z$-expansion to extend it to the full kinematic range.  Thus, in addition to the systematic uncertainties on the individual $q^2$-bins, we also need the correlations between $q^2$ values.  Although it is straightforward to obtain the statistical correlations further explanation is needed for the systematic error correlations. 

The chiral-continuum extrapolation error is estimated by varying the fit function and parametric inputs.  This procedure does not provide any information on correlations of the resulting systematic error between different $q^2$-bins.  Alternate chiral-continuum fits to our data with different fit functions do, however, exhibit highly similar statistical correlations between $q^2$-bins. Hence we take the (normalized) statistical correlation matrix from our preferred fit and multiply it by the estimated chiral-continuum extrapolation error at each $q^2$ value (For off-diagonal elements of the correlation matrix we use the product $\sigma_{q^2_i} \sigma_{q^2_j}$.)
We follow the same procedure to estimate the correlations between the $q^2$-dependent systematic error due to the light-quark mass uncertainty.  We take the remaining systematic errors for which we do not assume any $q^2$ dependence to be 100\% correlated.

Tables~\ref{tab:corre_mat_BtoPi} and \ref{tab:corre_mat_BstoK} present the normalized statistical and systematic correlation matrices, which enable the full reconstruction of the total covariance matrices using the values for $B\to\pi\ell\nu$ and $B_s\to K \ell\nu$ form factors and their errors from Table~\ref{tab:ErrorBudget}. 

\begin{table}[tb]
  \caption{\label{tab:corre_mat_BtoPi}  Normalized statistical (upper) and systematic (lower) correlation matrices 
  for the $B\to\pi\ell\nu$ form factors at three representative $q^2$ values.}
  \begin{ruledtabular}
    \begin{tabular}{cc|ccc|ccc}
& & \multicolumn{3}{c|}{$f_+^{B\pi}$}   & \multicolumn{3}{c}{$f_0^{B\pi}$} \\
& $q^2$~[GeV$^2$]  & 19.0  & 22.6  & 25.1  & 19.0  & 22.6  & 25.1  \\ \hline
                 &19.0  & 1.000 & 0.868 & 0.045 & 0.663 & 0.586 & 0.541 \\
$f_+^{B\pi}$  &22.6  & 0.868 & 1.000 & 0.239 & 0.591 & 0.654 & 0.616 \\
                 &25.1  & 0.045 & 0.239 & 1.000 & 0.176 & 0.188 & 0.283 \\ \hline
                 &19.0  & 0.663 & 0.591 & 0.176 & 1.000 & 0.822 & 0.836 \\
$f_0^{B\pi}$  &22.6  & 0.586 & 0.654 & 0.188 & 0.822 & 1.000 & 0.941 \\
                 &25.1  & 0.541 & 0.616 & 0.283 & 0.836 & 0.941 & 1.000 \\
    \end{tabular}
  \end{ruledtabular}
\vspace{0.2cm}
  \begin{ruledtabular}
    \begin{tabular}{cc|ccc|ccc}
& & \multicolumn{3}{c|}{$f_+^{B\pi}$}   & \multicolumn{3}{c}{$f_0^{B\pi}$} \\
& $q^2$~[GeV$^2$]  & 19.0  & 22.6  & 25.1  & 19.0  & 22.6  & 25.1  \\ \hline
                 &19.0  & 1.000 & 0.897 & 0.245 & 0.702 & 0.663 & 0.645 \\
$f_+^{B\pi}$  &22.6  & 0.897 & 1.000 & 0.427 & 0.639 & 0.725 & 0.719 \\
                 &25.1  & 0.245 & 0.427 & 1.000 & 0.289 & 0.342 & 0.448 \\ \hline
                 &19.0  & 0.702 & 0.639 & 0.289 & 1.000 & 0.840 & 0.840 \\
$f_0^{B\pi}$  &22.6  & 0.663 & 0.725 & 0.342 & 0.840 & 1.000 & 0.948 \\
                 &25.1  & 0.645 & 0.719 & 0.448 & 0.840 & 0.948 & 1.000 \\
    \end{tabular}
  \end{ruledtabular}
\end{table}
\begin{table}[tb]
    \caption{\label{tab:corre_mat_BstoK} Normalized statistical (upper) and systematic (lower) correlation matrices 
  for the $B_s\to K \ell\nu$ form factors at three representative $q^2$ values.}
  \begin{ruledtabular}
    \begin{tabular}{cc|ccc|ccc}
& & \multicolumn{3}{c|}{$f_+^{B_s K}$}   & \multicolumn{3}{c}{$f_0^{B_s K}$} \\
& $q^2$~[GeV$^2$]  & 17.6  & 20.8  &  23.4  & 17.6  & 20.8  & 23.4  \\ \hline
                 &17.6  &  1.000 & 0.868 & 0.828 & 0.799 & 0.754 & 0.702 \\
$f_+^{B_s K}$ &20.8  &  0.868 & 1.000 & 0.783 & 0.677 & 0.799 & 0.764 \\
                 &23.4  &  0.828 & 0.783 & 1.000 & 0.615 & 0.703 & 0.708 \\ \hline
                 &17.6  &  0.799 & 0.677 & 0.615 & 1.000 & 0.828 & 0.755 \\
$f_0^{B_s K}$  &20.8  &  0.754 & 0.799 & 0.703 & 0.828 & 1.000 & 0.974 \\
                 &23.4  &  0.702 & 0.764 & 0.708 & 0.755 & 0.974 & 1.000 \\
    \end{tabular}
  \end{ruledtabular}
\vspace{0.2cm}
  \begin{ruledtabular}
    \begin{tabular}{cc|ccc|ccc}
& & \multicolumn{3}{c|}{$f_+^{B_s K}$}   & \multicolumn{3}{c}{$f_0^{B_s K}$} \\
& $q^2$~[GeV$^2$]  & 17.6  & 20.8  & 23.4  & 17.6  & 20.8  & 23.4  \\ \hline
                 &17.6  &  1.000 & 0.939 & 0.921 & 0.865 & 0.843 & 0.808 \\
$f_+^{B_s K}$ &20.8  &  0.939 & 1.000 & 0.913 & 0.794 & 0.860 & 0.835 \\
                 &23.4  &  0.921 & 0.914 & 1.000 & 0.760 & 0.806 & 0.801 \\ \hline
                 &17.6  &  0.865 & 0.794 & 0.760 & 1.000 & 0.889 & 0.840 \\
$f_0^{B_s K}$ &20.8  &  0.843 & 0.860 & 0.806 & 0.889 & 1.000 & 0.983 \\
                 &23.4  &  0.808 & 0.835 & 0.801 & 0.840 & 0.983 & 1.000 \\
    \end{tabular}
  \end{ruledtabular}
\end{table}

\section{Form-factor results}
\label{Sec:FormFactors}

In this section we extrapolate our $B\to \pi$ ($B_s\to K$) form-factor results from large $q^2$, where we have our (synthetic) data, to $q^2=0$ using a model-independent parametrization based on the general properties of analyticity, unitarity, and crossing symmetry.  We first give the expressions for the $z$-parametrizations used in our analysis in Sec.~\ref{Sec:zExp}; we use the parametrization of  Bourrely, Caprini, and Lellouch (BCL) for our preferred results, but also consider the parametrization of Boyd, Grinstein, and Lebed (BGL) as a cross-check.  Next, in Sec.~\ref{Sec:LatzFit} we extrapolate our synthetic lattice data to $q^2=0$; we present our preferred results for $f_+$ and $f_0$ in Tables~\ref{tab:BtoPiBCLResults} and~\ref{tab:BstoKBCLResults} as coefficients of the $z$-expansion and the matrix of correlations between them. 

Use of the $z$-parametrization to describe semileptonic form factors has several advantages over other functional forms used in the literature~\cite{Becirevic:1999kt,Ball:2004ye}.    Because the absolute value of $|z|$ is small in the semileptonic region, and the $z$-coefficients are constrained to be small by unitarity and heavy-quark symmetry, one needs only the first few terms in the expansion to accurately describe the form factor shape with a negligible truncation error.    Moreover, as the precisions of both the lattice calculations and experimental measurements improve, one may easily include higher-order terms in $z$ as needed.  Finally, comparisons of the $z$-expansion parameters resulting from fits to different theoretical or experimental data sets  enable a meaningful quantitative comparison of the shapes, while a combined fit to lattice and experimental data enables a clean determination of $|V_{ub}|$.  The $z$-expansion has therefore been adopted as the preferred method for obtaining exclusive $|V_{ub}|$ by experimentalists on Babar and Belle, the Heavy Flavor Averaging Group, and the Particle Data Group~\cite{delAmoSanchez:2010af,Ha:2010rf,Lees:2012vv,Sibidanov:2013rkk,Beringer:2012zz,Amhis:2012bh}.

\subsection{\texorpdfstring{$z$-expansions of semileptonic form factors}{z-expansions of semileptonic form factors}}
\label{Sec:zExp}

The $B\to\pi\ell \nu$ and $B_s \to K \ell \nu$ form factors are analytic functions of $q^2$ except at physical poles and branch cuts above the production threshold.  Therefore, given a suitable change of variables, they can be expressed as a convergent power series (see, {\it e.g.},~\cite{Boyd:1994tt,Lellouch:1995yv,Boyd:1997qw,Bourrely:1980gp,Arnesen:2005ez,Bourrely:2008za}).  Unitarity and heavy-quark power counting bound the size of the series coefficients.  In the literature, the new variable is called $z$, and the class of functions are called $z$-parametrizations.  Two such parametrizations commonly used to extrapolate the $B\to\pi\ell\nu$ form factor are by Boyd, Grinstein, and Lebed (BGL)~\cite{Boyd:1994tt} and Bourrely, Caprini, and Lellouch (BCL)~\cite{Bourrely:2008za}.

The change of variables from $q^2$ to $z$ is given by
\begin{equation}
z(q^2, t_0) = \frac{\sqrt{1 - q^2/t_+}-\sqrt{1-t_0/t_+}}{\sqrt{1-q^2/t_+}+\sqrt{1-t_0/t_+}} ,
\label{eq:slep:z_trans}
\end{equation}
where $t_+\!\!\equiv\!\!(M_{B_{(s)}} + M_\pi )^2$ and $t_-\!\!\equiv\!\!(M_{B_{(s)}} -
M_\pi )^2$.  This transformation maps the semileptonic region $0 < q^2 < t_-$ onto a unit circle in the complex $z$ plane.  The $B\to\pi\ell \nu$ and $B_s \to K \ell \nu$ form factors can then be expanded as a simple power series in $z$:
\begin{equation}
P_i(q^2) \phi_i(q^2,t_0) f_i(q^2) = \sum_{k=0}^{\infty} a_i^{(k)}(t_0) z(q^2,t_0)^k ,
\label{eq:slep:z_exp}
\end{equation}
where $i = \{0, +\}$ for the scalar and vector form factors, respectively.  The free parameter $t_0$ in Eq.~(\ref{eq:slep:z_trans}) determines the range of $|z|$ in the semileptonic region, and hence can be chosen to accelerate the series convergence.  
The ``Blaschke factor'' $P_i(q^2)$ must be chosen to vanish at any subthreshold poles to preserve the correct analytic structure of $f_i(q^2)$.  For $f_+$, the relevant state is the $J^P = 1^-$ meson, while for $f_0$, the relevant state is the $J^P = 0^+$ meson.  As discussed earlier in Sec.~\ref{Sec:Extrapolations}, the scalar $B^*$ meson has not been observed experimentally, but is predicted to have a mass well above the $B\pi$ production threshold~\cite{Bardeen:2003kt,Gregory:2010gm}.  Thus the functions $P_i$ for $B\to\pi \ell\nu$ are typically taken in the literature to be
\begin{eqnarray}
	P^{B \pi}_+(q^2)  = z(q^2, M_{B^*}) \,, \quad
	P^{B \pi}_0(q^2) = 1\,.
\end{eqnarray}
Finally, the outer function $\phi_i(q^2,t_0)$ can be any analytic function of $q^2$; different choices for $\phi_i$ correspond to different $z$-parametrizations.   

The form factors that describe $B \to \pi \ell \nu$ ($B_s \to K \ell \nu$) in the range $0 < q^2 < t_-$, when analytically continued, also describe $B\pi$ ($BK$) production for $q^2 > t_+$.   The coefficients of the $z$-expansion are therefore bounded by the fact that the rate of production of $B\pi$ ($BK$) states is less than the production rate of all states coupling to the $b \to u$ current.   In Ref.~\cite{Boyd:1994tt}, Boyd, Grinstein, and Lebed choose the outer function $\phi_i$ so that the unitarity constraint on the series coefficients takes a particularly simple form:
\begin{equation}
\sum_{k=0}^{N} \left( a_i^{(k)} \right)^2 \ltapprox 1,
\label{eq:slep:a_const}
\end{equation}
where this holds for any value of $N$.
The explicit functions for $\phi_+^{\rm BGL}$ and $\phi_0^{\rm BGL}$ and their numerical values can be found in Ref.~\cite{Arnesen:2005ez}.  When using the BGL parametrization for subsequent $z$-fits, we use $t_0=0.65t_-$ as in Ref.~\cite{Arnesen:2005ez}, such that $- 0.341 <  z < 0.216$ ($-0.144 <  z < 0.148$) for $B\to\pi l \nu$ ($B_s \to K \ell\nu$) decay.   

In Ref.~\cite{Bourrely:2008za}, Bourrely, Caprini, and Lellouch (who only discuss $f_+$) choose a simpler outer function $\phi_0^{\rm BCL} = 1$.  They also point out that the BGL form-factor parametrization does not obey the known asymptotic behavior near the $B\pi$ production threshold ${\rm Im} f_+(q^2) \sim \left( q^2 - t_+ \right)^{3/2}$ (which is due to angular momentum conservation).  Therefore, at $q^2 = t_+$ ($z=-1$), the derivative of the form factor must satisfy 
\begin{equation}
	\left. \left[ \frac{d f_+}{dz} \right] \right|_{z=-1}= 0 \,.
\end{equation}
BCL use this constraint on the derivative of the form factor to remove an independent degree of freedom from the series expansion in $z$.  Thus they arrive at the following parametrization for the vector form factor:
\begin{equation}
	f_+(q^2) = \frac{1}{1-q^2/m_{B^*}^2} \sum_{k=0}^{K-1} b_+^{(k)} \left[ z^k - (-1)^{k-K} \frac{k}{K} z^K  \right] \,, \label{eq:BCLConstraint}
\end{equation}
where we label the BCL series coefficients $b_k$ to distinguish them from the BGL coefficients $a_k$.  There is no analogous constraint to Eq.~(\ref{eq:BCLConstraint}) on the value or derivative of $f_0$ at any $z$, so one cannot remove a further degree of freedom in the series expansion for the scalar form factor.  We therefore use the following functional forms for the scalar form factors:
\begin{eqnarray}
	f_0^{B\pi}(q^2) &=&  \sum_{k=0}^{K-1} b_0^{(k)} z^k \,, \label{eq:f0BtoPiBCL} \\
	f_0^{B_sK}(q^2) &=&  \frac{1}{1-q^2/m_{B^*}^2} \sum_{k=0}^{K-1} b_0^{(k)} z^k \,, \label{eq:f0BstoKBCL}
\end{eqnarray}
where we include a pole at the theoretically predicted value $M_{B^*}(0^+) = 5.63$~GeV for $B_s \to K \ell\nu$~\cite{Bardeen:2003kt}.   Equation~(\ref{eq:f0BtoPiBCL}) has been called the ``simplified series expansion'' in the literature~\cite{Bharucha:2010im}.  To minimize the error from truncating the $z$-expansion for the $B\to\pi\ell\nu$ form factor, BCL choose $t_0 = t_{\rm opt} \equiv \left( M_B + M_\pi \right) \left( \sqrt{M_B} - \sqrt{M_\pi} \right)^2$, such that the magnitude of $|z| \leq 0.280$ is minimized in the semileptonic range. With the analogous choice for $B_s \to K \ell\nu$, $|z| \leq 0.146$ for the semileptonic range.

Although the functional form of the BCL parametrization is simpler than that of BGL, 
the unitarity constraint on the coefficients is more complicated~\cite{Bourrely:2008za}:
\begin{align}
	& \sum _{j,k=0}^{K} B_{jk}(t_0) b_i^{(j)}(t_0) b_i^{(k)} (t_0) \ltapprox 1\,, \\
	& B_{jk}(t_0) = \sum_{n=0}^\infty \eta_{n}(t_0)\eta_{n+|j-k|}(t_0),
\end{align}
where  $\eta_i$ is the Taylor coefficients  in the expansion of the outer function
\begin{eqnarray}
   \Psi(z) &=& \frac{M_{B^*}^2}{4(t_+-t_0)} \phi_i(q^2(z),t_0) \frac{(1-z)^2(1-z_*)^2}{(1-zz_*)^2}, \\
   z_* &=& z(M_B^2, t_0), 
\end{eqnarray}
around $z=0$.
The values of $B_{jk}$ for the $B\to\pi\ell\nu$ and $B_s \to K \ell \nu$ form factors
with the choice $t_0 = t_{\rm opt}$ are given in Table~\ref{tab:Bjk}.

\begin{table}[tb]\caption{\label{tab:Bjk} Matrix elements $B_{jk}(t_0)$ that enter the unitarity bound on the BCL series coefficients for the choice $t_0 = t_{\rm opt}$. The remaining  coefficients can be obtained from the relations $B_{j(j+k)}=B_{0k}$ and the symmetry property $B_{jk}=B_{kj}$.  To derive these results we use the outer functions $\phi_+$ and $\phi_0$ in Eq.~(7) of Ref.~\cite{Arnesen:2005ez} with inputs from Ref.~\cite{Bourrely:2008za}, giving $\chi^{(0)}_+ = 5.03 \times 10^{-4}$ and $\chi^{(0)}_0 = 1.46\times 10^{-2}$.}
\begin{ruledtabular}
\begin{tabular}{lccccccc}
& $B_{00}$&$B_{01}$&$B_{02}$  &$B_{03}$&$B_{04}$ &$B_{05}$ \\\hline
$f_+^{B \pi}$  & 0.0197 & 0.0042  & -0.0109 & -0.0059 & -0.0002  &\ 0.0012 \\
$f_0^{B \pi}$  & 0.1062 & 0.0420  & -0.0368 & -0.0406 & -0.0201  &  -0.0057 \\\hline
$f_+^{B_s  K}$ & 0.0115 & 0.0004  & -0.0076 & -0.0007 &\ 0.0018  &\ 0.0004  \\
$f_0^{B_s  K}$ & 0.0926 & 0.0137  & -0.0484 & -0.0174 & -0.0003  &\ 0.0024  \\
\end{tabular}
\end{ruledtabular}
\end{table}

For the $B\to\pi \ell\nu$ vector form factor, Becher and Hill~\cite{Becher:2005bg} use heavy-quark power counting to provide an estimate for the sum of the coefficients:
\begin{equation}
        \sum_{k=0}^{N} \left( a_+^{(k)} \right)^2 \sim \left( \frac{\Lambda}{m_b}\right)^3  \,,  \label{eq:HQConstraint}
\end{equation}
where $\Lambda$ is a typical hadronic scale.  Taking $\Lambda \sim 1000$~MeV, this would imply $\sum a_k^2 \sim 0.01$, which is well below the bound from unitarity.  Experimental measurements~\cite{delAmoSanchez:2010af,Ha:2010rf,Lees:2012vv,Sibidanov:2013rkk} and previous lattice calculations~\cite{Bailey:2008wp} confirm this expectation.  This argument also applies to the $B_s \to K \ell\nu$ vector form factor, where we emphasize that Eq.~(\ref{eq:HQConstraint}) is only a rough constraint due to the imprecise scale $\Lambda$ and omitted higher-order corrections in the OPE and $1/m_b$.

\subsection{\texorpdfstring{Extrapolation of lattice form factors to $q^2=0$}{Extrapolation of lattice form factors to q2=0}}
\label{Sec:LatzFit}

We now extrapolate our results for the $B\to\pi\ell\nu$ and $B_s\to K\ell\nu$ form factors to $q^2=0$ using the $z$-expansion.   We first generate synthetic data points in the range of simulated data from the output of the chiral-continuum extrapolation.   Recall that the continuum, physical quark-mass form factors are obtained from fits to Eqs.~(\ref{eq:fpar_ChPT}) and~(\ref{eq:fperp_ChPT}) by fixing $M_\pi^2$ to the physical value and $a^2 \to 0$.  After these replacements, the physical form factors depend upon three independent functions of the pion or kaon energy $E_P$.  We therefore generate three synthetic data points each for $f_0$ and $f_+$ in order to ensure that the covariance matrix is not singular.   In anticipation of the $z$-fit, we choose the points to be evenly spaced in $z$ (rather than $q^2$).  The $q^2$ values and error budgets for the synthetic lattice data are given in Table~\ref{tab:ErrorBudget}.

We fit our synthetic lattice data for the $B\to\pi\ell\nu$ and $B_s\to K\ell\nu$ form factors including statistical and systematic correlations between $q^2$ values.  For our preferred fit we use the BCL parametrization with the kinematic constraint $f_+(0)=f_0(0)$ and use the theoretical estimate from heavy-quark power counting to constrain the sum of the coefficients of the vector form factor via Bayesian priors.  We study the central values and errors of the series coefficients as a function of the truncation $K$ such that our final form-factor results include the truncation error.  The complete $z$-fit results are given in Appendix~\ref{app:zfits}.   We also compare to results using the BGL parametrization as a check.

\begin{figure*}[tb]
\includegraphics[width=.49\textwidth]{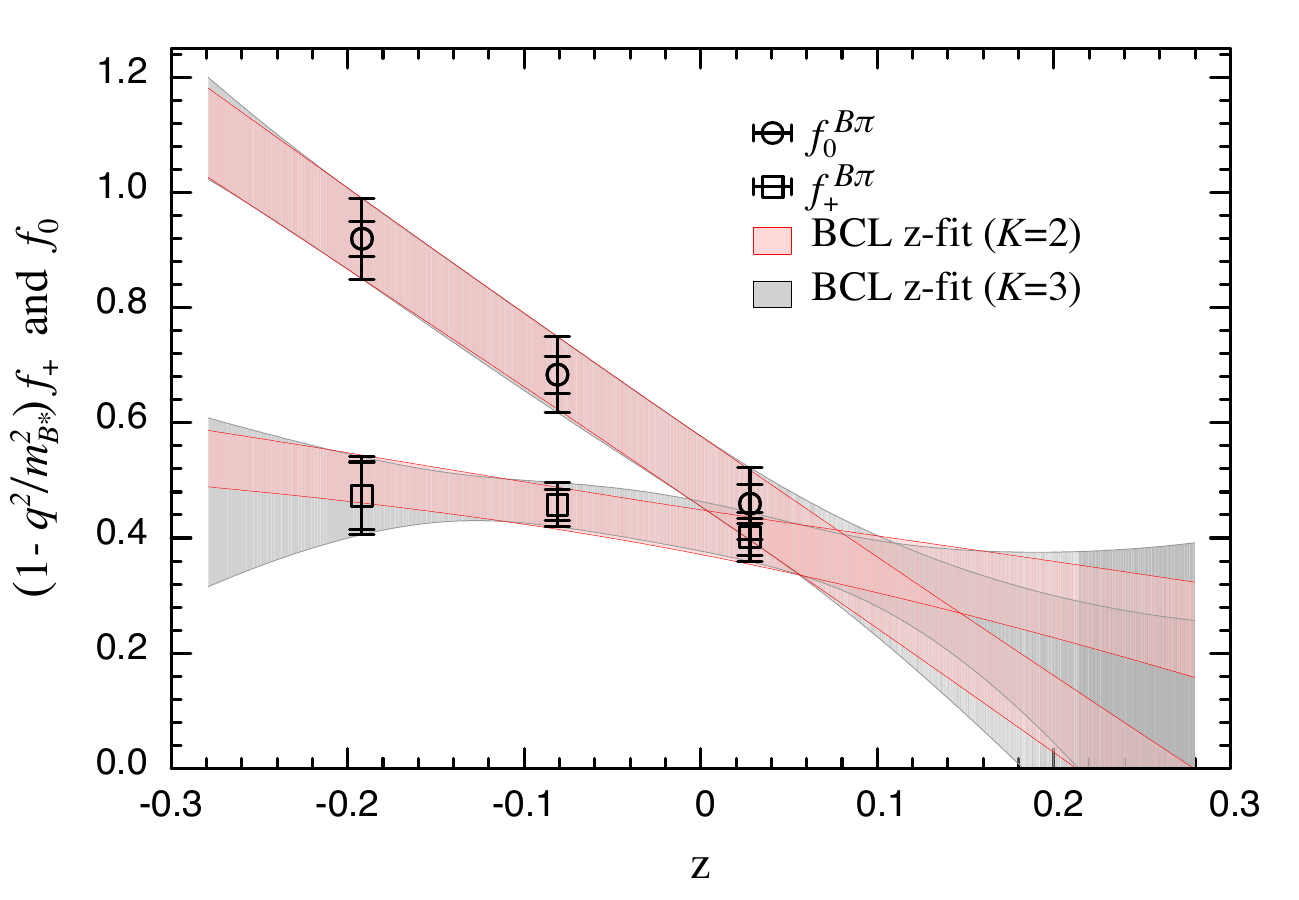}
\includegraphics[width=.49\textwidth]{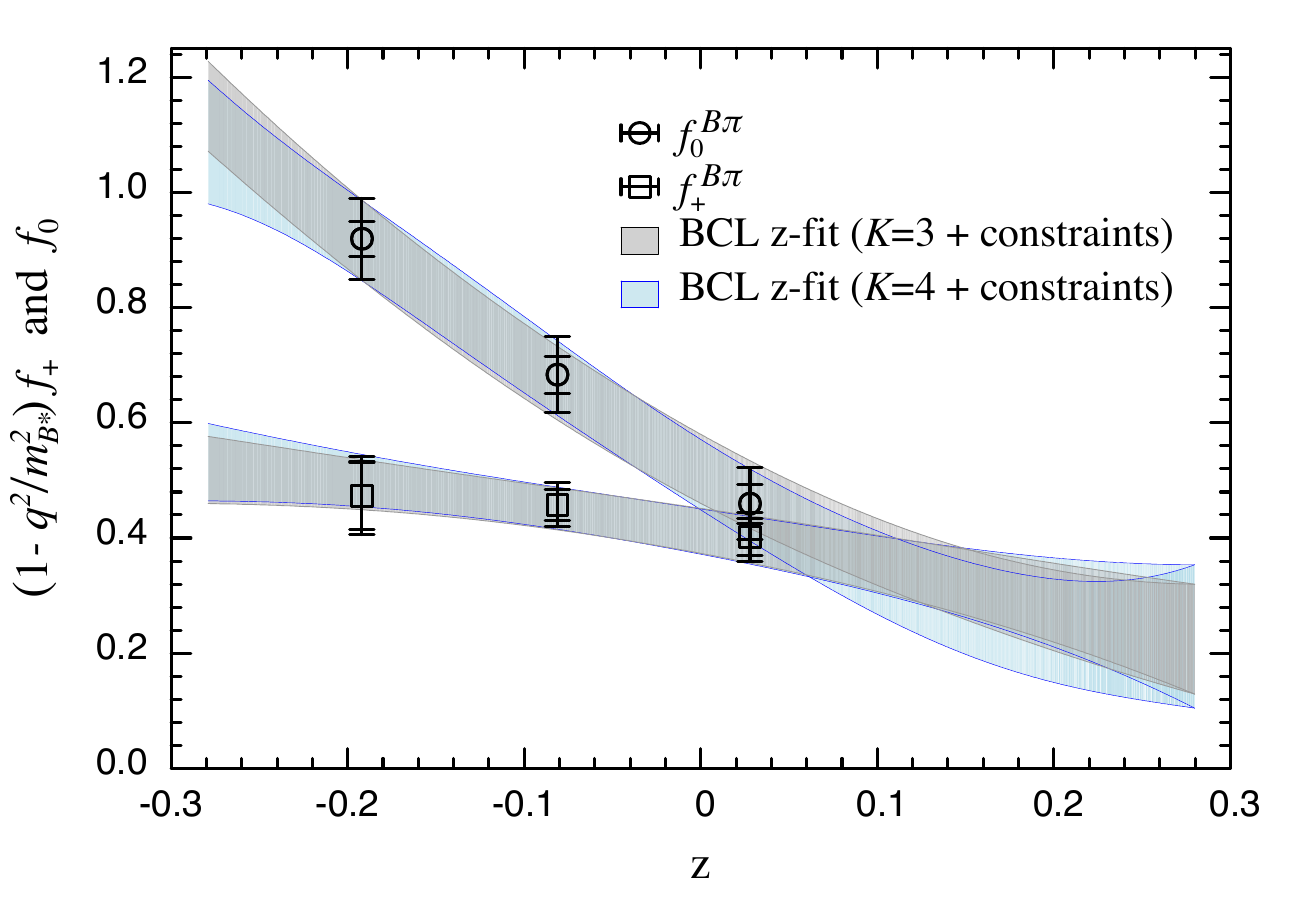} 
\includegraphics[width=.49\textwidth]{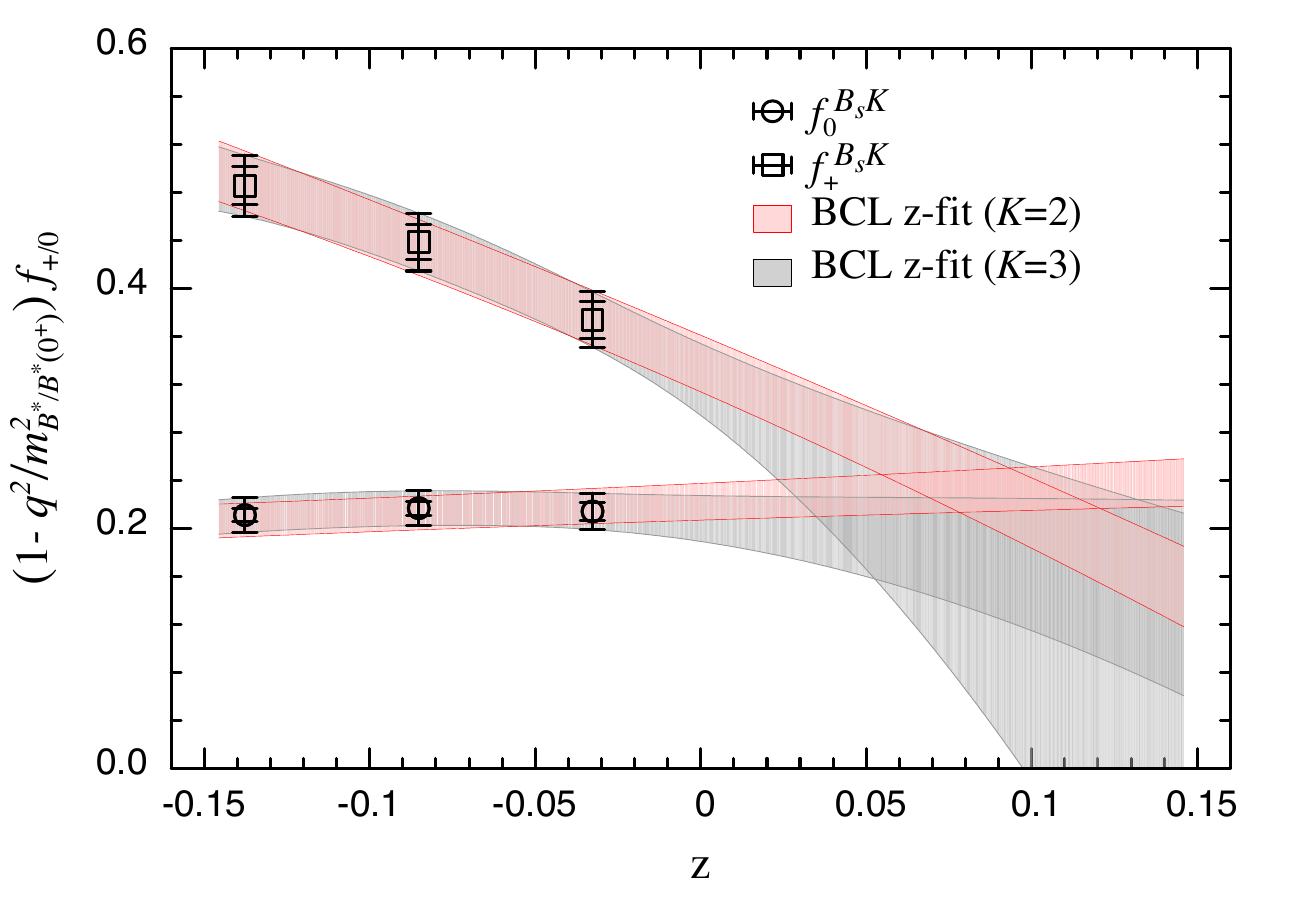} 
\includegraphics[width=.49\textwidth]{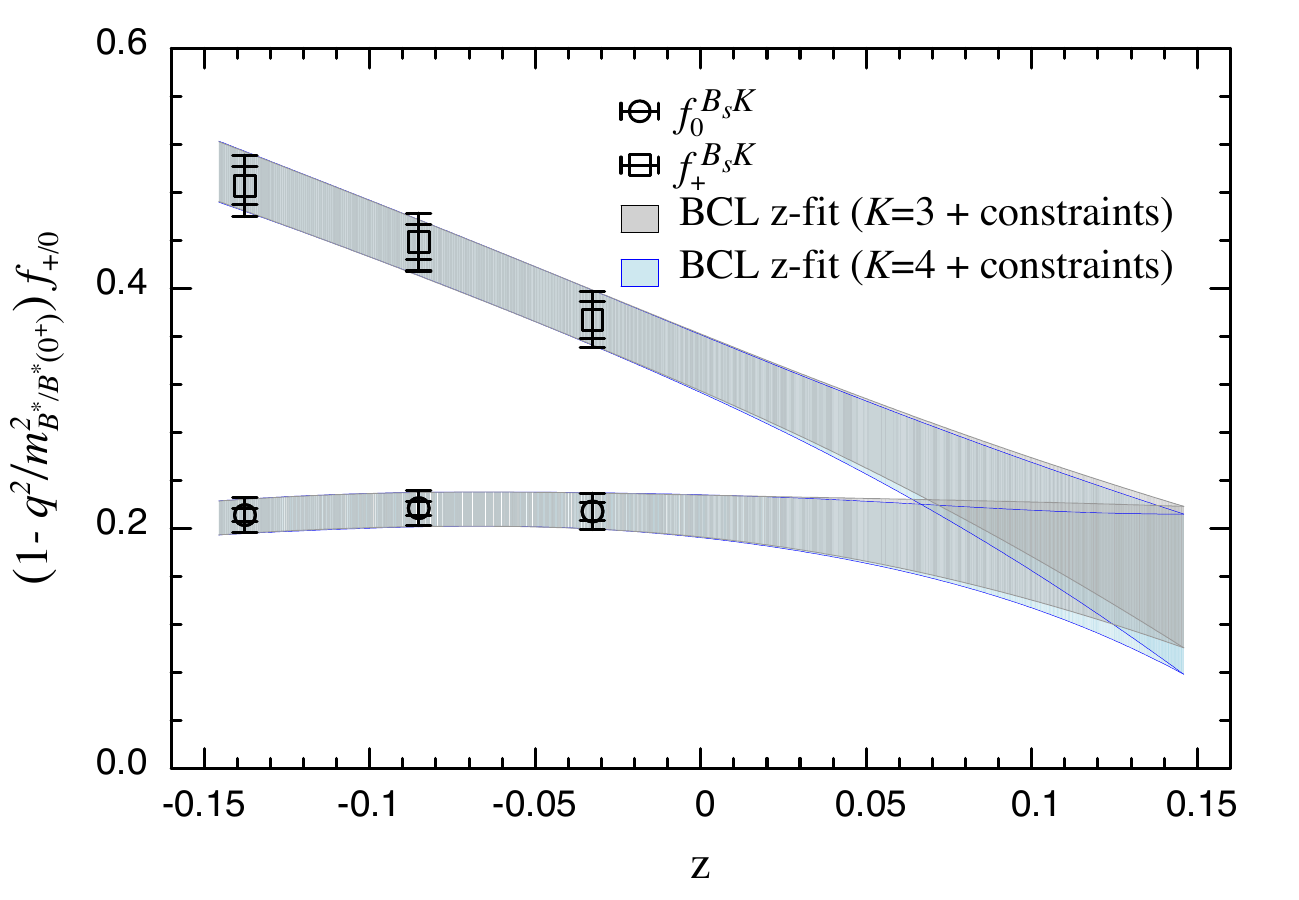} 
\caption{Fits of the $B\to\pi\ell\nu$ (upper plots) and $B_s\to K\ell\nu$ (lower plots) lattice form factors to the $z$-expansion versus truncation $K$ without constraints (left) and with the kinematic and heavy-quark constraints (right).  The black open symbols show the synthetic data points with statistical (inner) and statistical $\oplus$ systematic (outer) error bars.  The curves with colored bands show the fit results with errors for different truncations $K$.  We do not show unconstrained fits with $K=4$ in the left-hand plot because we only have three synthetic data points;  the inclusion of the kinematic and heavy-quark constraints allows us to perform the $K=4$ fit shown in the right-hand plot.  In the right-hand plot, we do not show the $K=2$ combined fit because of the poor fit quality.}
\label{fig:BtoPiLatzFits}
\end{figure*}

We first perform separate fits of $f_+$ and $f_0$ without imposing any constraints on the sum of coefficients.  The results for $B\to\pi\ell\nu$ are given in the top two panels of Table~\ref{tab:BtoPiLatzFits}, and for $B_s \to K \ell \nu$ in the upper two panels of Table~\ref{tab:BstoKLatzFits}.  The separate fits of $f_+$ and $f_0$ for $K=2,3$ are shown in the left-hand plots of Fig.~\ref{fig:BtoPiLatzFits} for $B\to\pi\ell\nu$ (upper) and $B_s\to K\ell\nu$ (lower).    The synthetic lattice data points are correlated, and one must include a term quadratic in $z$ to obtain a good fit (recall that for $f_+$ the expression with $K=2$ includes a term proportional to $z^2$ that is related to the $z^0$ and $z^1$ terms).  The normalizations $b_i^{(0)}$ are well determined by the lattice data, with central values that are stable within errors when going from $K=2$ to $K=3$.  This is important because the normalization of the vector form factor plays a key role in the determination of $|V_{ub}|$ (see Sec.~\ref{Sec:Vub}).  We cannot go beyond $K=3$ because we have only three synthetic data points.

In the separate fits to $f_+$ and $f_0$ with $K=3$, the kinematic constraint $f_+(0)=f_0(0)$ is automatically satisfied within uncertainties, but with large errors.  We can therefore impose the kinematic constraint $f_+(0)=f_0(0)$.  The results of the combined fits are given in the third panels of Tables~\ref{tab:BtoPiLatzFits} and~\ref{tab:BstoKLatzFits}.  As expected, the constrained fits with $K=2$ for both $f_+$ and $f_0$ have poor $p$-values, but the remaining fits tried are all of good quality.  Adding the kinematic constraint (and only considering the good fits) has little impact on the results for the normalizations and even on the slopes ($b^{(1)}_i/b^{(0)}_i$).  It reduces the error on the curvatures ($b^{(2)}_i/b^{(0)}_i$) as compared to the separate fits with $K=3$, however, and consequently improves the determination of $f_+(q^2=0)$.

Even with the kinematic constraint, however, the slopes and curvatures of the form factors are still not well determined by the lattice data, with errors ranging from 25\% to as much as 300\%.  For all fits considered, the sum of the coefficients $\sum B_{jk}b_jb_k$ satisfy the unitarity constraint.   Further, for $f_+$, the sum $\sum B_{jk}b_jb_k$ is also consistent with expectations from heavy-quark power counting, Eq.~(\ref{eq:HQConstraint}), but with large uncertainties.   We can therefore use theoretical guidance from heavy-quark power counting to further improve our lattice form-factor determination.  Keeping the kinematic constraint, we also constrain the sum of the coefficients of the $B\to\pi\ell\nu$ and $B_s \to K\ell\nu$ vector form factors with Bayesian priors based on their estimated size from heavy-quark power counting.  For the hadronic scale in the heavy-quark estimate we take 1000~MeV, with a generous uncertainty of $\pm 500$~MeV.  Thus for the prior central value we use $\bar{B} = 0.01$, and for the Gaussian prior width we use $\sigma_B = 0.03$.  We implement the Bayesian fit by minimizing the augmented $\chi^2$~\cite{Lepage:2001ym},
\begin{equation}
	\chi^2_{\rm aug} = \chi^2 + \chi^2_{\rm prior} \,, \label{eq:chi2Aug}
\end{equation}
where
\begin{equation}
	\chi^2_{\rm prior} = (\bar{B} - \sum B_{jk}b_jb_k)^2 / \sigma_B^2 \,.
\end{equation}
The results for different truncations $K$ are given in the bottom panels of Tables~\ref{tab:BtoPiLatzFits} and~\ref{tab:BstoKLatzFits}.  The inclusion of the heavy-quark constraint improves the determinations of the slopes and curvatures, and leads to a reduction in the absolute error on $f_+^{B\pi}(0)$ by about a factor of 2 for $B\to\pi\ell\nu$ for $K=3$.  The improvement in the error on $f_+^{B_sK}(0)$ is smaller but non-negligible, about 25\%.  

After implementing the heavy-quark constraint, we are able to include an additional parameter in our fits and can consider expansions with $K=4$.  This enables us to study the stability of the central values and errors of the parameters with truncation $K$, and thus assess the systematic uncertainty associated with truncating the $z$-expansion.  The central values and errors for the normalizations and slopes are stable when increasing the truncation from $K=3$ to $K=4$, in most cases changing only in the last decimal place (except for the slope of $f_0^{B_s K}(q^2)$, for which the results are still consistent within uncertainties).  The combined fits of $f_+$ and $f_0$ imposing the kinematic and heavy-quark constraints are shown versus the truncation $K$ in the right-hand plots of Fig.~\ref{fig:BtoPiLatzFits} for $B\to\pi\ell\nu$ (upper) and $B_s\to K\ell\nu$ (lower).   The central fit curves for $K=3$ and $K=4$ lie almost on top of each other, while the widths of the error bands and the uncertainties in $f_+(0)$ increase only slightly in going to $K=4$.  Thus we conclude that the $K=3$ constrained fit includes the systematic uncertainty due to truncating the series in $z$.

\begin{figure*}[tb]
\includegraphics[width=.49\textwidth]{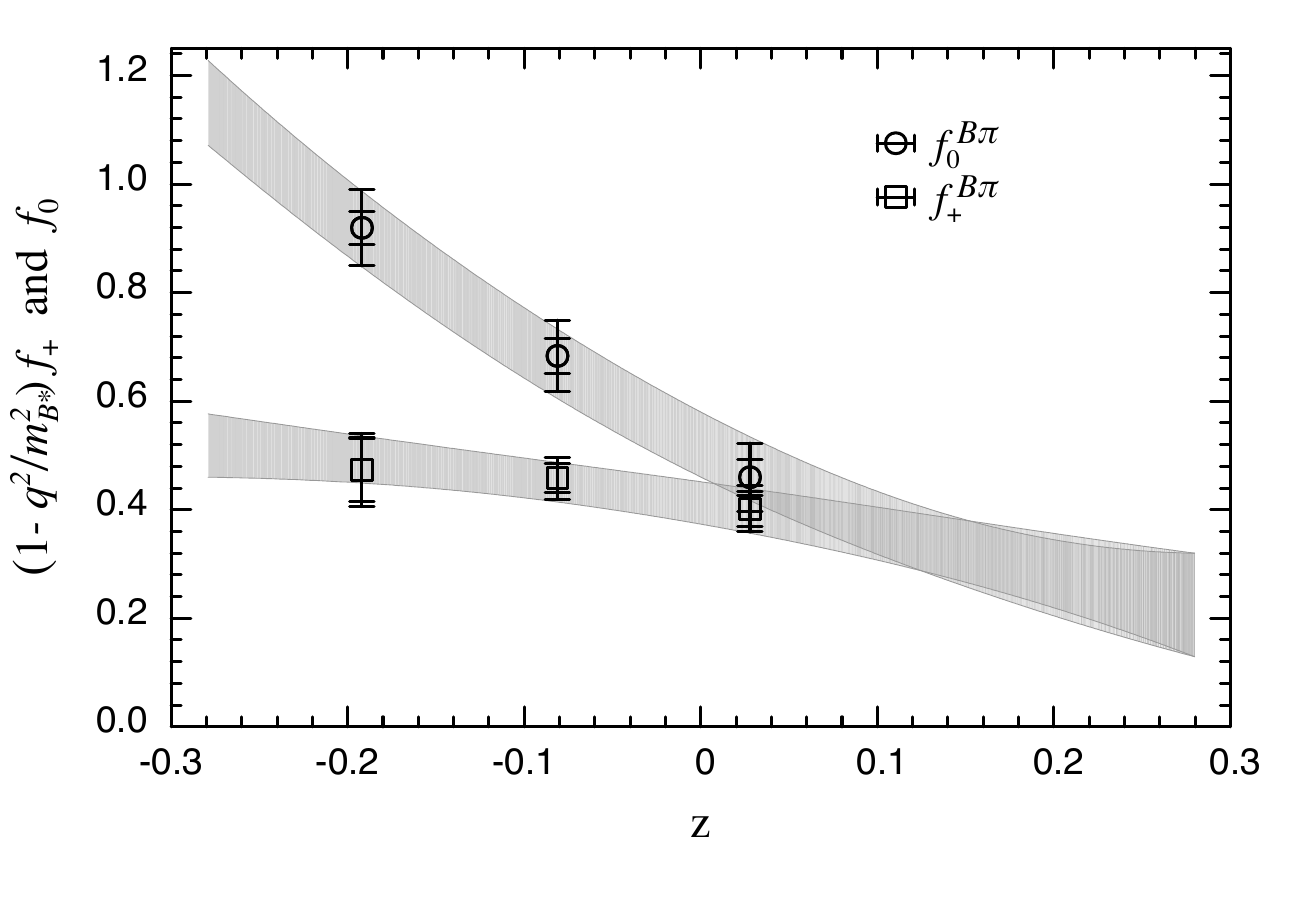}
\includegraphics[width=.49\textwidth]{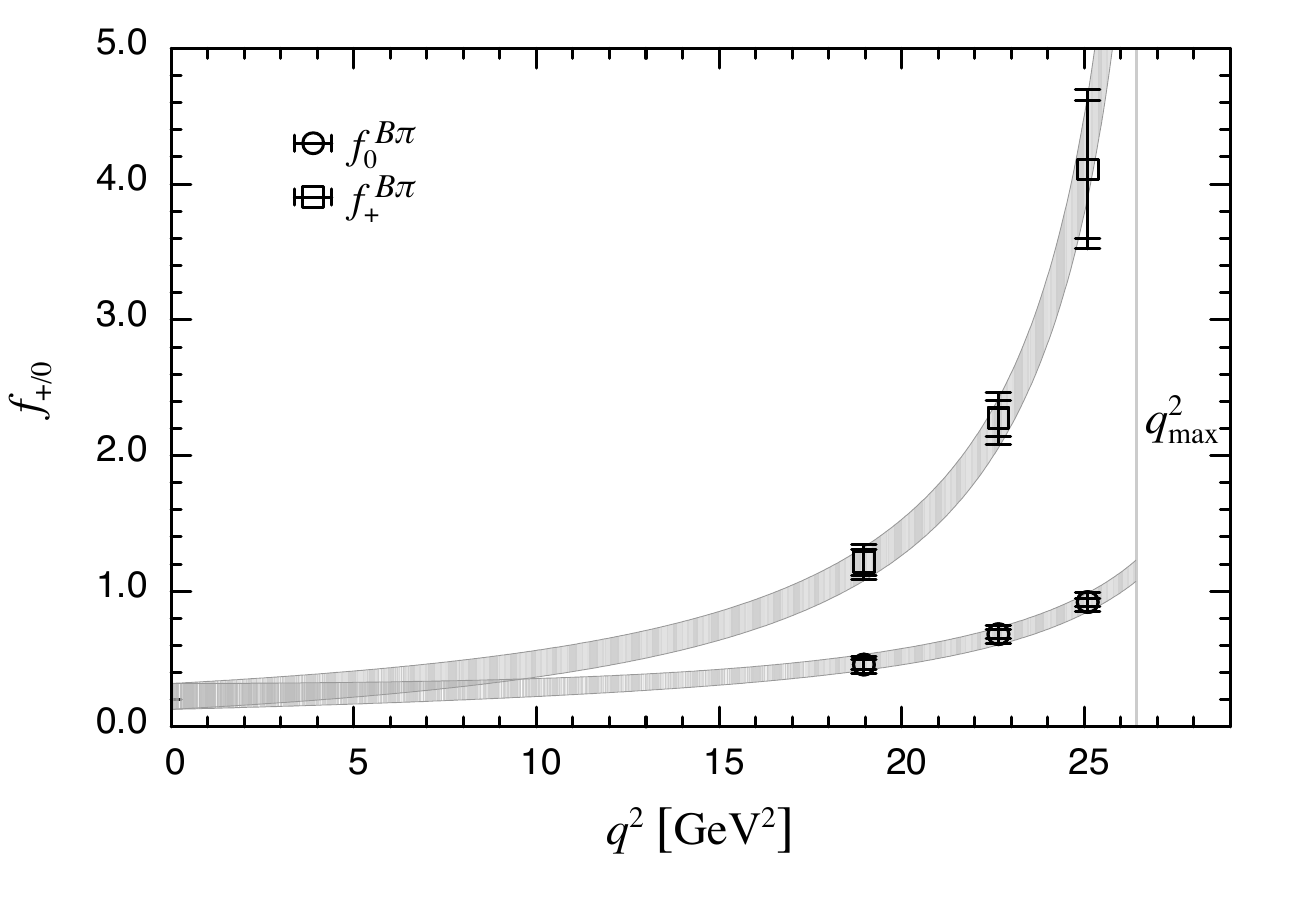} 
\includegraphics[width=.49\textwidth]{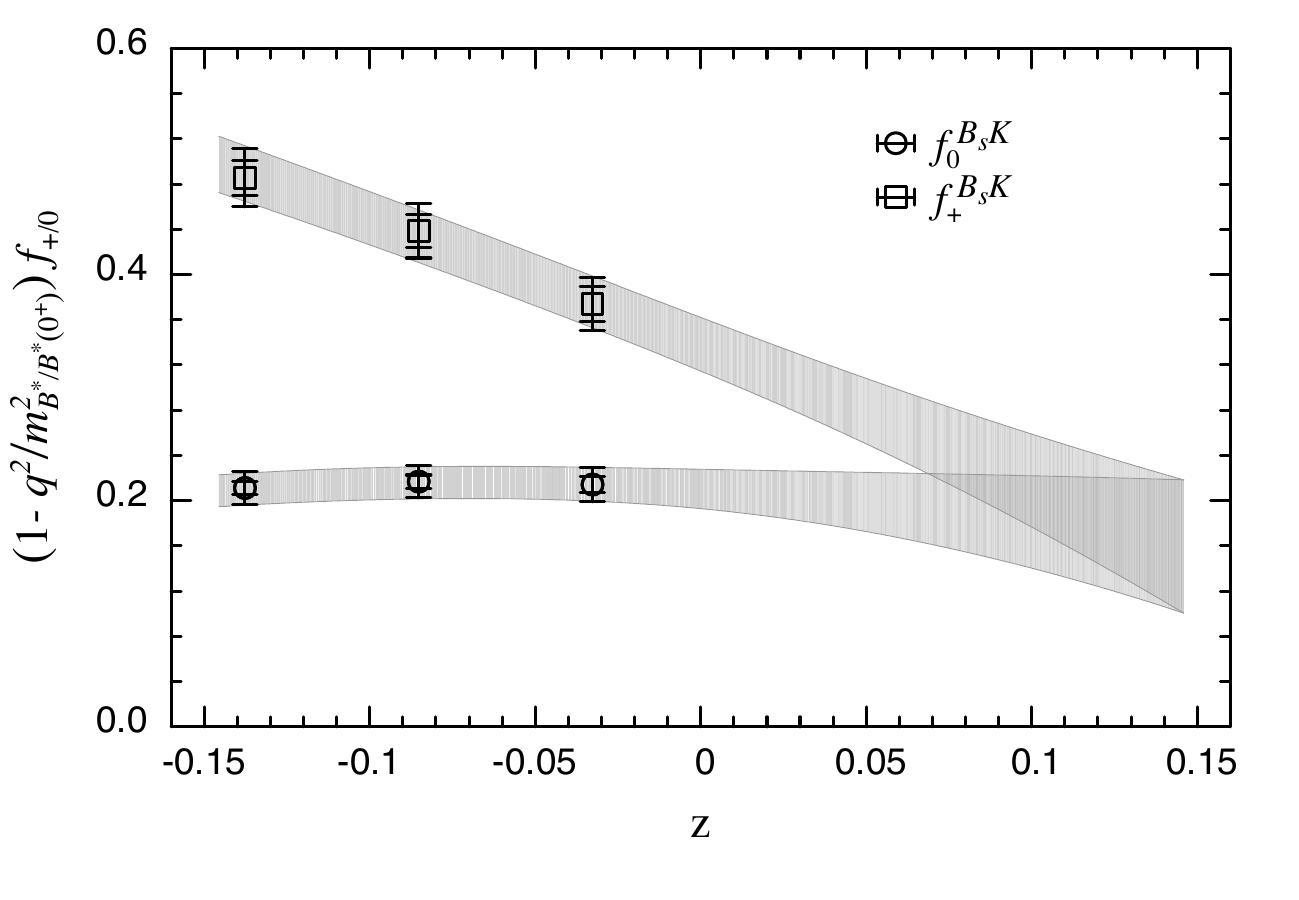}
\includegraphics[width=.49\textwidth]{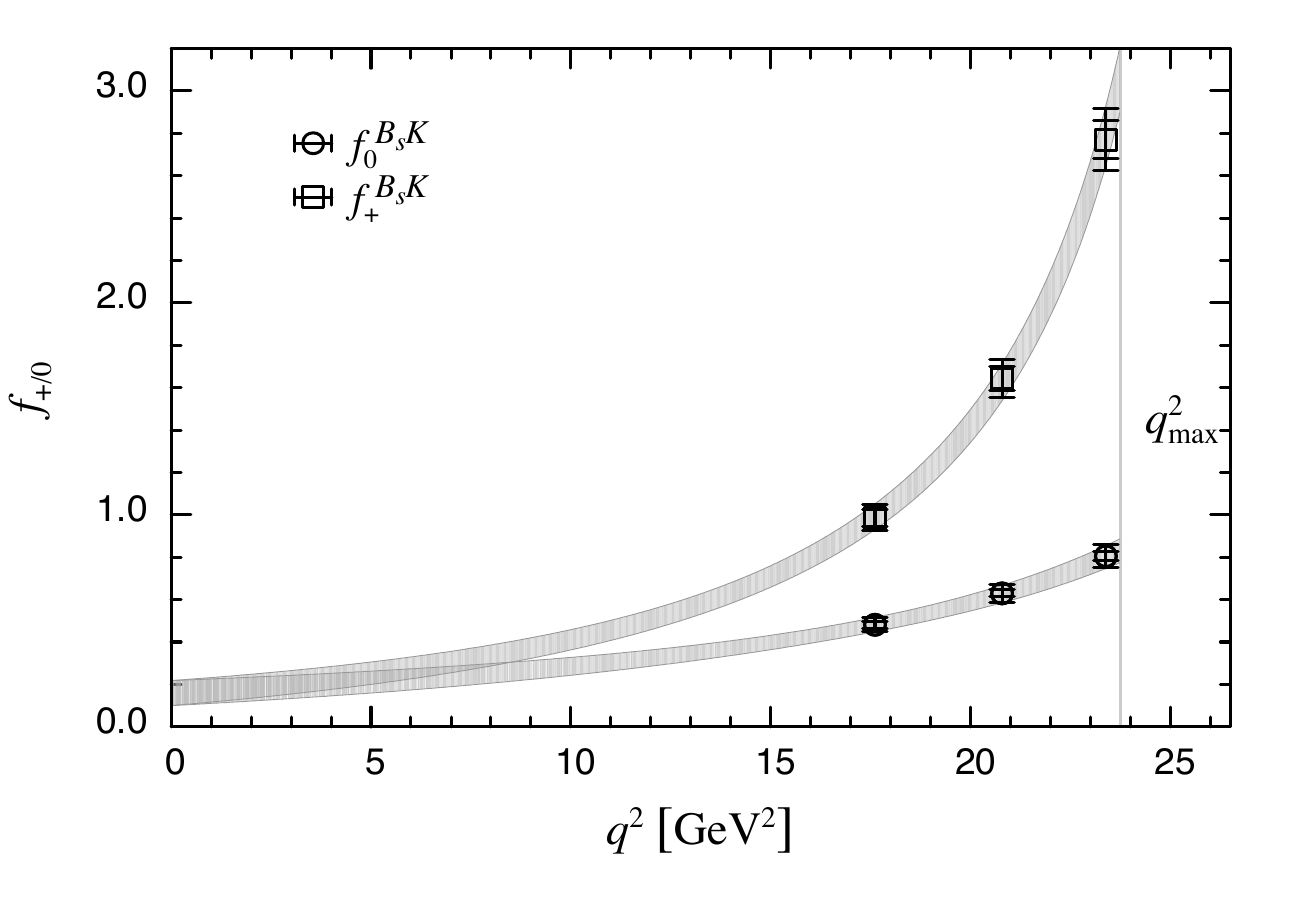} 
\caption{Preferred $K=3$ fit of the $B\to\pi\ell\nu$ (upper plots) and $B_s\to K\ell\nu$ (lower plots) lattice form factors to the $z$-expansion including the kinematic and heavy-quark constraints versus $z$ (left) and versus $q^2$ (right).  The black open symbols show the synthetic data points with statistical (inner) and statistical $\oplus$ systematic (outer) error bars.  The solid curves with error bands show the fit results for $f_+(q^2)$ and $f_0(q^2)$.}
\label{fig:BtoPiPreferredFits}
\end{figure*}

We therefore take as our preferred fits for $B\to\pi\ell\nu$ and $B_s\to K \ell\nu$ the results from the fit with $K=3$ for both $f_+$ and $f_0$ including the kinematic and heavy-quark constraints.  This is the highest truncation $K$ for which we still have more data points than fit parameters, and the uncertainties are comparable to the $K=4$ fits.  Figure~\ref{fig:BtoPiPreferredFits} shows our preferred fits for $B\to\pi\ell\nu$ (upper plots) and $B_s \to K \ell\nu$ (lower plots) plotted versus $z$ (left) and versus $q^2$ right.

\begin{figure*}[tb]
\includegraphics[width=.49\textwidth]{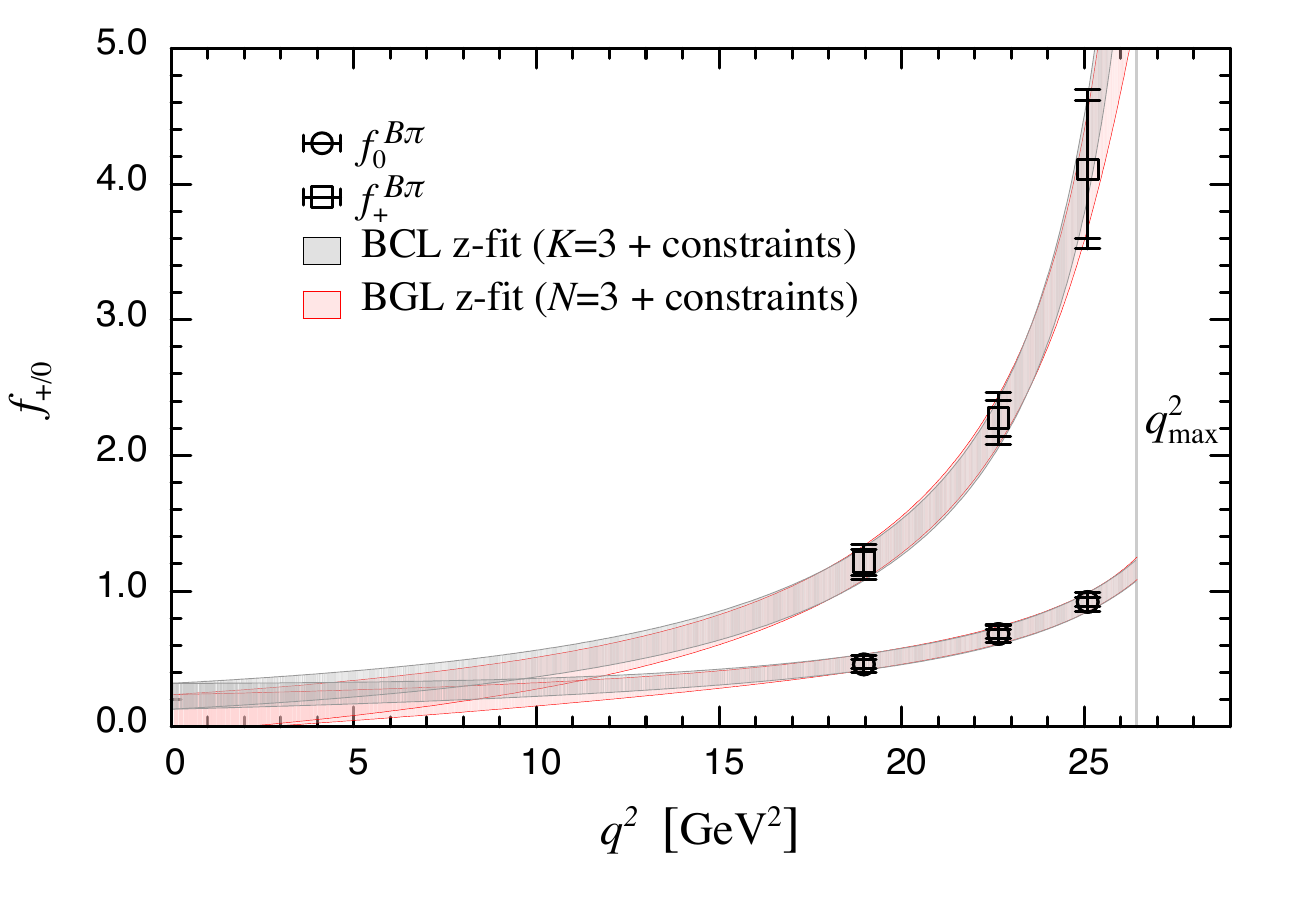} 
\includegraphics[width=.49\textwidth]{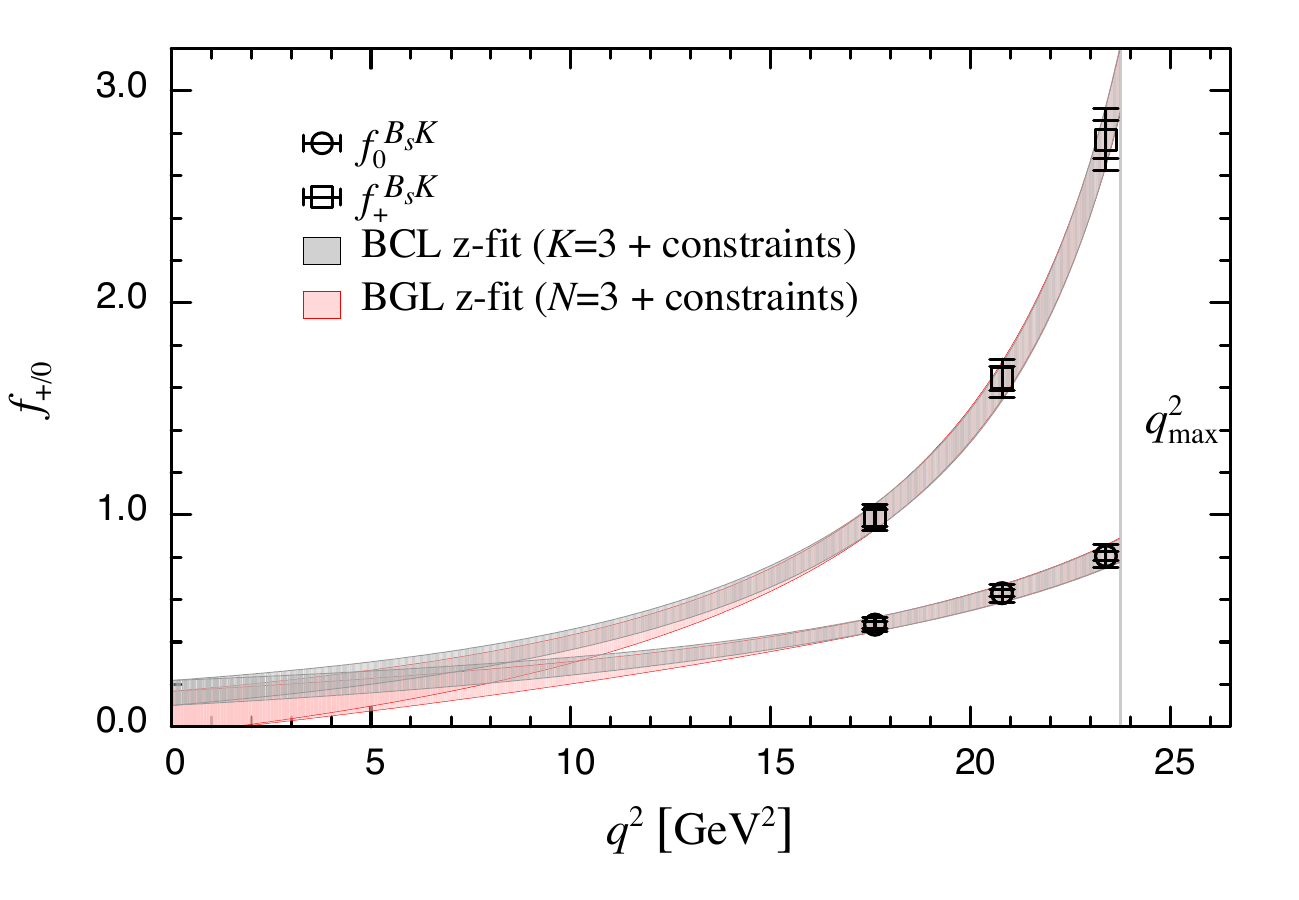} 
\caption{Comparison of fits to the $B \to \pi\ell\nu$ (left) and $B_s\to K\ell\nu$ (right) lattice form-factor data using the BCL and BGL parametrizations.  The black open symbols show the synthetic data points with statistical (inner) and statistical $\oplus$ systematic (outer) error bars.  The gray and red curves with error bands show the BCL and BGL fits, respectively. }
\label{fig:BCLvsBCLFits}
\end{figure*}

As a cross-check, we compare our preferred fit using the BCL parametrization to the analogous fit (also imposing the kinematic and heavy-quark constraints, and to the same order in $z$) using the BGL parametrization.  Figure~\ref{fig:BCLvsBCLFits} overlays the results of the BCL and BGL fits for $B\to\pi\ell\nu$ (left) and $B_s \to K \ell\nu$ (right).  The fits to the different series expansions are consistent, indicating that our quoted form-factor uncertainties encompass the error due to truncating the $z$-expansion.  The error bands from the BCL fits are narrower because the BCL form for $f_+$ relates the coefficient of highest-order term in $z$ to the coefficients of the lower-order terms.  

Tables~\ref{tab:BtoPiBCLResults} and~\ref{tab:BstoKBCLResults} present our final results for the $B\to\pi\ell\nu$ and $B_s \to K \ell\nu$ form factors as coefficients of the $z$-expansion and the matrix of correlations between them.  These results are model independent and valid over the entire semileptonic region of $q^2$.  As we illustrate in the next section, they can be used in combined fits with experimental data to obtain the CKM matrix element $|V_{ub}|$, or to make predictions for Standard-Model observables for these decay processes.

\begin{figure*}[tb]
\includegraphics[width=.49\textwidth]{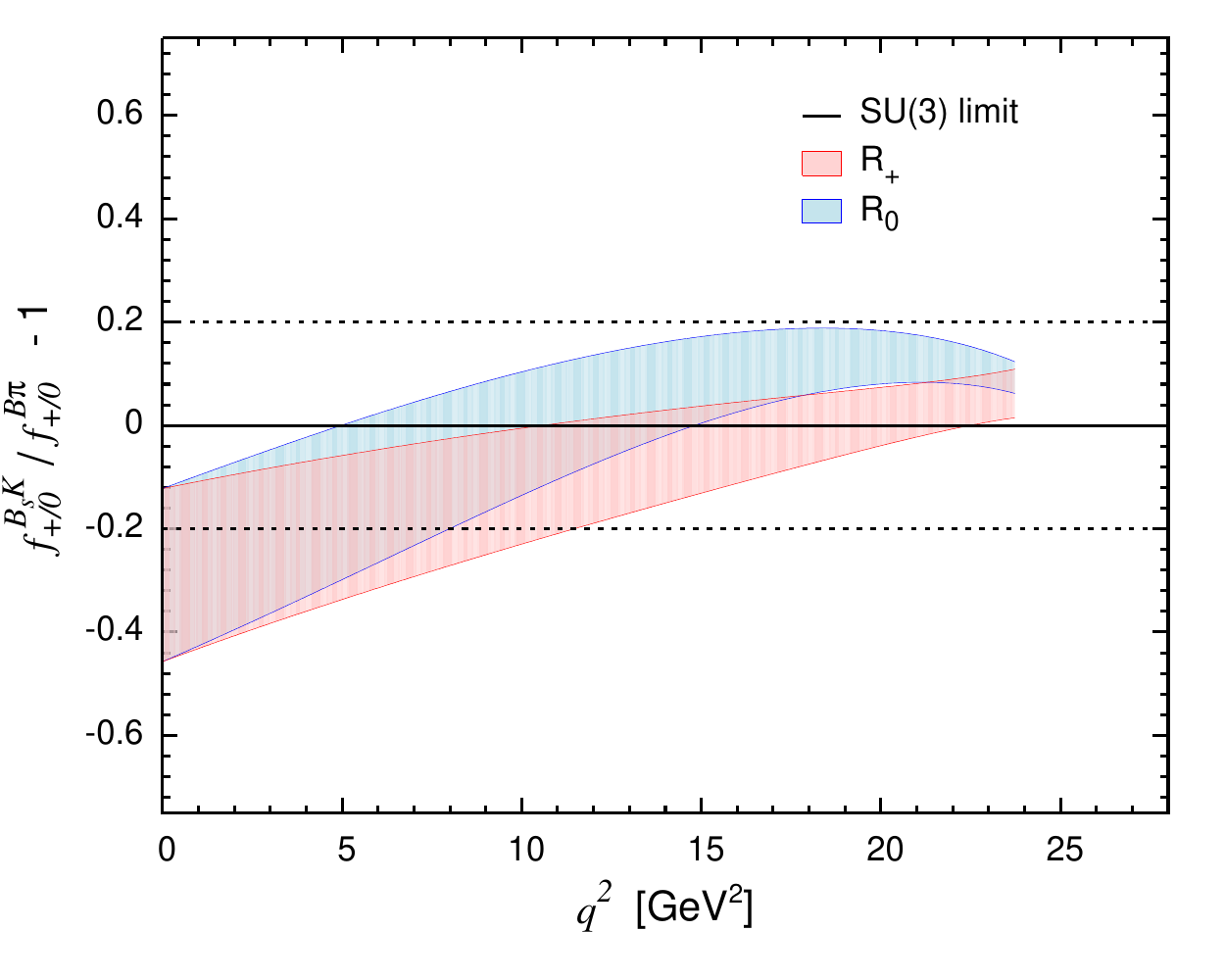} 
\includegraphics[width=.49\textwidth]{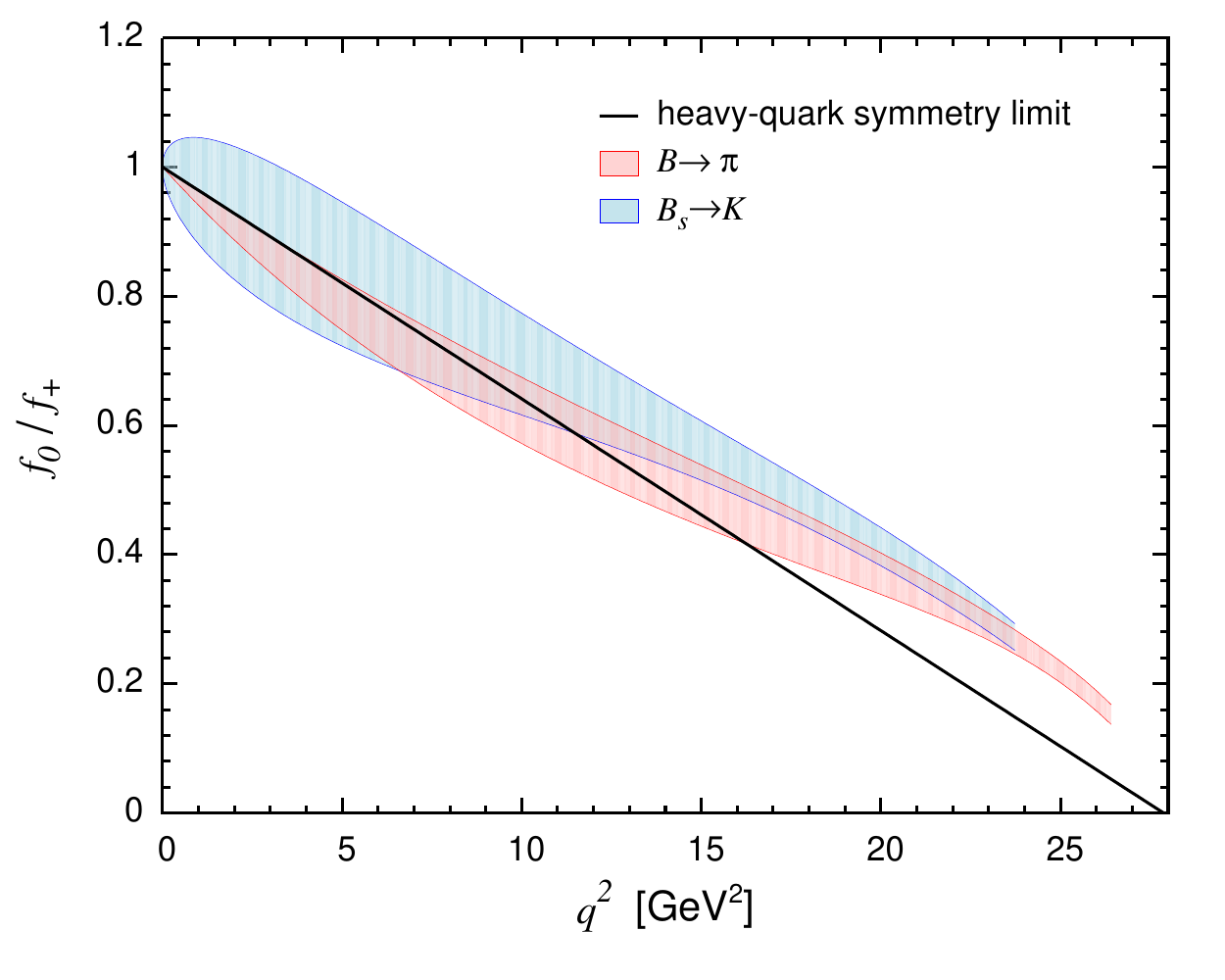} 
\caption{Left:  SU(3)-breaking ratios $R_+(q^2)$ (red) and $R_0(q^2)$ (blue).  The dashed horizontal lines show the expected size of SU(3) breaking from simple power counting. 
Right:  ratios of scalar to vector form factor $f_0(q^2)/f_+(q^2)$ for $B\to\pi\ell\nu$ (red) and $B_s \to K \ell\nu$ (blue).  The black diagonal line shows the prediction in the heavy-quark symmetry limit.   
Errors are shown in shaded bands. In the left plot, errors shown are statistical only.
\label{fig:SU3_and_HQS_Ratios}}
\end{figure*}

It is interesting to compare ratios of these form factors to predictions from approximate symmetries of QCD.  In the SU(3) limit ($m_d = m_s$), the form factors for $B\to\pi\ell\nu$ and $B_s \to K \ell\nu$ should be identical.  Thus the ratios $R_i(q^2) = f_i^{BK}(q^2)/f_i^{B\pi }(q^2) - 1$, for $i=\{+,0\}$, provide a measure of SU(3)-breaking in $B \to {\rm light}$ semileptonic form factors.  Figure~\ref{fig:SU3_and_HQS_Ratios}, left, plots these ratios for the full kinematic range.  The results for $f_+$ and $f_0$ are similar.  The deviations from unity are consistent with expectations from simple power counting of $(m_s - m_d) / \Lambda_{\rm QCD} \sim 20\%$, but with large uncertainties.   

At large recoil (low $q^2$) and in the heavy-quark symmetry limit ($m_b / \Lambda_{QCD} \to \infty$), the $B \to\pi\ell\nu$ and $B_s \to K \ell\nu$ processes are each described by a single independent form factor as follows~\cite{Beneke:2000wa}:
\begin{eqnarray}
	f_0(q^2) &=& \frac{ m_{B_{(s)}}^2 - q^2 }{m_{B_{(s)}}^2} f_+(q^2) \,. \label{eq:HQS_f0}
\end{eqnarray}
This expression reduces to the kinematic constraint $f_+(q^2=0) = f_0(q^2=0)$ at $q^2=0$.  Figure~\ref{fig:SU3_and_HQS_Ratios}, right, plots the ratio $f_0(q^2) / f_+ (q^2)$ for the full kinematic range.  The results are similar for $B \to\pi\ell\nu$ and $B_s \to K \ell\nu$.  They agree exactly with the  prediction from Eq.~(\ref{eq:HQS_f0}) at $q^2=0$ by construction because we imposed the kinematic constraint in our preferred $z$-fit, but are consistent with heavy-quark-symmetry expectations throughout the low $q^2$ region.  

\begin{table}[tb]
  \caption{\label{tab:BtoPiBCLResults} Central values, errors, and correlation matrix for the parameters of our preferred fit to the $B\to\pi\ell\nu$ form factors $f_+$ and $f_0$ to the BCL $z$-parametrization in Eqs.~(\ref{eq:BCLConstraint}) and~(\ref{eq:f0BtoPiBCL}) using $t_0 = t_{\rm opt} \equiv \left( M_B + M_\pi \right) \left( \sqrt{M_B} - \sqrt{M_\pi} \right)^2$ in Eq.~(\ref{eq:slep:z_trans}) and pole mass $M_{B^*}=5.3252(4)$~GeV in $f_+(q^2)$.}
  \begin{ruledtabular}\begin{tabular}{llrrrrrr}
      & & \multicolumn{6}{c}{Correlation matrix}  \\[1mm]
      & value& $b^{(0)}_+$ & $b^{(1)}_+$ & $b^{(2)}_+$ & $b^{(0)}_0$ & $b^{(1)}_0$ & $b^{(2)}_0$ \\ \hline
      $b^{(0)}_+$ &\  0.412(39)  &  1.000 & 0.337  &-0.076 &  0.679 &  0.045 &  0.100 \\
      $b^{(1)}_+$ &  -0.511(184) &  0.337 & 1.000  & 0.150 &  0.222 &  0.698 &  0.581 \\
      $b^{(2)}_+$ &  -0.524(612) & -0.076 & 0.150  & 1.000 &  0.029 &  0.436 &  0.659 \\
      $b^{(0)}_0$ &\  0.520(60)  &  0.679 & 0.222  & 0.029 &  1.000 & -0.258 & -0.224 \\
      $b^{(1)}_0$ &  -1.657(182) &  0.045 & 0.698  & 0.436 & -0.258 &  1.000 &  0.564 \\
      $b^{(2)}_0$ &\  2.146(682) &  0.100 & 0.581  & 0.659 & -0.224 &  0.564 &  1.000 \\
    \end{tabular}
  \end{ruledtabular}
\end{table}

\begin{table}[tb]
\caption{\label{tab:BstoKBCLResults} Central values, errors, and correlation matrix for the parameters of our preferred fit to the $B_s\to K\ell\nu$ form factors $f_+$ and $f_0$ to the BCL $z$-parametrization in Eqs.~(\ref{eq:BCLConstraint}) and~(\ref{eq:f0BstoKBCL}) using $t_0 = t_{\rm opt} \equiv \left( M_{B_s} + M_K \right) \left( \sqrt{M_{B_s}} - \sqrt{M_K} \right)^2$ in Eq.~(\ref{eq:slep:z_trans}) and pole masses $M_{B^*}=5.3252(4)$~GeV and $M_{B^*}(0^+)=5.63$~GeV in $f_+(q^2)$ and $f_0(q^2)$, respectively.}
  \begin{ruledtabular}\begin{tabular}{llrrrrrr}
      & & \multicolumn{6}{c}{Correlation matrix}  \\[1mm]
      & value& $b^{(0)}_+$ & $b^{(1)}_+$ & $b^{(2)}_+$ & $b^{(0)}_0$ & $b^{(1)}_0$ & $b^{(2)}_0$ \\ \hline
$b^{(0)}_+$ &\  0.338(24)    & 1.000 & 0.255 & 0.146 & 0.873 & 0.603 & 0.423 \\
$b^{(1)}_+$ &  -1.161(192)   & 0.255 & 1.000 & 0.823 & 0.311 & 0.954 & 0.770 \\
$b^{(2)}_+$ &  -0.458(1.009) & 0.146 & 0.823 & 1.000 & 0.346 & 1.060 & 0.901 \\
$b^{(0)}_0$ &\  0.210(17)    & 0.873 & 0.311 & 0.346 & 1.000 & 0.556 & 0.479 \\
$b^{(1)}_0$ &  -0.169(202)   & 0.603 & 0.954 & 1.060 & 0.556 & 1.000 & 0.965 \\
$b^{(2)}_0$ &  -1.235(880)   & 0.423 & 0.770 & 0.901 & 0.479 & 0.965 & 1.000 \\
\end{tabular}
\end{ruledtabular}
\end{table}

\section{Phenomenological applications}
\label{Sec:Pheno}

In this section we present two phenomenological applications of our form-factor results.

First, in Sec.~\ref{Sec:Vub}, we use our results for the $B\to\pi\ell\nu$ form factors to determine the CKM matrix element $|V_{ub}|$.  We fit recent  experimental measurements of the $B\to\pi\ell\nu$  differential branching fraction to the $z$-parametrization to obtain the slope $b^{(1)}_+/b^{(0)}_+$ and curvature $b^{(2)}_+/b^{(0)}_+$.  Confirming that the lattice and experimental shapes are indeed consistent, we then perform a combined $z$-fit of our numerical $B\to\pi\ell\nu$ form-factor data with the experimental measurements to obtain a model-independent determination of $|V_{ub}|$.  This method can also be applied to the decay $B_s \to K \ell \nu$, once it has been observed experimentally, to provide an alternate determination of $|V_{ub}|$.

Next, in Sec.~\ref{Sec:BstoKPheno}, we make predictions for Standard-Model observables for the decay processes $B\to\pi\ell\nu$ and $B_s \to K \ell \nu$ for both $\ell=\mu,\tau$ final-state charged leptons.  (Here we use $\mu$ to indicate both muon and electron final states, for which the Standard-Model predictions are indistinguishable at the current level of precision.)  We show results for the differential branching fractions, forward-backward asymmetries, and $\mu/\tau$ ratios (which are independent of $|V_{ub}|$).  We only calculate observables that depend upon $|V_{ub}|$ for $B_s\to K \ell\nu$ decays, using the value determined previously in Sec.~\ref{Sec:Vub}.  Once the experimental error on the branching fraction is commensurate with the theoretical form-factor uncertainties, our $B_s \to K \ell \nu$ form-factor results will enable a sufficiently precise determination of $|V_{ub}|$ to illuminate the discrepancy between $|V_{ub}|$ from inclusive $B \to X_u \ell\nu$ and exclusive $B\to\pi\ell\nu$ semileptonic decays.  

\subsection{\texorpdfstring{Determination of $|V_{ub}|$ from $B\to\pi\ell\nu$}{Determination of Vub from B->pi l nu}}
\label{Sec:Vub}

For the determination of $|V_{ub}|$, we include the two most recent experimental measurements from BaBar, which are the untagged 6-bin (``BaBar 2010'') and 12-bin (``BaBar 2012'') analyses in Refs.~\cite{delAmoSanchez:2010af,Lees:2012vv}.  Because the 12-bin analysis uses more data and different candidate selections and cuts than the 6-bin analysis, the statistical correlations between the two data sets are quite small, and we treat the two data sets as statistically uncorrelated.  There are some correlations between the systematic uncertainties in the two analyses, but these are estimated to be sufficiently small that they have a tiny impact on $|V_{ub}|$~\cite{DingfelderPrivateComm}.  We therefore treat the two BaBar analyses as fully independent.  We also include the two most recent experimental measurements from Belle, which are the untagged analysis in Ref.~\cite{Ha:2010rf} (``Belle 2010'') and the full-reconstruction tagged analysis in Ref.~\cite{Sibidanov:2013rkk} (``Belle 2013'').   The tagged and untagged data sets have little overlap.  Further, the dominant systematic error in the tagged analysis is from the uncertainty in the tagging calibration, which is not present for the untagged analysis.  Thus we treat the Belle tagged and untagged analyses as independent.  The BaBar and Belle data sets are statistically independent.  The only commonality to the BaBar and Belle analyses is the use of the same event generation~\cite{DingfelderPrivateComm}.  Because the event generation is not a significant source of uncertainty in the analyses, we treat the systematic uncertainties as uncorrelated between the BaBar and Belle data sets.

\begin{figure}[tb]
\includegraphics[width=.49\textwidth]{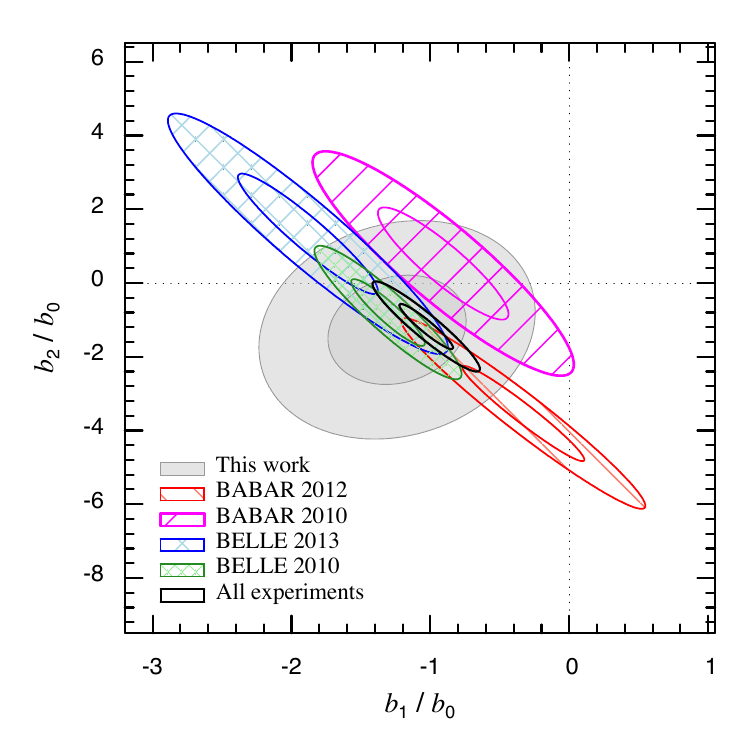}
\caption{Shape parameters $b_+^{(1)}/b_+^{(0)}$ and $b_+^{(2)}/b_+^{(0)}$ from $K=3$ BCL fits to the $B\to\pi\ell\nu$ form factor $f_+^{B\pi}(q^2)$ (filled ellipse) and from experimental measurements of the $B\to\pi\ell\nu$ branching fraction~\cite{delAmoSanchez:2010af,Ha:2010rf,Lees:2012vv,Sibidanov:2013rkk} (patterned and empty ellipses).  The colored ellipses show the constraints from the individual experiments, while the black ellipse shows the constraint from all experiments.  For each determination, the inner and outer contours show the 68\% and 95\% allowed confidence limits, respectively.}
\label{fig:ShapeParams}
\end{figure}

We first fit the experimental measurements to the BCL $z$-parametrization to obtain the shape parameters ($b_+^{(i)}/b_+^{(0)}$).  For these fits, we do not impose any constraint on the sum of the coefficients $\sum B_{mn} b_m b_n$.  Fits with truncation $K=2$ are sufficient to obtain good $\chi^2/{\rm dof}$ for three of the four data sets, but we perform fits with $K=3$ in order to enable comparison of both the slopes and curvatures with those of the form factor $f_+^{B\pi}(q^2)$ obtained in the previous section.  The numerical results for the $K=3$ fits to the individual experimental data sets, as well from a combined fit to all experiments, are given in Table~\ref{tab:BtoPiExpZFits}.   The fit to the BaBar 2010 data set has a somewhat large $\chi^2/{\rm dof}$ that stems from the highest $q^2$ bin, for which the error on the measured differential branching fraction is small but the central value is low with respect to the other points.  The inconsistency of the BaBar~2010 data leads the fit to all four experimental measurements to have a somewhat low, but still reasonable, $p$-value of 5\%.  Figure~\ref{fig:ShapeParams} shows the constraints on the slope ($b_+^{(1)}/b_+^{(0)}$) versus curvature ($b_+^{(2)}/b_+^{(0)}$) from the different experimental measurements, as well as from the combined fit to all four measurements.   The three most recent measurements agree at the 2$\sigma$ level, but display some tension with the BaBar~2010 result.   Combining the information from all four experimental analyses improves the determination of the shape parameters significantly.  

Because we do not impose any constraint on the sum of the coefficients $\sum B_{mn} b_m b_n$, we can check to see whether the experimental data is compatible with expectations from heavy-quark power counting for the size of the series coefficients.  Taking the determination of $|V_{ub}| = 3.63(12) \times 10^{-3}$ from CKM unitarity~\cite{UTfit}, we find a value for $\sum B_{mn} b_m b_n \sim 0.02$ from the fit to all experimental data.  This is consistent with the prediction from Eq.~(\ref{eq:HQConstraint}) taking a reasonable value for the heavy-quark scale $\Lambda \sim 1.1$~GeV, and validates the prior central value and width that we used to constrain $\sum B_{mn} b_m b_n$ in our preferred $z$-fit of the lattice form factors in the previous section.
 
Finally, before we fit the experimental and lattice data together to obtain $|V_{ub}|$, it is important to check that their shapes are consistent.  Figure~\ref{fig:ShapeParams} also shows the determination of the slope and curvature from our calculation of $f_+^{B\pi}(q^2)$ (see Table~\ref{tab:BtoPiBCLResults}).   The shapes of the lattice form factors and the experimental data are in good agreement, but the shape (as well as the overall normalization) is determined more precisely by experiment.  This suggests that the error on $|V_{ub}|$ can be minimized by performing a combined fit to the lattice and experimental data, as we now show. 

\begin{figure*}[tb]
\includegraphics[width=.49\textwidth]{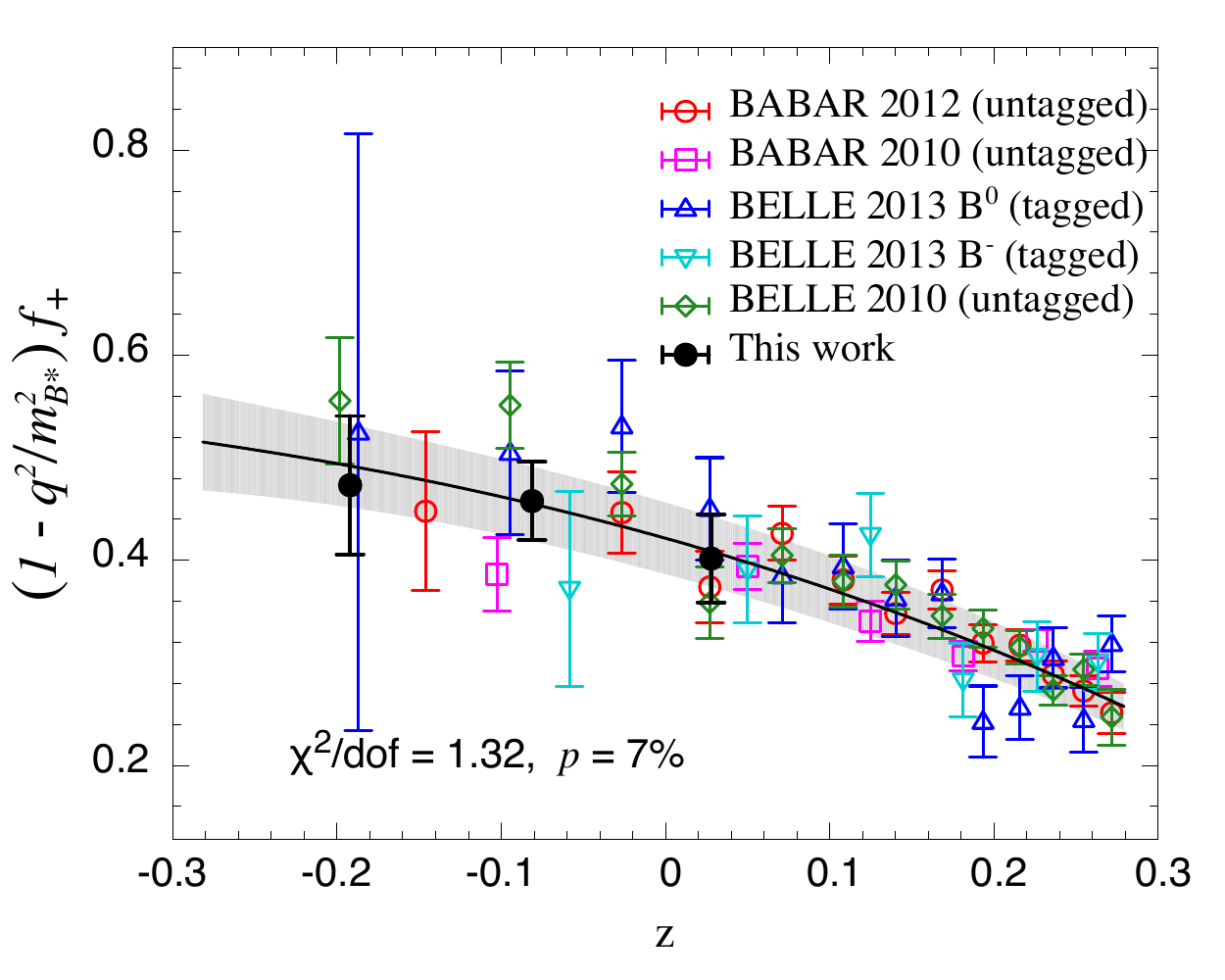}
\includegraphics[width=.49\textwidth]{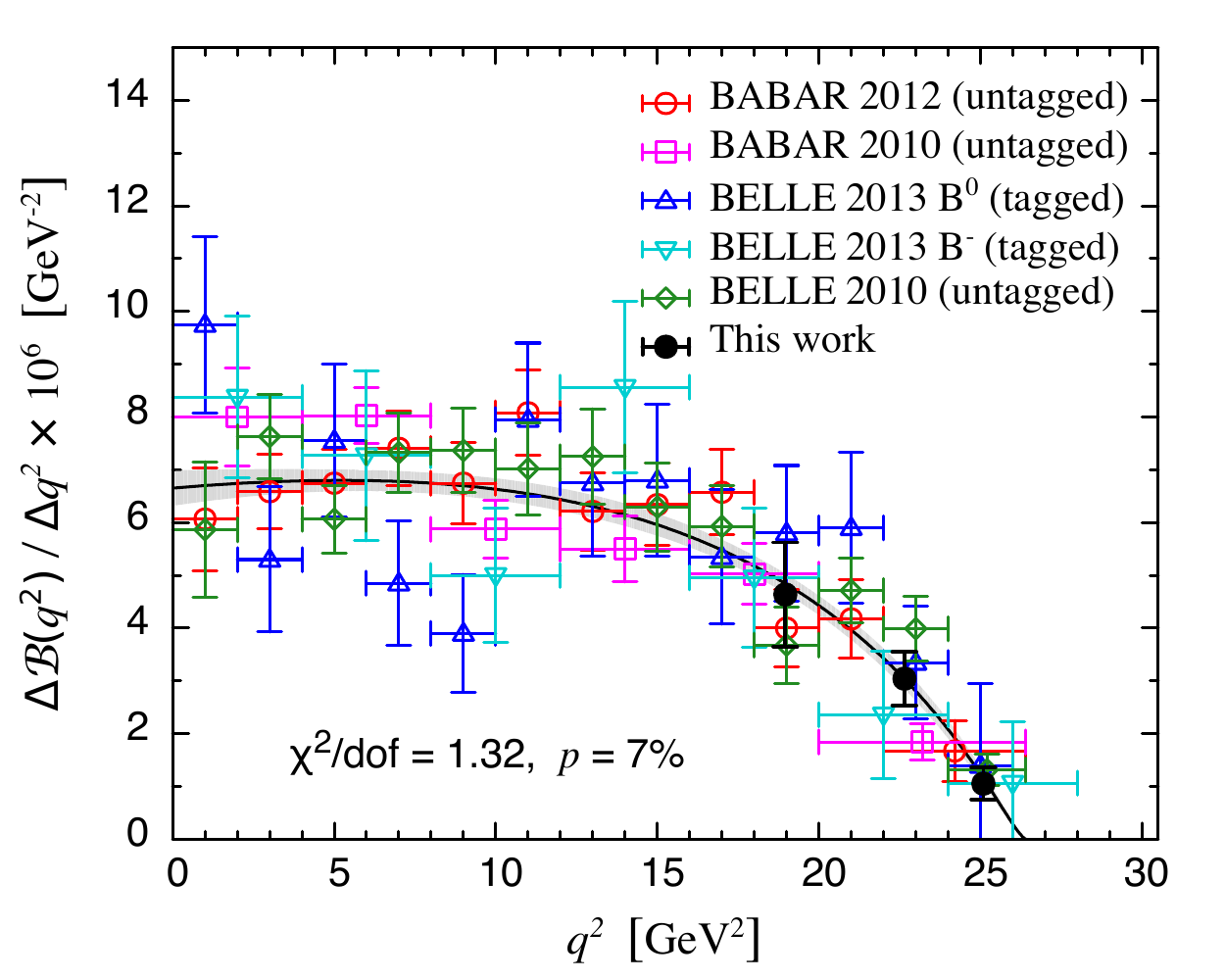}
\caption{Model-independent determination of $|V_{ub}|$ from a combined fit of experimental measurements of the $B\to\pi\ell\nu$ branching fraction~\cite{delAmoSanchez:2010af,Ha:2010rf,Lees:2012vv,Sibidanov:2013rkk}  and our lattice result for the $B\to\pi\ell\nu$ form factor $f_+(q^2)$ to the BCL $z$ parametrization, Eqs.~(\ref{eq:BCLConstraint}) and~(\ref{eq:f0BtoPiBCL}), with $K=3$.  The left plot shows $(1 - q^2/m_{B^*}^2) f_+(q^2)$ vs. $z$ (where the experimental data have been rescaled by the value of $|V_{ub}|$ determined in the fit), while the right plot shows $\Delta {\mathcal B} / \Delta q^2$ vs. $q^2$ (where the lattice points have been rescaled by $|V_{ub}|$).  In both plots, the filled black circles show the lattice data, while the open colored symbols show the experimental data.  The black curve with gray error band shows the fit result. }
\label{fig:VubZFit}
\end{figure*}

Table~\ref{tab:VubZFits} shows the results for the BCL coefficients and $|V_{ub}|$ obtained from a combined fit of the experimental measurements for the $B\to\pi\ell\nu$ differential branching fraction and the lattice determination of the form factor $f_+^{B\pi}(q^2)$, leaving the relative normalization $|V_{ub}|$ as a free parameter to be determined in the fit.  As in the experiment-only $z$-fits above, we do not constrain the sum of the coefficients $\sum B_{mn} b_m b_n$.  We present results from separate fits to each experimental data set, as well as from a fit including all experimental data.   The results for $|V_{ub}|$ from fits to the different experimental data sets agree within about $1\sigma$, and the $p$-value of the $K=3$ fit to all data is 6\%.  We also show results for truncations $K=3,4,5$ to study the uncertainty due to truncating the expansion in $z$.  The errors on $|V_{ub}|$ remain the same size as the number of fit parameters increase, and the central value for the fit including all experimental data is unchanged.  We take our final result 
\begin{equation}
	|V_{ub}| = 3.61(32) \times 10^{-3} \label{eq:VubResult}
\end{equation}
from the fit to all experimental data with $K=3$.
The quoted error on $|V_{ub}|$ is the total uncertainty, and includes both the theoretical error from the form factor and the experimental error (as well as the uncertainty from truncating the $z$-expansion).  Figure~\ref{fig:VubZFit} shows the preferred $K=3$ BCL $z$-fit used to obtain $|V_{ub}|$ plotted as $(1 - q^2/m_{B^*}^2) f_+(q^2)$ vs. $z$ (left) and as $\Delta {\mathcal B} / \Delta q^2$ vs. $q^2$ (right).  

Although we cannot precisely disentangle the error contributions, we can estimate the contribution to the error on $|V_{ub}|$ from the lattice form-factor determination.   Our most precise synthetic data point has a total statistical plus systematic uncertainty of 8.4\%.  If we assume that this is the lattice contribution to the 8.9\% error $|V_{ub}|$ in Eq.~(\ref{eq:VubResult}), this suggests that the experimental error contribution is approximately 2.8\%. 

The combined $z$-fit optimally combines the available information from lattice and experiment in a model-independent manner, thereby providing a determination of $|V_{ub}|$ that is both reliable and precise.  We can quantify this statement by comparing the error on $|V_{ub}|$ from the simultaneous fit to the error obtained from the previously standard approach.  One can determine $|V_{ub}|$ by relating the measured partial branching fraction in an interval $[ q^2_\text{min}, q^2_\text{max} ]$ to the normalized partial decay rate calculated from the form factor as follows:
\begin{equation}
 \Delta {\mathcal B}(q^2_{\rm min}, q^2_{\rm max}) / |V_{ub}|^2 = \tau_0 \Delta\zeta(q^2_{\rm min}, q^2_{\rm max}), \label{eq:VubTrad}
\end{equation}
where 
\begin{equation}
\Delta\zeta (q^2_{\rm min}, q^2_{\rm max}) \equiv
\frac{G_F^2}{24\pi^3}\int\limits_{q^2_\textrm{min}}^{q^2_\textrm{max}}dq^2 {\vec{p}_\pi}^{\,3} |f_+(q^2)|^2 \,.
\label{eq:zeta}
\end{equation}
The momentum range $q^2>16$~GeV$^2$ is typically used for comparison with lattice-QCD calculations because it is directly accessible in simulations and avoids (or at least minimizes) the need for an extrapolation.  From our preferred BCL parametrization ({\it cf.} Table~\ref{tab:BtoPiBCLResults}) we obtain 
\begin{equation}
	\Delta\zeta^{B\pi} (16~{\rm GeV}^2, q^2_{\rm max}) = 1.77(34) \ {\rm ps}^{-1}\,. \label{eq:DeltaZeta}
\end{equation}
Table~\ref{tab:VubTrad} shows the determinations of $|V_{ub}|$ obtained from combining $\Delta\zeta (16~{\rm GeV}^2, q^2_{\rm max})$ above with the different experimental measurements of $\Delta {\mathcal B}(16~{\rm GeV}^2, q^2_{\rm max})$, and with their average, via Eq.~(\ref{eq:VubTrad}).  The results agree with those from the simultaneous $z$-fits, but with larger errors.  In particular, the error on $|V_{ub}|$ obtained using the average $\Delta {\mathcal B}(16~{\rm GeV}^2, q^2_{\rm max})$ from all experiments is 10.0\%, to be compared with the 8.5\% error from our combined $z$-fit.  Separate $z$-fits to the lattice and experimental data lead to a similar error on $|V_{ub}|$.  The error on the normalization of the form factor $b_0$ in Table~\ref{tab:BtoPiBCLResults} is 9.4\%, while the error on the normalization of the experimental branching fraction from the $K=3$ fit to all experimental data $b_0 |V_{ub}|$ is 2.2\%.  Adding these in quadrature leads to a total error of 9.7\%.  Thus we conclude that the combined $z$-fit of all lattice and experimental data is indeed the best approach for minimizing the uncertainty on $ |V_{ub}|$.

\begin{table}[tb]
\caption{\label{tab:VubTrad} Determinations of $|V_{ub}|$ from a comparison of the measured $B\to\pi\ell\nu$ partial branching fractions with the normalized partial decay rate $\Delta\zeta^{B\pi}(16~{\rm GeV}^2, q^2_{\rm max}) = 1.77(34)$ calculated from our preferred BCL paramterization of the vector form factor $f_+^{B\pi}(q^2)$.}
\begin{ruledtabular}\begin{tabular}{lcr}
 & $ \Delta {\mathcal B}(16~{\rm GeV}^2, q^2_{\rm max}) \times 10^7 $ & $|V_{ub}| \times 10^3 $  \\ \hline
All                                     & 368(19)    & 3.69(37) \\ \hline
BaBar 2010~\cite{delAmoSanchez:2010af}  & 319(34)    & 3.44(38) \\
BaBar 2012~\cite{Lees:2012vv}           & 369(32)    & 3.70(39) \\
Belle 2010~\cite{Ha:2010rf}             & 398(30)    & 3.84(40) \\
Belle 2013~\cite{Sibidanov:2013rkk}     & 386(51)    & 3.78(44) \\ 
\end{tabular}
\end{ruledtabular}
\end{table}

\subsection{\texorpdfstring{Standard-Model predictions for $B\to\pi\ell\nu$ and $B_s\to K \ell \nu$ observables}{Standard-Model predictions for B->pi l nu and Bs -> K l nu observables}}
\label{Sec:BstoKPheno}

\begin{figure*}[tb]
\includegraphics[width=.49\textwidth]{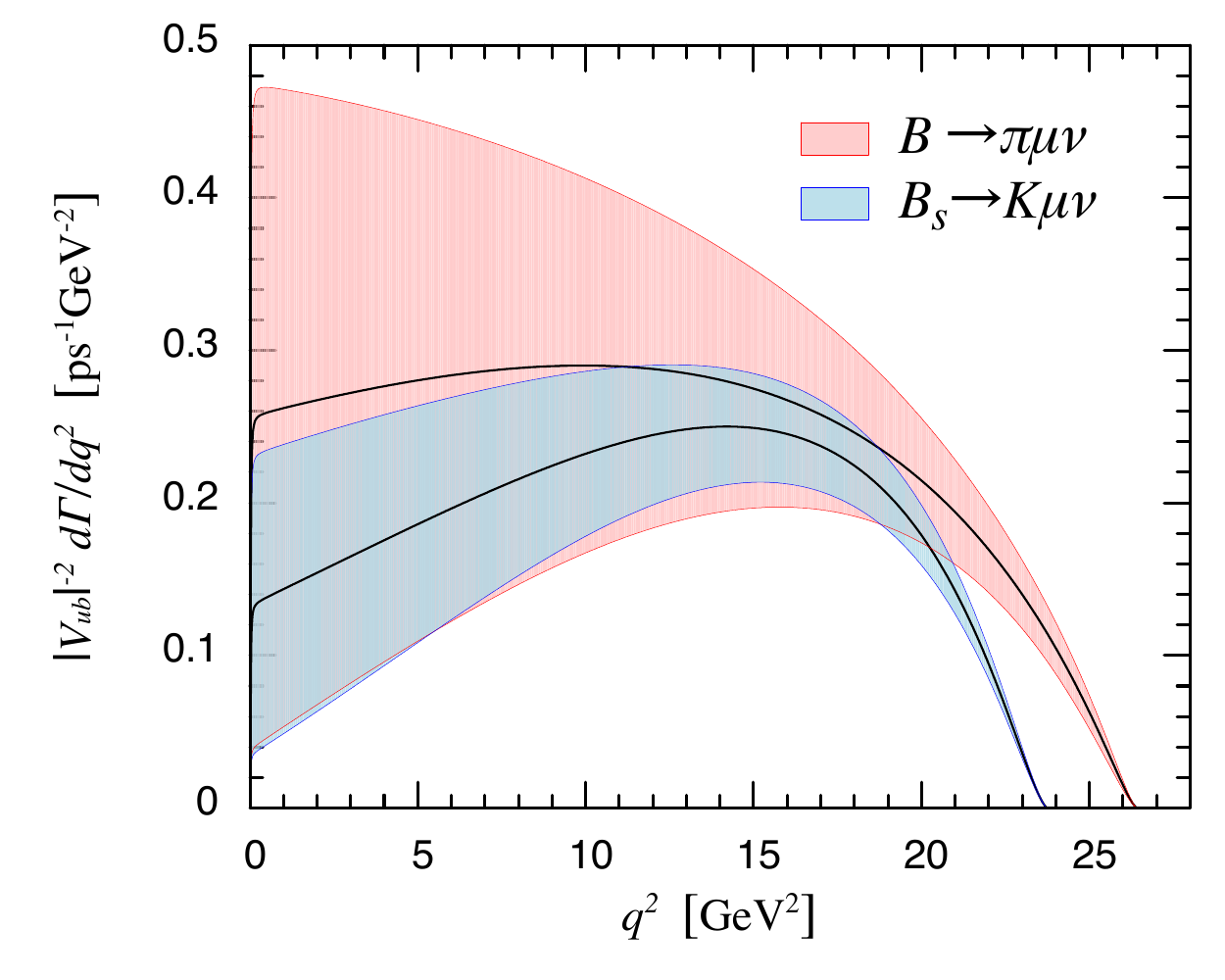}
\includegraphics[width=.49\textwidth]{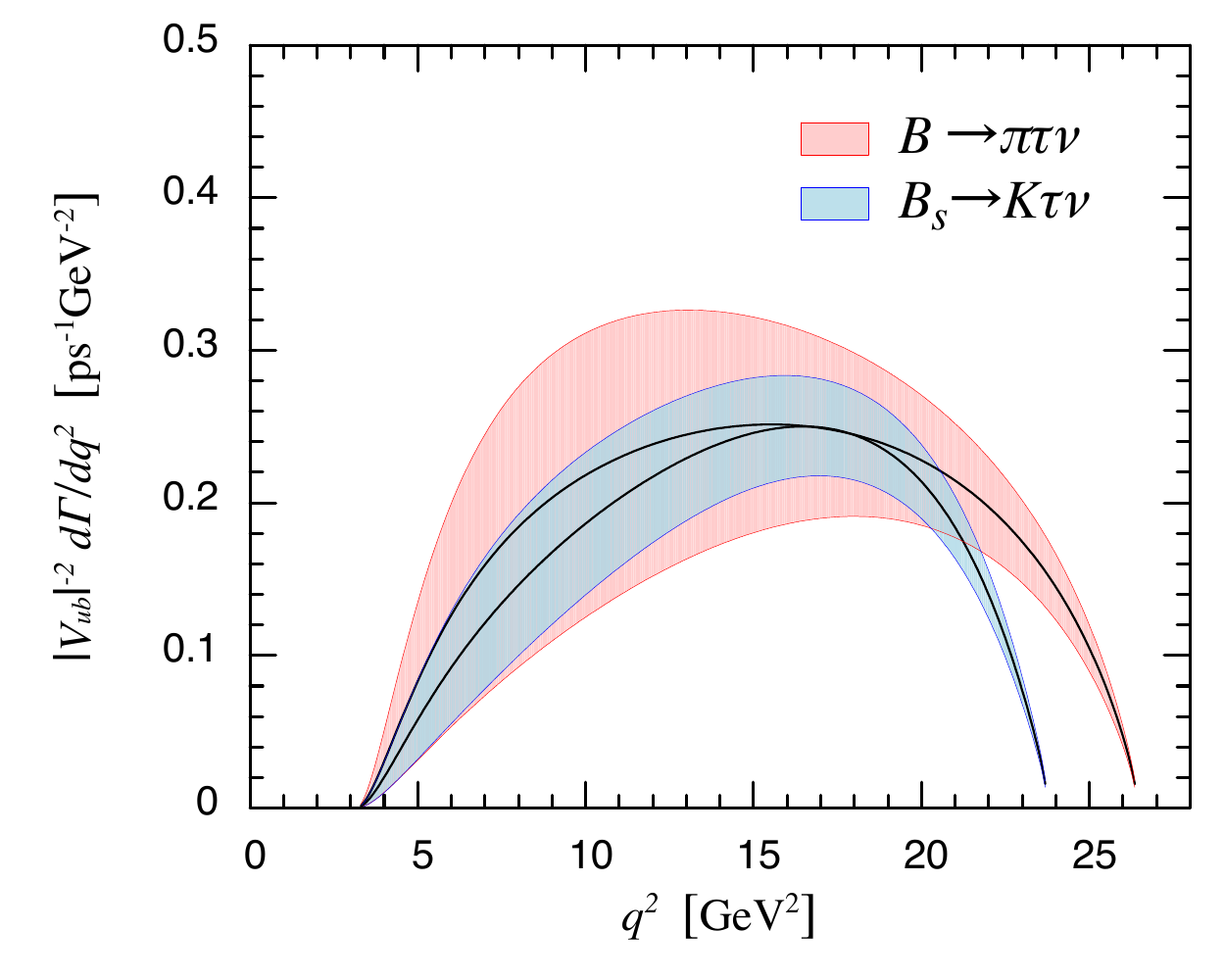}
\caption{Standard-Model predictions for the differential decay rate divided by $|V_{ub}|^2$ for $B_{(s)}\to P \ell\nu$ decays to muon (left) and $\tau$-lepton (right) final states using our form-factor determinations in Tables~\ref{tab:BtoPiBCLResults} and~\ref{tab:BstoKBCLResults}. }
\label{fig:BtoPDecayRates}
\end{figure*}

\begin{figure*}[tb]
\includegraphics[width=.49\textwidth]{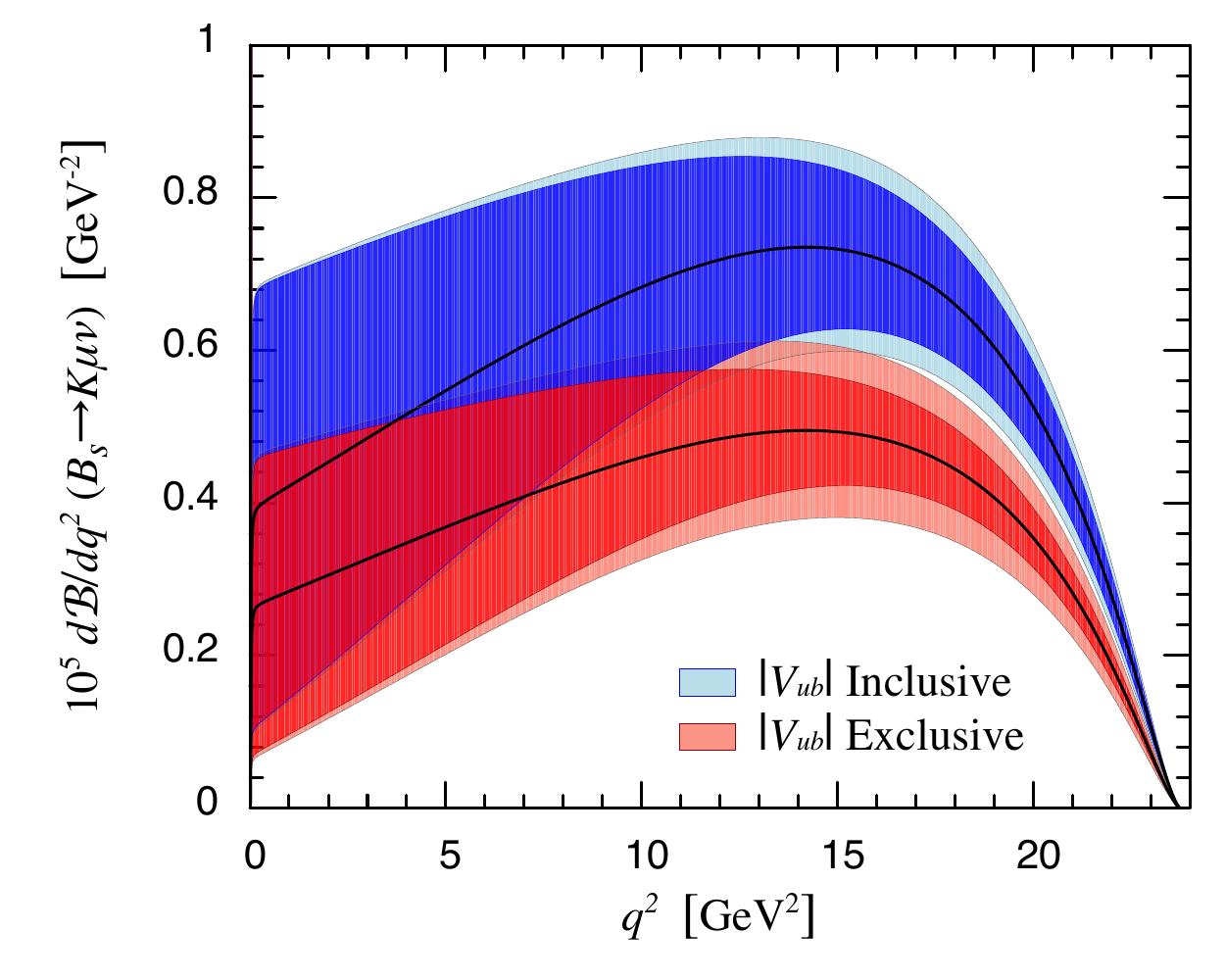} 
\includegraphics[width=.49\textwidth]{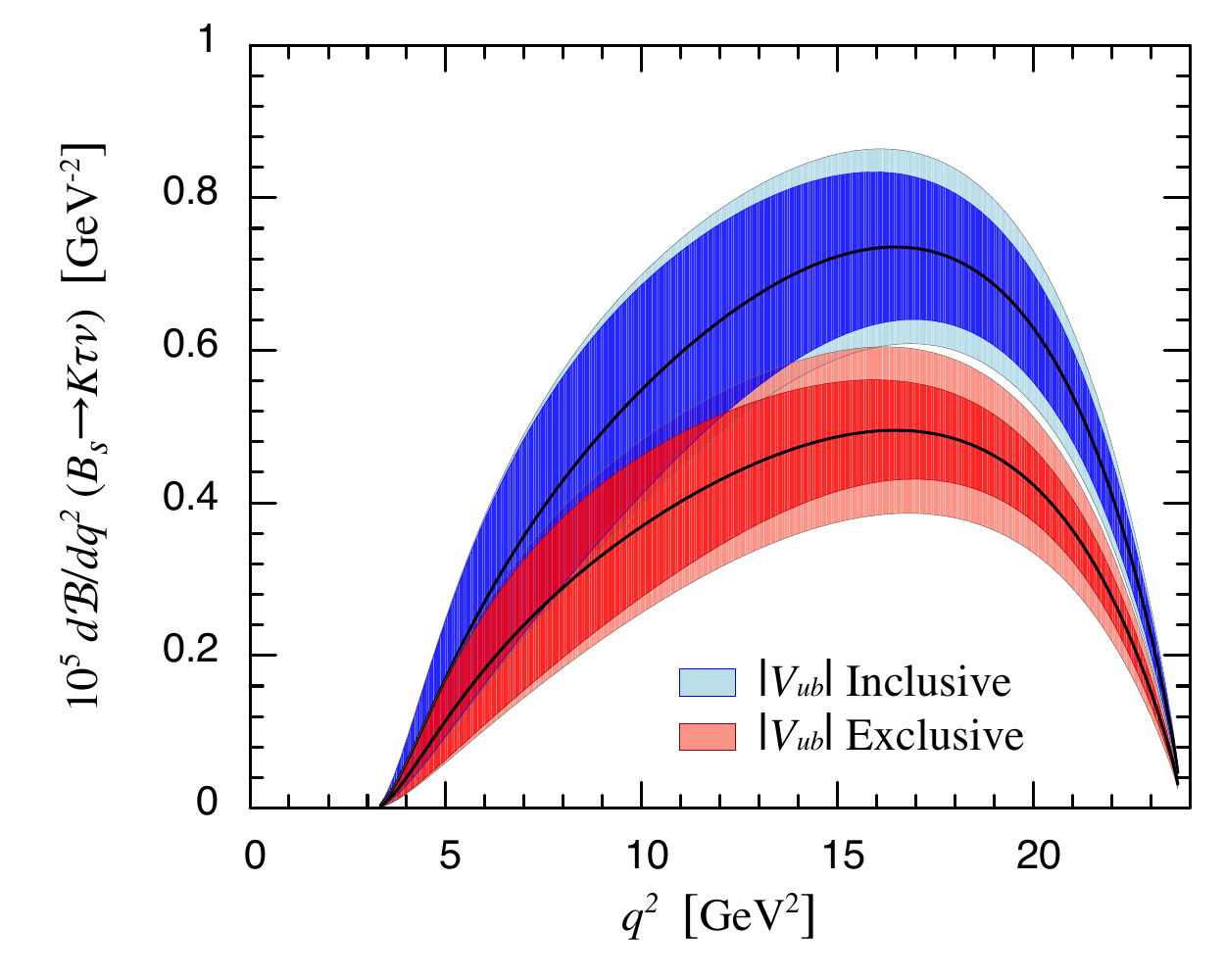}
\caption{Standard-Model predictions for the differential branching fraction for $B_s \to K \mu\nu$~(left) and $B_s \to K \tau\nu$~(right) using our determinations of the $B_s\to K \ell\nu$ form factors in Table~\ref{tab:BstoKBCLResults}.  Each plot shows predictions for $d{\mathcal B} /d q^2$ using our determination of $|V_{ub}|$ in Eq.~(\ref{eq:VubResult}) from exclusive $B\to\pi\ell\nu$ decay as well as using the determination of $|V_{ub}|$ from inclusive $B \to X_u \ell \nu$ decay~\cite{HFAGSemiPDG13}.  In these plots, the outer error band denotes the total uncertainty in $d{\mathcal B} /d q^2$, while the inner error band removes the error contribution from $|V_{ub}|$.  }
\label{fig:BstoKDecayRates}
\end{figure*}

The Standard-Model differential decay rate for $B_{(s)} \to P \ell \nu$ is given in Eq.~(\ref{eq:B_semileptonic_rate}).  Using the experimentally measured lepton and meson masses~\cite{Beringer:2012zz}, we obtain predictions for the differential decay rate divided by $|V_{ub}|^2$.  These are plotted for the muon and $\tau$-lepton final states in Fig.~\ref{fig:BtoPDecayRates}, where we use ``muon'' to denote decays to either of the light charged leptons ($\ell=\mu,e$) throughout this section.   Integrating the differential decay rates over the kinematically-allowed $q^2$ range gives\footnote{In practice, the full kinematic range may not be accessible experimentally, in which case the limits of integration here and throughout this section will need to be changed accordingly.}
 \begin{eqnarray}
{\Gamma(B \to \pi \mu\nu)/|V_{ub}|^2}  &=&  {6.2(2.5)} \ {\rm ps}^{-1} \,, \\
{ \Gamma(B \to \pi \tau\nu)/|V_{ub}|^2} &=&  {4.3(1.2)} \ {\rm ps}^{-1} \,, \\
 \Gamma(B_s\to K \mu\nu)/|V_{ub}|^2  &=&  4.55(1.08) \ {\rm ps}^{-1} \,, \\
 \Gamma(B_s\to K \tau\nu)/|V_{ub}|^2 &=&  3.52(0.60) \ {\rm ps}^{-1} \,,
 \end{eqnarray}
with errors of about {25--40}\% and {15--30}\% for the $\mu$ and $\tau$ final states, respectively.  We also use the determination of $|V_{ub}|$ from our calculation of the $B\to\pi\ell\nu$ form factors (Eq.~(\ref{eq:VubResult})) to make predictions for the $B_s \to K \ell \nu$ differential branching fractions for $\ell=\mu, \tau$.  These are plotted in Fig.~\ref{fig:BstoKDecayRates}.  For comparison, we also show the prediction for $d{\mathcal B} /d q^2$ using the determination of $|V_{ub}|$ from inclusive $B \to X_u \ell \nu$ decay~\cite{HFAGSemiPDG13}.  The form-factor uncertainties are sufficiently small for $q^2 \gtapprox 13 {\rm \ GeV}^2$ that, given an experimental measurement of the branching fraction in this region with commensurate precision, one can distinguish between the curves corresponding to $|V_{ub}|_{\rm excl.}$ and $|V_{ub}|_{\rm incl.}$.  Thus we anticipate that $B_s \to K \ell \nu$ semileptonic decay will eventually play an important role in addressing the current ``$|V_{ub}|$ puzzle.''

\begin{figure}[tb]
\includegraphics[width=.49\textwidth]{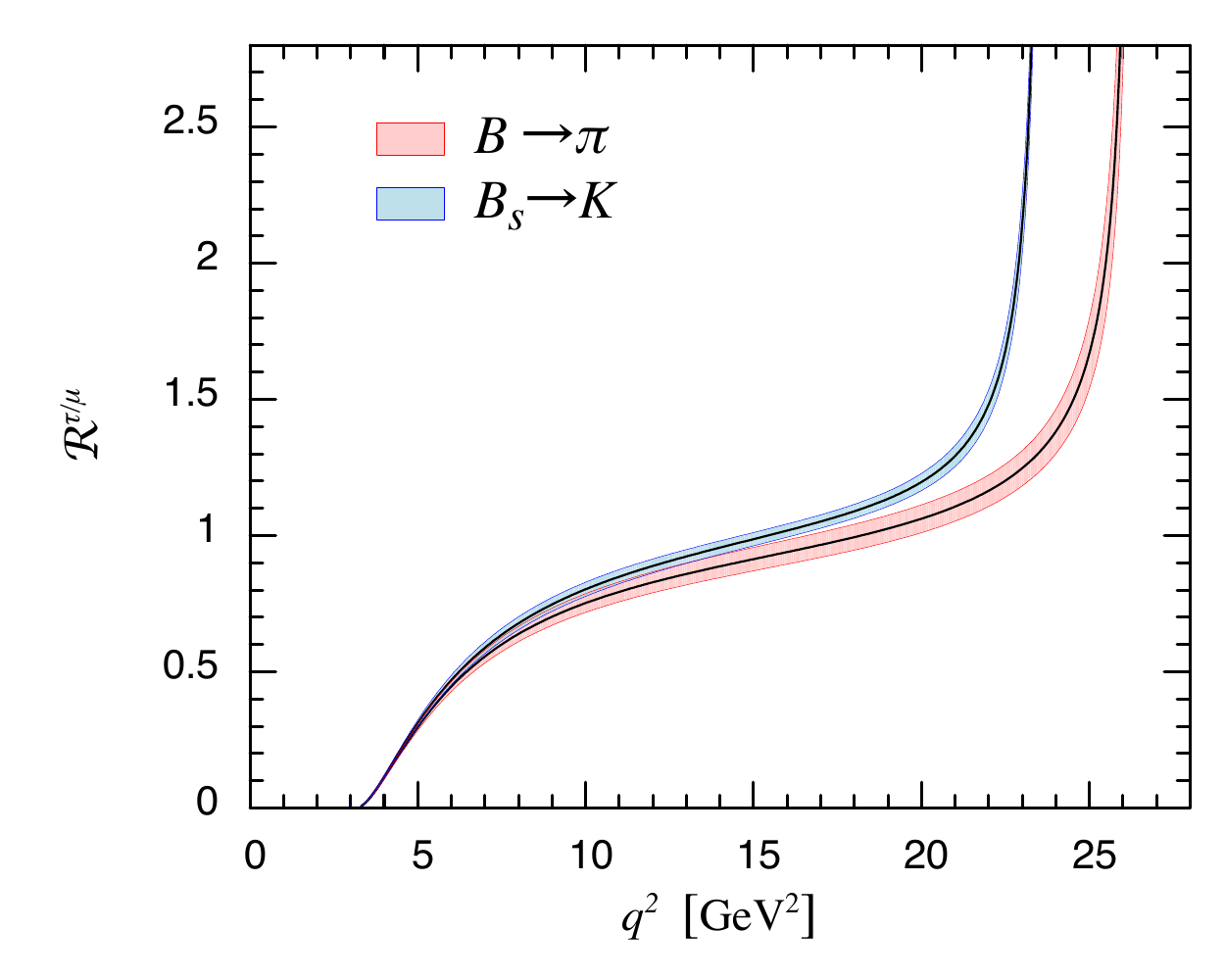} 
\caption{Standard-Model ratio of differential branching fractions {${\mathcal R}^{\tau/\mu}_P(q^2)$} using our determinations of the $B \to \pi \ell\nu$ and $B_s\to K \ell\nu$ form factors in Tables~\ref{tab:BtoPiBCLResults} and~\ref{tab:BstoKBCLResults}. }
\label{fig:BstoKdBdq2_Rmutau}
\end{figure}

Semileptonic decays to $\tau$ leptons may be particularly sensitive to new physics associated with electroweak symmetry breaking due to the large $\tau$ mass, or more generally sensitive to any Standard-Model extensions with new scalar currents.  Moreover, the ratio of $\mu/\tau$ differential decay rates~\cite{Meissner:2013pba}
\begin{equation}
 {\mathcal R}^{\tau/\mu}_{{P}}(q^2) 
{\equiv} \frac{d\Gamma ({B_{(s)}\to P} \tau\nu)/dq^2}
       { d\Gamma ({B_{(s)}\to P} \mu\nu)/dq^2}
\end{equation}
provides a precise test of the Standard Model that is independent of the CKM matrix element $|V_{ub}|$.  Figure~\ref{fig:BstoKdBdq2_Rmutau} shows the predictions for the ratios of differential branching fractions using our determinations of the $B\to\pi\ell\nu$ and $B_s\to K \ell\nu$ form factors in Tables~\ref{tab:BtoPiBCLResults} and~\ref{tab:BstoKBCLResults}.  Integrating over the kinematically allowed ranges, we obtain the following Standard-Model predictions for $ R{^{\tau/\mu}_P} \equiv \Gamma ({B_{(s)}\to P} \tau\nu)/\Gamma ({B_{(s)}\to P} \mu\nu)$:
\begin{eqnarray}
 R{^{\tau/\mu}_\pi} &=&  {0.69(19)} \,, \\
 R{^{\tau/\mu}_K} &=&  {0.77(12)} \,.
\end{eqnarray}

\begin{figure*}[tb]
\includegraphics[width=.49\textwidth]{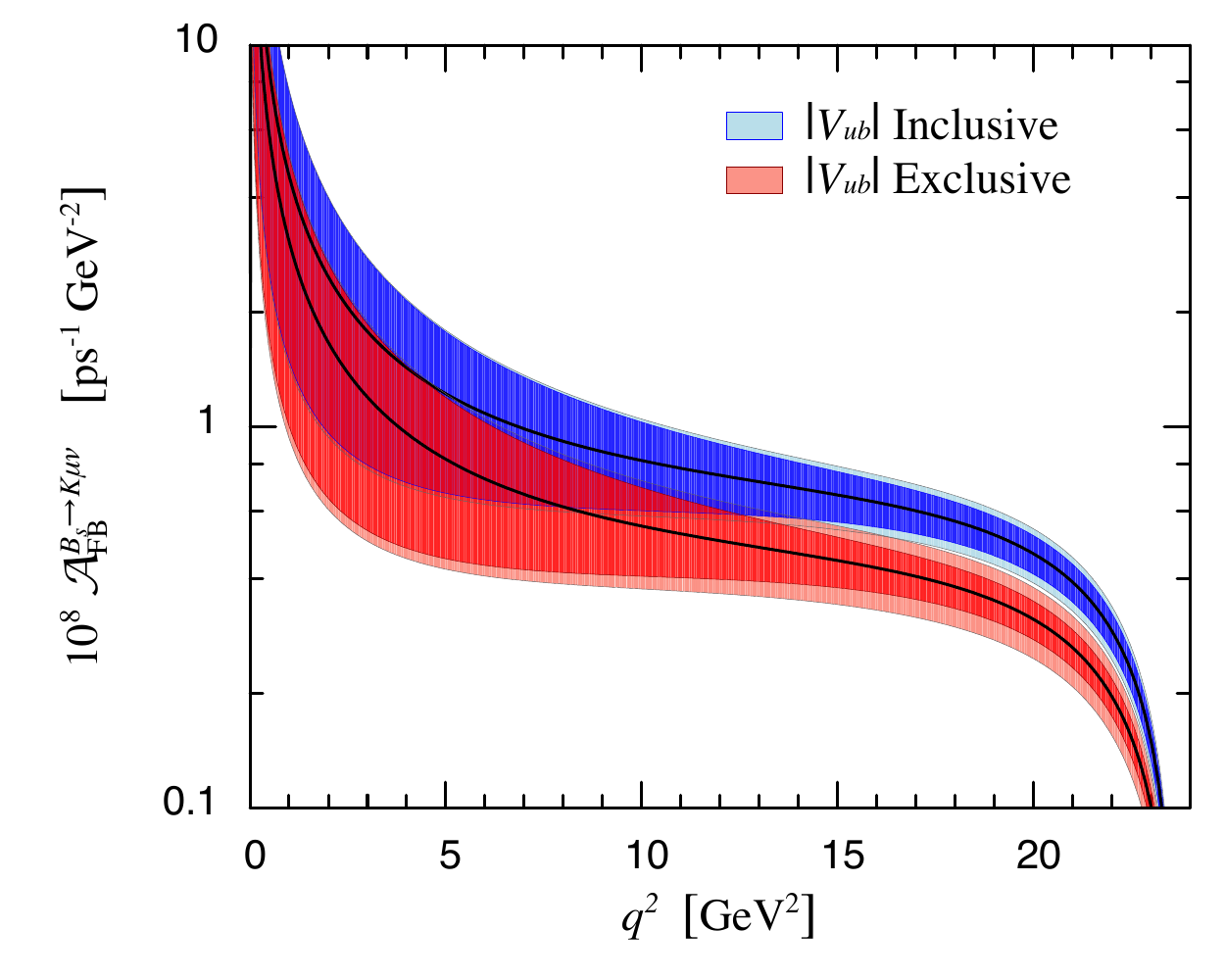}
\includegraphics[width=.49\textwidth]{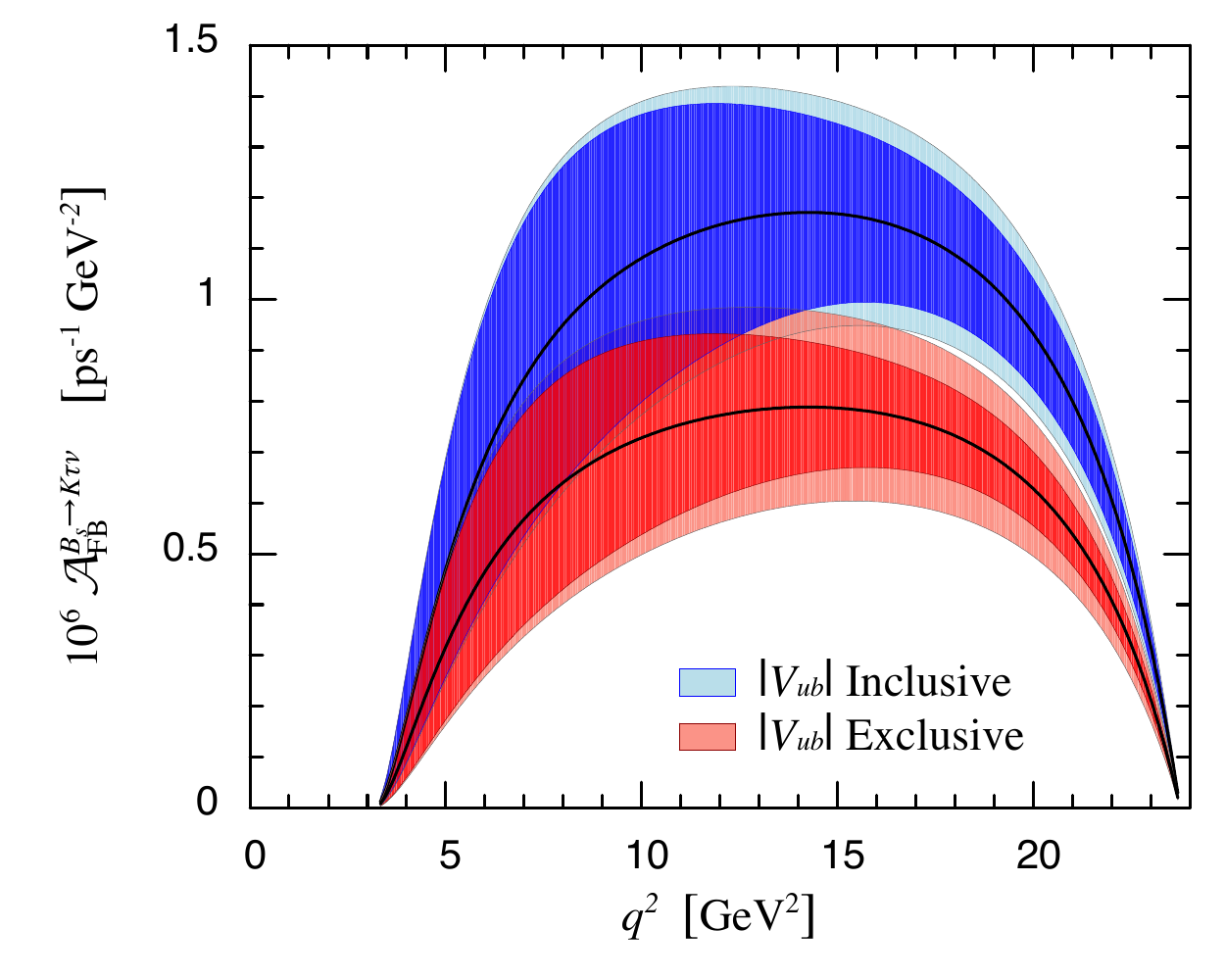}
\caption{Standard-Model predictions for the forward-backward asymmetry for $B_s \to K \mu\nu$~(left) and $B_s \to K \tau\nu$~(right) using our determinations of the $B_s\to K \ell\nu$ form factors in Table~\ref{tab:BstoKBCLResults}.  The plots show predictions for ${\cal A}_{FB}$ using our determination of $|V_{ub}|$ in Eq.~(\ref{eq:VubResult}) from exclusive $B\to\pi\ell\nu$ decay as well as using the determination of $|V_{ub}|$ from inclusive $B \to X_u \ell \nu$ decay~\cite{HFAGSemiPDG13}.  In these plots, the outer error band denotes the total uncertainty in $d{\mathcal B} /d q^2$, while the inner error band removes the error contribution from $|V_{ub}|$.}
\label{fig:BstoKFBAsymmetry}
\end{figure*}

The three-body final state in ${B_{(s)}\to P} \ell\nu$ decay also enables one to construct and study observables that depend on the kinematics of the decay products.  Such angular observables are particularly sensitive to possible right-handed currents. An example is the forward-backward difference\footnote{This quantity is sometimes referred to as an ``asymmetry" in the literature, but does not satisfy the convention that its magnitude is bounded by unity.}
\begin{equation}
{\cal A}_{FB}^{{B_{(s)}\to P \ell\nu}}(q^2) \equiv \left[\int^{1}_{0}-\int^{0}_{-1}\right]d\cos\theta_\ell  \frac{ d^2\Gamma({\overline B_{(s)}\to P} \ell\nu )}{dq^2d\cos\theta_\ell } \,
\end{equation}
where $\theta_\ell$ is the angle between the charged-lepton and $B_{(s)}$-meson momenta in the $q^2$ rest frame.  In the Standard Model, the forward-backward difference is given by~\cite{Meissner:2013pba}
\begin{eqnarray}
{\cal A}_{FB}^{{B_{(s)}\to P \ell\nu}}(q^2)&=& \frac{G_F^2|V_{ub}|^2}{32\pi^3 M_{B_s}}
  \left(1-\frac{m_\ell^2}{q^2}\right)^2 |\vec{p}_K|^2 \nonumber \\
    	&{\times}& \frac{m_\ell^2}{q^2} (M_{B_s}^2 - M_K^2) f_+(q^2)f_0(q^2) \,.
\label{eq:AFB}\end{eqnarray}
Figure~\ref{fig:BstoKFBAsymmetry} shows the Standard-Model predictions for ${\cal A}_{FB}^{{B_s \to K \ell\nu}}(q^2)$ with $\ell=\mu,\tau$ using our determination of $|V_{ub}|$ from $B\to\pi\ell\nu$ decay (Eq.~(\ref{eq:VubResult})) and the inclusive determination of $|V_{ub}|$ from $B \to X_u \ell \nu$ decay~\cite{HFAGSemiPDG13}.  Again, the theoretical errors are sufficiently small at large $q^2$ that an experimental measurement with commensurate precision could distinguish between the two predictions.  Integrating over the full kinematic ranges we obtain
\begin{eqnarray}
{\int_{m_\mu^2}^{ q_{\rm max}^2} dq^2 {\mathcal A}^{B\to\pi \mu\nu}_{\rm FB}(q^2)  /|V_{ub}|^2} &=& {0.028(19)  \ {\rm ps}^{-1}} \,, \\
{\int_{m_\tau^2}^{ q_{\rm max}^2} dq^2 {\mathcal A}^{B\to\pi \tau\nu}_{\rm FB}(q^2)  /|V_{ub}|^2} &=& {1.08(35)   \ {\rm ps}^{-1}} \,, \\
\int_{m_\mu^2}^{ q_{\rm max}^2} dq^2 {\mathcal A}^{B_s \to K \mu\nu}_{\rm FB}(q^2)  /|V_{ub}|^2 &=& 0.0175(87)  \ {\rm ps}^{-1}  \,, \\
\int_{m_\tau^2}^{ q_{\rm max}^2} dq^2 {\mathcal A}^{B_s \to K \tau\nu}_{\rm FB}(q^2)  /|V_{ub}|^2 &=& 0.93(18)   \ {\rm ps}^{-1} \,,
 \end{eqnarray}
 where the $\mu$ results are much smaller than the $\tau$ results due to helicity suppression from the small muon mass.

\begin{figure*}[tb]
\includegraphics[width=.49\textwidth]{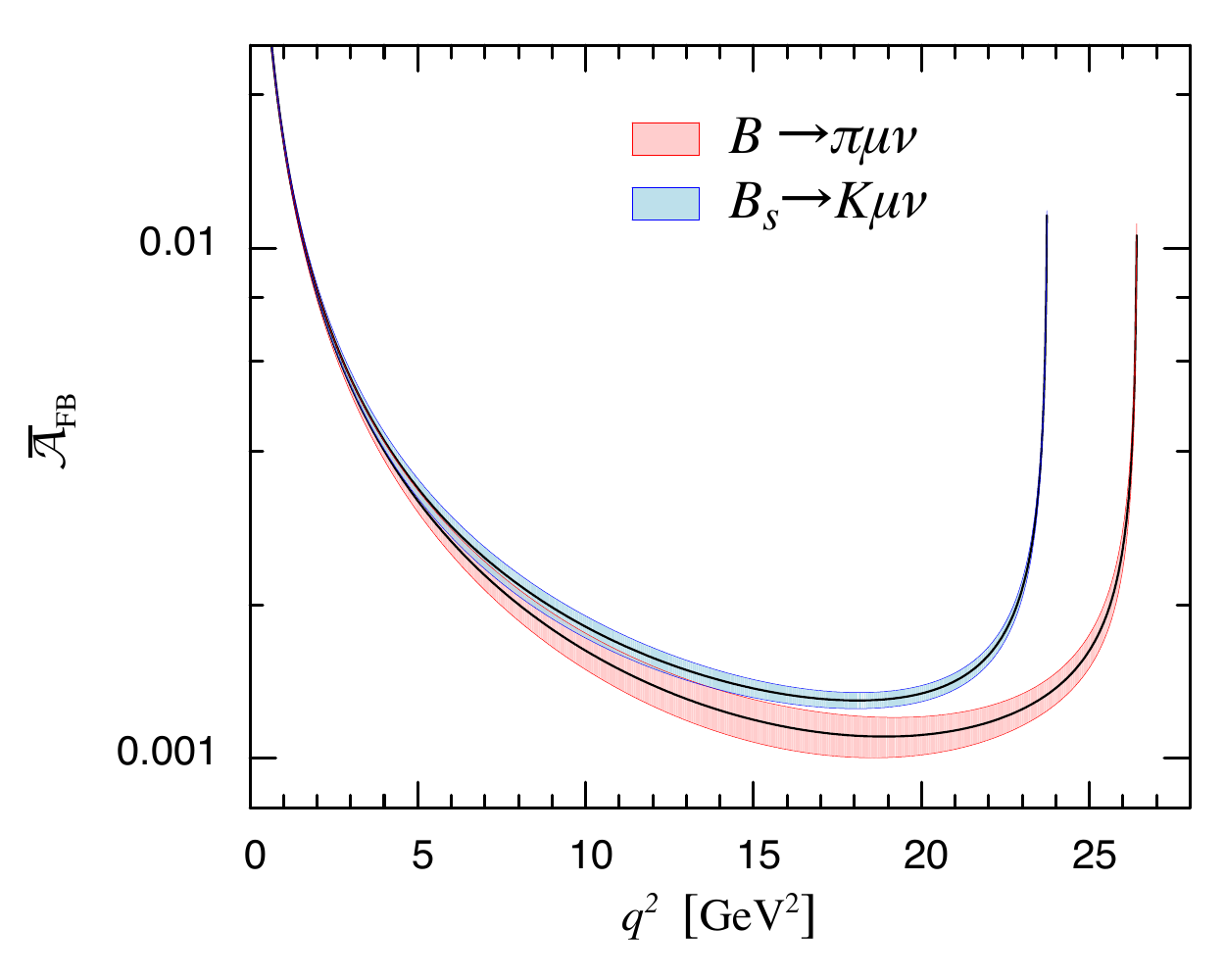}  
\includegraphics[width=.49\textwidth]{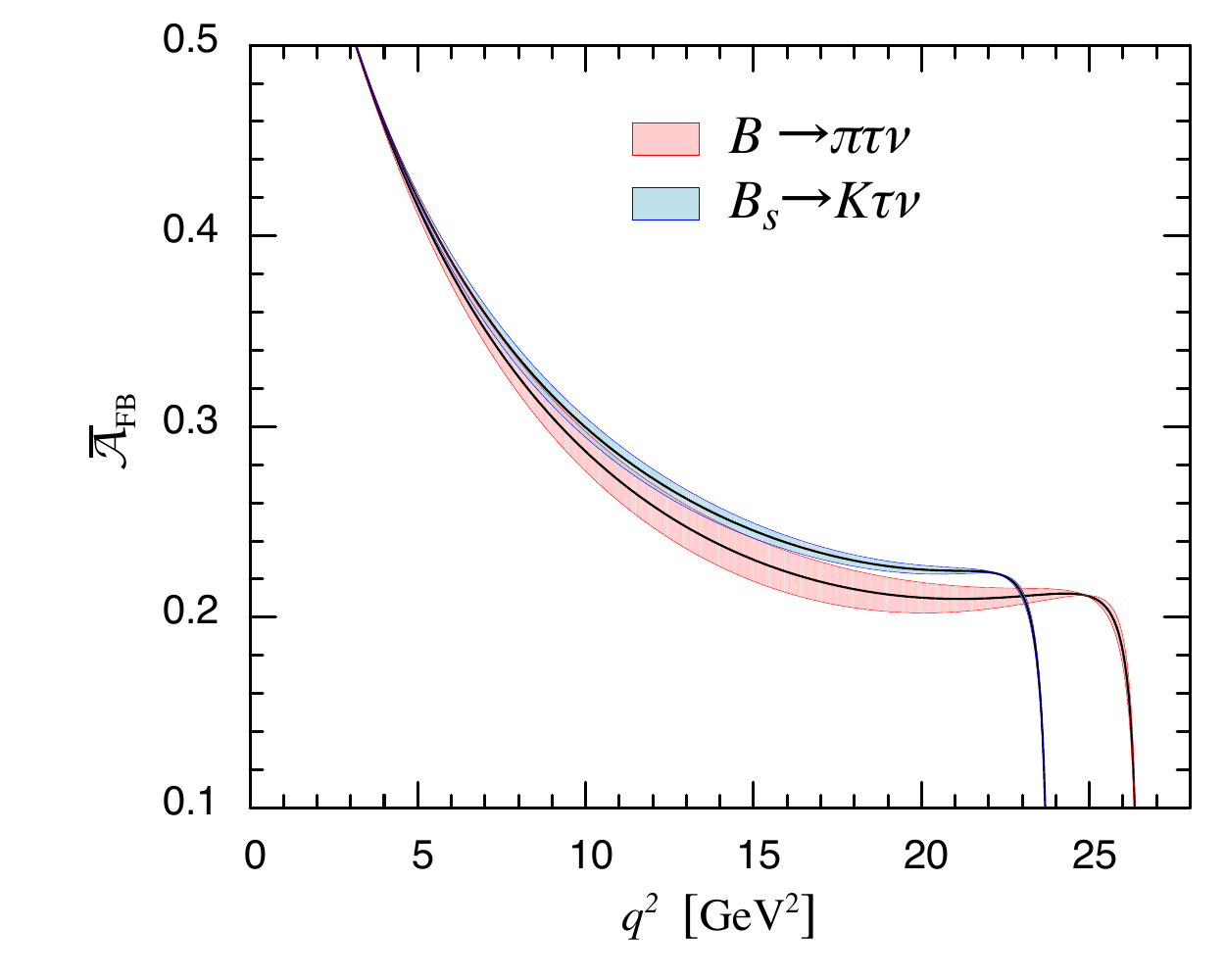}
\caption{Standard-Model predictions for the normalized forward-backward asymmetry $\bar{\cal A}_{FB}$ for decays to muon (left) and $\tau$-lepton~(right) final states using our form-factor determinations in Tables~\ref{tab:BtoPiBCLResults} and~\ref{tab:BstoKBCLResults}. }
\label{fig:NormFBAsymmetry}
\end{figure*}

The normalized forward-backward asymmetry is particularly interesting because it removes the ambiguity from $|V_{ub}|$, and the theoretical form-factor uncertainties cancel to some degree:
\begin{equation}
  \bar{\mathcal A}^{{B_{(s)}\to P \ell\nu}}_{\rm FB} 
{\equiv} \frac{\int_{m^2_\ell}^{ q_{\rm max}^2} dq^2  {\mathcal A}^{{B_{(s)}\to P \ell\nu}}_{\rm FB}(q^2)}{\int_{m^2_\ell}^{ q_{\rm max}^2} dq^2 \, d\Gamma({B_{(s)}\to P}  \ell\nu)/dq^2} \,.
\end{equation}
The Standard-Model predictions for $\bar{\cal A}_{FB}^{{B_{(s)}\to P \ell\nu}}(q^2)$ with $\ell=\mu,\tau$ are given in Fig.~\ref{fig:NormFBAsymmetry}.  The percentage uncertainty in $\bar{\cal A}_{FB}^{{B_{(s)}\to P \ell\nu}}(q^2)$ is indeed smaller than in ${\cal A}_{FB}^{{B_{(s)}\to P \ell\nu}}(q^2)$.  Integrating $\bar{\cal A}_{FB}^{{B_{(s)}\to P \ell\nu}}(q^2)$ over the full kinematic range we obtain
\begin{eqnarray}
 \bar{\mathcal A}^{{B \to \pi\mu\nu}}_{\rm FB}  &=& 0.0044(13) \,, \\
 \bar{\mathcal A}^{{B \to \pi \tau\nu}}_{\rm FB} &=& 0.252(12) \,, \\
 \bar{\mathcal A}^{{B_s \to K \mu\nu}}_{\rm FB}  &=& 0.0039(11)\,  \\
 \bar{\mathcal A}^{{B_s \to K \tau\nu}}_{\rm FB}&=& 0.2650(79) \,,
\end{eqnarray}
with errors of about 30\% and 3--5\% for the $\mu$ and $\tau$ final states, respectively. 

\section{Results and conclusions}
\label{Sec:Conc}

We have calculated the form factors for $B\to\pi\ell\nu$ and $B_s\to K \ell \nu$ semileptonic decay in dynamical lattice QCD using (2+1) flavors of domain-wall light quarks and relativistic $b$ quarks.  We then extended our results from the simulated range of lattice momenta to $q^2=0$ using the model-independent $z$-expansion based on analyticity and unitarity.   We obtain our preferred results for the form factors $f_+(q^2)$ and $f_0(q^2)$ using the BCL form of the $z$-expansion \cite{Bourrely:2008za} and imposing the kinematic constraint $f_+(0) = f_0(0)$ and a constraint on the sum of the coefficients for $f_+(q^2)$ based on heavy-quark power counting~\cite{Becher:2005bg}.  The resulting BCL $z$-coefficients for $f_+(q^2)$ and $f_0(q^2)$ for $B\to\pi\ell\nu$ are given, along with their correlation matrix, in Table~\ref{tab:BtoPiBCLResults}, while the BCL $z$-coefficients for $B_s \to K \ell\nu$ are in Table~\ref{tab:BstoKBCLResults}.   These results can be combined with current and future experimental measurements of the $B \to \pi\ell\nu$ and $B_s \to K \ell\nu$ branching fractions to obtain the CKM matrix element $|V_{ub}|$.

\begin{figure*}[tb]
\includegraphics[width=.49\textwidth]{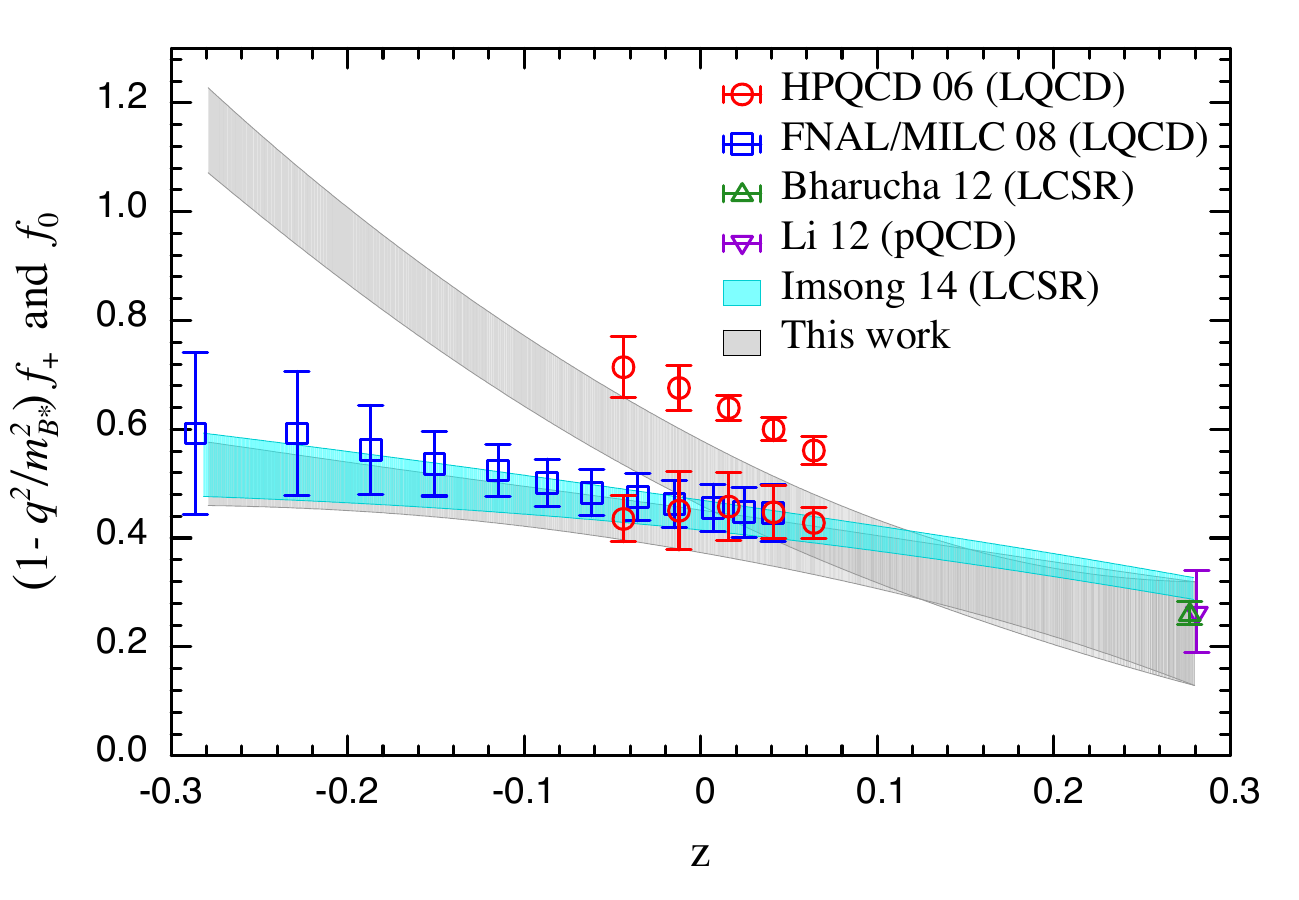}
\includegraphics[width=.49\textwidth]{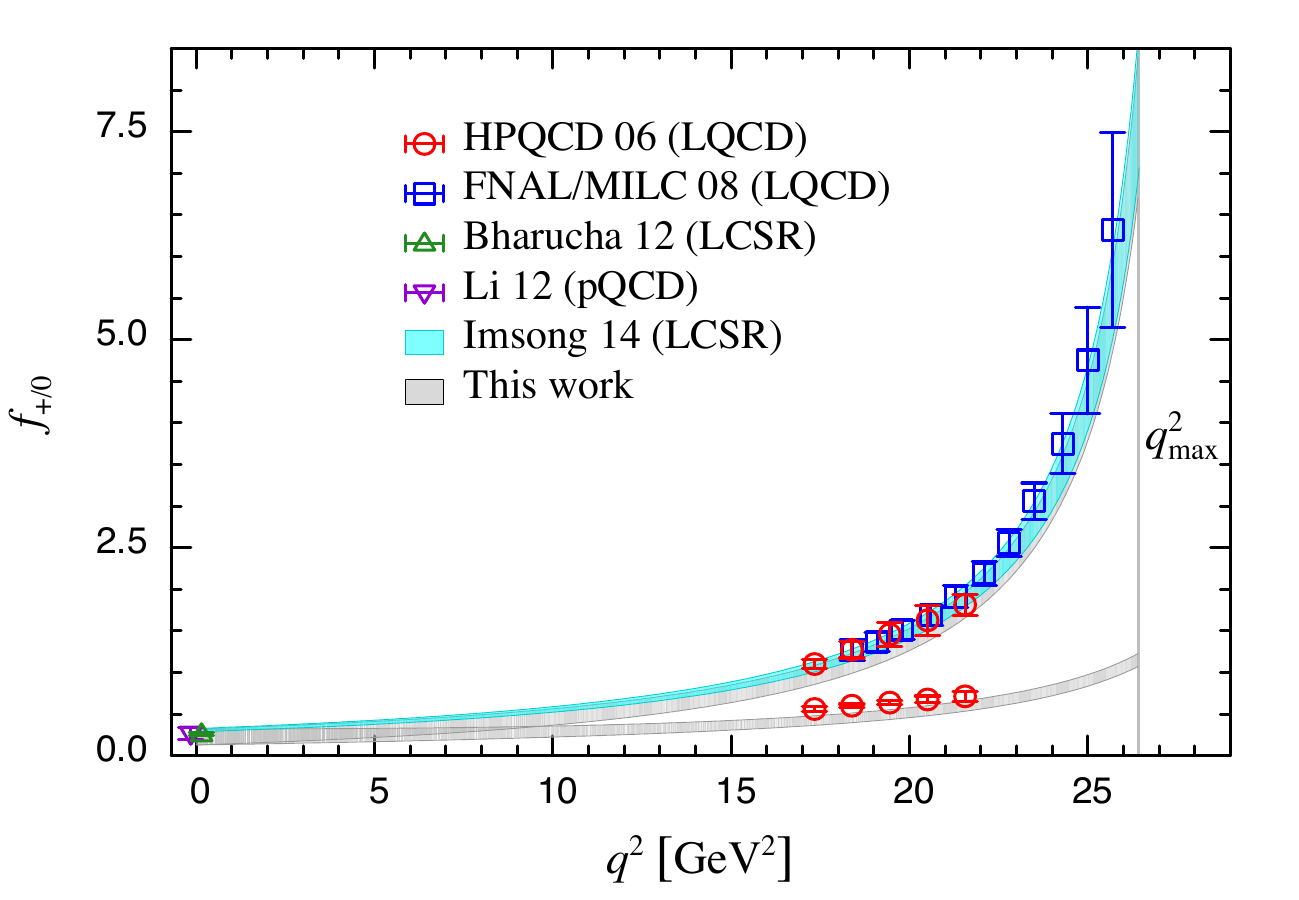} 
\includegraphics[width=.49\textwidth]{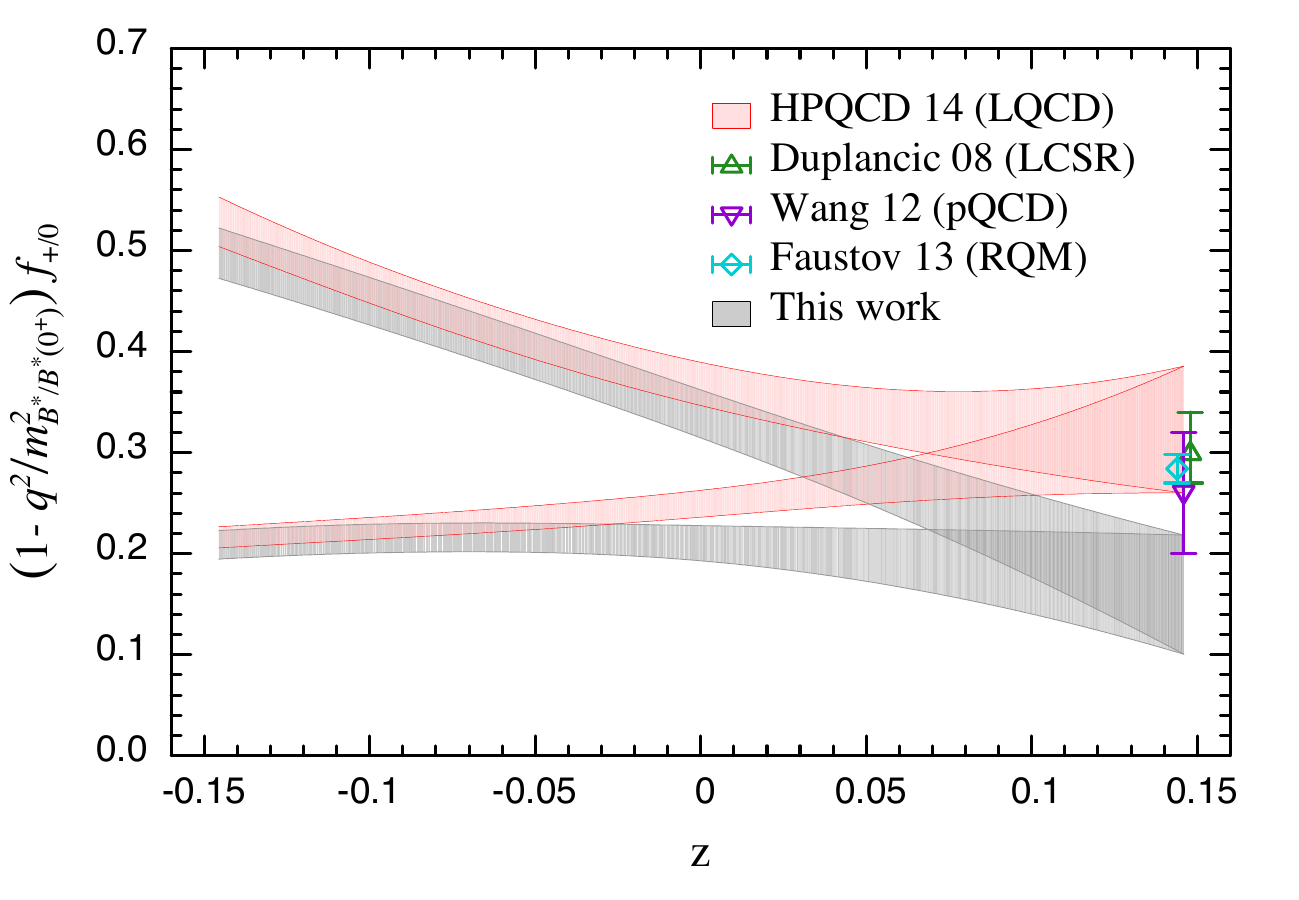}
\includegraphics[width=.49\textwidth]{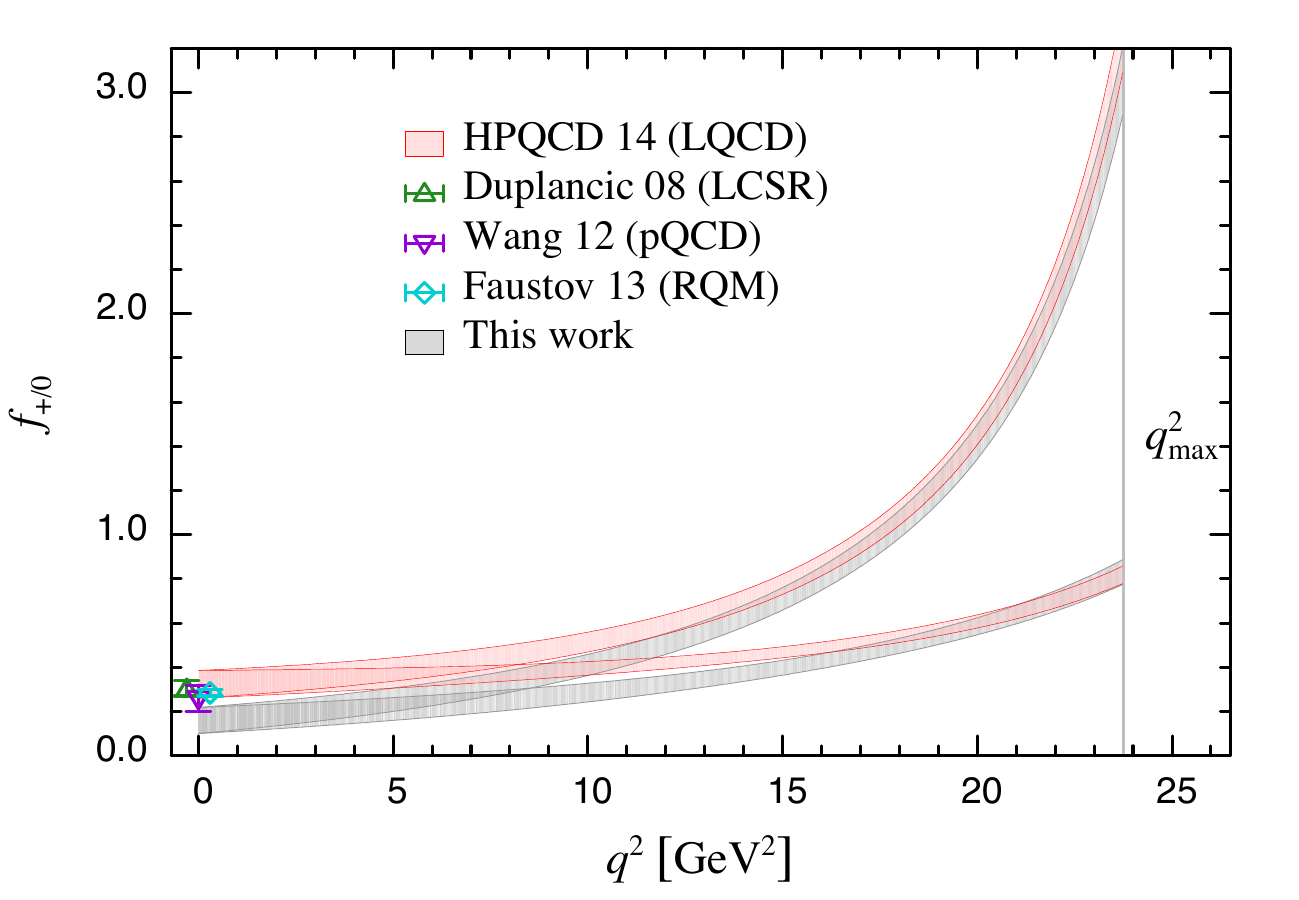} 
\caption{Top: theoretical calculations of the $B\to\pi\ell\nu$ form factors from {light-cone sum rules~\cite{Bharucha:2012wy,Imsong:2014oqa}, NLO perturbative QCD~\cite{Li:2012nk}}, and (2+1)-flavor lattice QCD~\cite{Dalgic:2006dt,Bailey:2008wp}.  Bottom: theoretical calculations of the $B_s \to K \ell\nu$ form factors from QCD models~\cite{Duplancic:2008tk, Wang:2012ab,Faustov:2013ima} and (2+1)-flavor lattice QCD~\cite{Bouchard:2014ypa}. {In all plots, the predictions for $f_+(q^2=0)$ are displayed with a slight horizontal offset for clarity.}}
\label{fig:FFComparisons}
\end{figure*}

Figure~\ref{fig:FFComparisons}, top, compares our $B \to \pi\ell\nu$ form-factor determinations with other theoretical calculations from {light-cone sum rules (LCSR)~\cite{Bharucha:2012wy,Imsong:2014oqa}, NLO perturbative QCD (pQCD)~\cite{Li:2012nk},} and (2+1)-flavor lattice QCD {(LQCD)}~\cite{Dalgic:2006dt,Bailey:2008wp}.   Both of the earlier lattice calculations use staggered light quarks.  The HPQCD collaboration uses NRQCD $b$ quarks, while the Fermilab Lattice \& MILC collaborations use relativistic $b$ quarks with the Fermilab interpretation.  Our result for the form factor $f_+^{B\pi}$ agrees with earlier determinations, and is slightly more precise.  Our scalar form factor $f_0^{B\pi}$ is lower than the HPQCD result (although by less than $2\sigma$), but we note that their calculation used only a single lattice spacing.   {We also agree with the most recent light-cone-sum-rule prediction for the $B\to\pi\ell\nu$ form factor from Imsong {\it et al.}~\cite{Imsong:2014oqa}, who present the first extrapolation of LCSR results from low $q^2$ to $q^2_{\rm max}$ using the $z$ expansion.}

\begin{figure}[tb]
\centering
    \includegraphics[width=0.49\textwidth]{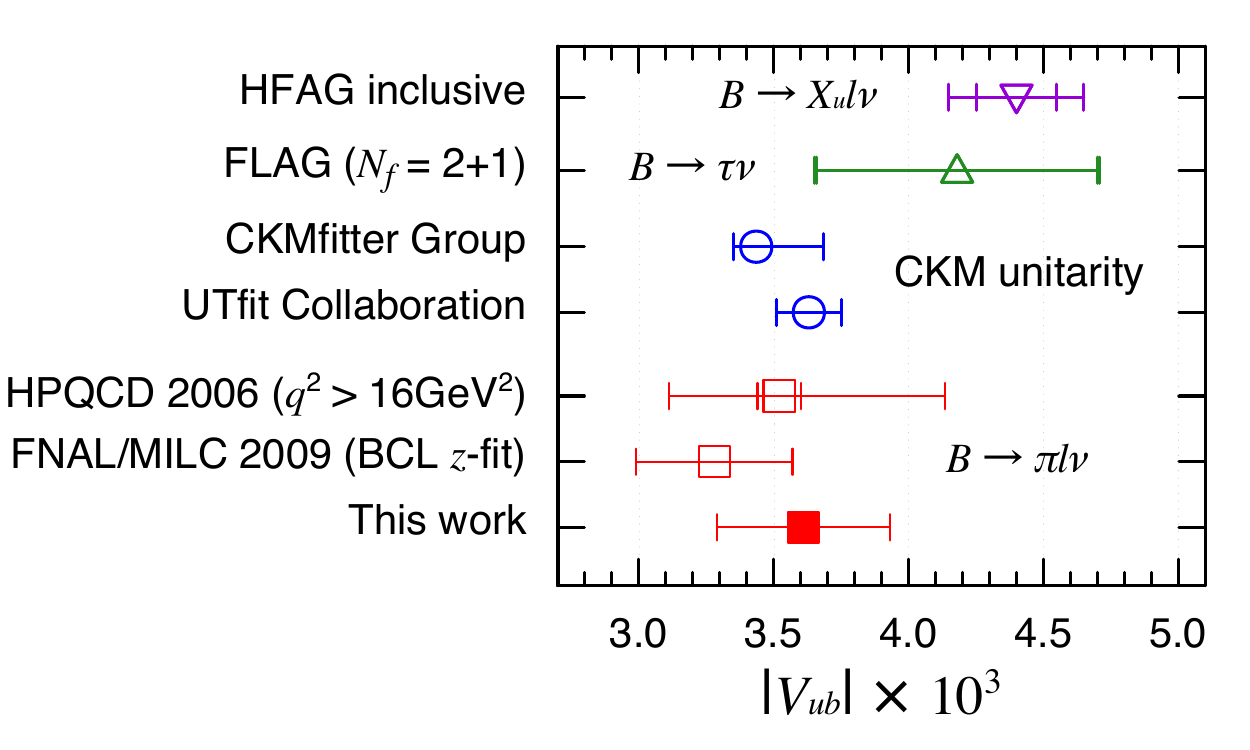} \hfill \vspace{-18pt}
\caption{Determinations of $|V_{ub}|$ from Table~\ref{tab:VubCompare}.  For points with double error bars, the inner error bars are experimental while the outer error bars show the total experimental plus theoretical uncertainty added in quadrature.  
}
\label{fig:VubCompare}
\end{figure}

We fit our results for the $B \to \pi\ell\nu$ form factors together with the experimentally measured decay rates from BaBar~\cite{delAmoSanchez:2010af,Lees:2012vv} and Belle~\cite{Ha:2010rf,Sibidanov:2013rkk}, leaving the relative normalization as a free parameter, to determine the CKM matrix element $|V_{ub}|$.  We obtain 
\begin{equation}
	|V_{ub}| = 3.61(32) \times 10^{-3} \,,
\end{equation}
where the error includes both theoretical and experimental uncertainties.
Table~\ref{tab:VubCompare} and Figure~\ref{fig:VubCompare} compare the determination of $|V_{ub}|$ using our lattice form-factor result with determinations using other theoretical calculations of the $B\to\pi\ell\nu$ form factor, as well as with determinations from inclusive $B \to X_u \ell \nu$ decay, $B\to\tau\nu$ leptonic decay, and predictions from CKM unitarity.  Our $|V_{ub}|$ result agrees with other lattice determinations, as well as with the less precise determination from $B\to\tau\nu$ decay, and with the more precise predictions from CKM unitarity.  The central value is higher, however, than the one obtained from the FNAL/MILC calculation so the tension between our result and the determination from inclusive $B \to X_u \ell \nu$ decay is less, only about 2$\sigma$.

\begin{table*}[tb]
\caption{Determinations of $|V_{ub}|$.  Top panel: results from inclusive $B \to X_u \ell \nu$ decay~\cite{HFAGSemiPDG13} and $B\to\tau\nu$ leptonic decay~\cite{Aoki:2013ldr}.  Middle panel: predictions from CKM unitarity~\cite{CKMfitterWinter2014,UTfitSummer2014}.  Bottom panel:  results from exclusive $B\to\pi\ell\nu$ decay using form factors from (2+1)-flavor lattice QCD~\cite{HFAGSemiPDG13,Dalgic:2006dt,Bailey:2008wp}.  Errors shown are either the total uncertainty or the experimental and theoretical uncertainties, respectively.  
}
\label{tab:VubCompare}
    \begin{tabular}{lll}
    \hline\hline
      & From & $|V_{ub}|\times 10^3$ \\[2pt] \hline \\[-2ex]
      HFAG inclusive average  \cite{Amhis:2012bh}                          & $B\to X_u \ell\nu$ & $4.40(15)(20)$          \\
      FLAG ($N_f = 2+1$)~\cite{Aoki:2013ldr}                               & $B\to\tau\nu$      & $4.18(52)(9)$           \\[2pt] \hline \\[-2ex]
      CKMfitter Group~\cite{CKMfitterWinter2014}                           & CKM unitarity      & $3.435(^{+250}_{-84})$  \\
      UTfit collaboration~\cite{UTfitSummer2014}                           & CKM unitarity      & $3.63(12)$              \\[2pt] \hline \\[-2ex]
      HPQCD (HFAG $q^2>16 {\rm GeV}^2$)~\cite{Dalgic:2006dt,HFAGSemiPDG13} & $B\to\pi\ell\nu$   & $3.52(8)(^{+61}_{-40})$ \\ 
      FNAL/MILC (HFAG BCL $z$-fit)~\cite{Bailey:2008wp,HFAGSemiPDG13}      & $B\to\pi\ell\nu$   & $3.28(29)$         \\
      This work                                                            & $B\to\pi\ell\nu$   & $3.61(32)$ \\
      \hline\hline 
    \end{tabular}
\end{table*}

Figure~\ref{fig:FFComparisons}, bottom, compares our $B_s \to K \ell\nu$ form-factor determinations with the theoretical calculation by the HPQCD collaboration using (2+1)-flavor lattice QCD with staggered light quarks and NRQCD $b$ quarks~\cite{Bouchard:2014ypa}, as well as with predictions at $q^2=0$ from the light-cone sum rules~\cite{Duplancic:2008tk}, NLO perturbative QCD~\cite{Wang:2012ab} and the relativistic quark model {(RQM)}~\cite{Faustov:2013ima}.  The lattice results agree in the range of simulated lattice data, and have similar precision, but diverge slightly when extrapolated to lower $q^2$.  Even at $q^2=0$, however, the predicted form factors $f_+^{B_s K}(0)$ differ by only $1.9\sigma$.  

The semileptonic decay $B_s \to K \ell\nu$ has not yet been observed experimentally.  We can therefore make predictions for Standard-Model observables for the decay processes $B_s \to K \mu\nu$ and $B_s \to K \tau \nu$.  Using our results for the $B_s\to K \ell\nu$ form factors in Table~\ref{tab:BstoKBCLResults} and our determination of $|V_{ub}|$ from $B\to\pi\ell\nu$ given above, we calculate the differential branching fractions and  forward-backward asymmetries.   Our lattice form-factor determinations at $q^2 \gtapprox 13 {\rm GeV}^2$ are sufficiently precise that future experimental measurements of the $B_s\to K \ell\nu$ differential branching fraction in this range with similar uncertainties will be able to distinguish between Standard-Model predictions using $|V_{ub}|$ from inclusive $B \to X_u \ell\nu$ and exclusive $B\to\pi\ell\nu$ semileptonic decays, and thereby weigh in on the current $\sim 3\sigma$ disagreement between the two determinations.  

We also calculate the ratio of $\mu$-to-$\tau$ differential decay rates, and the normalized forward-backward asymmetries, for both $B\to\pi\ell\nu$ and $B_s \to K \ell\nu$.  Because these quantities are independent of $|V_{ub}|$, they potentially provide more stringent tests of Standard-Model extensions such as ones that give rise to new scalar or right-handed currents.  In practice, it will likely be difficult for LHCb to measure $B_s$ decays with $\tau$ leptons in the final state, so the $B_s \to K \mu\nu$ predictions in Sec.~\ref{Sec:BstoKPheno} may be most useful for the foreseeable future.   Belle~II, however, should observe the decay $B \to \pi \ell\nu$ with both $\ell=\mu,\tau$ final states, and we anticipate that they will measure the forward-backward asymmetries and eventually the $\mu/\tau$ ratio for this decay.  Future measurements of these observables will be especially important given the current $\gtapprox 3\sigma$ discrepancy observed in $R(D)$ and $R(D^*)$ for the similar decays $B \to D^{(*)} \ell \nu$~\cite{Lees:2012xj,Bailey:2012jg}.

Our results for the $B \to \pi\ell\nu$ and $B_s \to K \ell\nu$ provide important independent checks of existing calculations using staggered light quarks.  Such confirmation is especially important given the present approximately 3$\sigma$ tension between $|V_{ub}|$ obtained from inclusive and exclusive semileptonic $B$ decays.  Currently the precision of our determination of the form factors is limited by statistics and by the relatively large chiral extrapolation to the physical light-quark mass.  To address the chiral-extrapolation error, we are presently analyzing the RBC/UKQCD M{\"o}bius domain-wall + Iwasaki ensemble{~\cite{Brower:2004xi,Brower:2012vk,Blum:2014tka}} with a  lattice spacing close to the coarser value $a\approx 0.11$~fm used in our current analysis, but with $M_\pi \approx 140$~MeV.   We are also using all-mode averaging~\cite{Blum:2012uh,Shintani:2014vja} to reduce the statistical errors on the individual numerical data points, and expect some reduction in the statistical errors.  With these improvements we anticipate a reduction in the current form-factor errors.  Further, our future physical-mass results will also include a determination of the tensor form factor, which will enable a calculation of the Standard-Model rate for the rare decay $B\to\pi\ell^+\ell^-$ and similar processes.

Because we present our results for the $B \to \pi\ell\nu$ and $B_s \to K \ell\nu$ form factor as coefficients of the BCL $z$-parametrization and the matrix of correlations between them, our form factors can be combined with new experimental measurements (and even with other lattice form-factor calculations) to further improve $|V_{ub}|$ in the future.  In particular, our $B_s \to K \ell\nu$ form-factor results will enable an alternate determination of  $|V_{ub}|$ once the process has been observed in experiment.  More generally, our form-factor results in Tables~\ref{tab:BtoPiBCLResults} and~\ref{tab:BstoKBCLResults} can be used to compute all possible Standard-Model observables for these decays whenever they are needed for comparison with experiment.

\section*{Acknowledgments}

Computations for this work were carried out in part on facilities of the USQCD collaboration, which are funded by the Office of Science of the 
U.S. Department of Energy.  We thank BNL, Columbia University, Fermilab, RIKEN, and the U.S. DOE for providing the facilities essential 
for the completion of this work.
This work was supported in part by the U.K. Science and Technology 
Facilities Council (STFC) Grants No. ST/J000396/1 and No.~ST/L000296/1 (J.M.F.),  and by the Grant-in-Aid of the Ministry of Education, Culture, Sports, Science and 
Technology, Japan (MEXT Grant) No.~22540301, No.~23105715, and No.~26400261 (T.I).   T.K. is supported by the JSPS Strategic Young Researcher Overseas 
Visits Program for Accelerating Brain Circulation (No. R2411).  O.W.~acknowledges support at Boston University by the U.S. NSF Grant No. OCI-0749300.  This manuscript has been authored by employees of Brookhaven Science Associates, LLC under Contract No. DE-SC0012704 with the U.S. Department of Energy.  Fermilab is operated by Fermi Research Alliance, LLC, under Contract No. DE-AC02-07CH11359 with the U.S. Department of Energy.

\appendix
\section{Correlator fit results}
\label{App:results_2pt3pt}

Here we summarize the results for pion, kaon, $B$ and $B_s$ meson masses and three-point ratios from the correlator fits described in Secs.~\ref{Sec:2pt} and~\ref{Sec:3pt}.

\begin{table*}[h]
  \begin{ruledtabular}
    \caption{Pion and kaon masses on all ensembles.  See Sec.~\ref{Sec:2pt} for details.}
    \label{tab:data_2pt_pion}
    \begin{tabular}{c|cccc|cccc} 
      & \multicolumn{4}{c|}{Pion} & \multicolumn{4}{c}{Kaon} \\
      $m_l$  & [$t_{\rm min}$, $t_{\rm max}$] & $aM_\pi$& $\chi^2/{\rm dof}$ & $p$  & [$t_{\rm min}$, $t_{\rm max}$] & $aM_K$& $\chi^2/{\rm dof}$ & $p$\\ \hline
      0.005   & [12:23] & 0.18959(53) & 0.83 & 61\% & [12:23] & 0.31287(45) & 1.31 & 21\% \\
      0.01    & [12:23] & 0.24305(48) & 0.76 & 68\% & [12:23] & 0.33352(44) & 0.82 & 62\% \\
      0.004   & [16:30] & 0.12611(51) & 1.08 & 37\% & [16:30] & 0.23249(42) & 1.35 & 17\% \\
      0.006   & [16:30] & 0.15207(36) & 0.41 & 97\% & [16:30] & 0.24189(33) & 0.58 & 88\% \\
      0.008   & [16:30] & 0.17265(42) & 1.05 & 40\% & [16:30] & 0.24854(39) & 1.06 & 39\% \\
    \end{tabular}
  \end{ruledtabular}    
\end{table*}

\begin{table*}[h]
  \begin{ruledtabular}
    \caption{$B$- and $B_s$-meson masses on all ensembles.  See Sec.~\ref{Sec:2pt} for details. }
    \label{tab:data_2pt_B}
    \begin{tabular}{c|cccc|cccc} 
          & \multicolumn{4}{c|}{$B$ meson} & \multicolumn{4}{c}{$B_s$ meson} \\
      $m_l$  & [$t_{\rm min}$, $t_{\rm max}$] & $aM_{B}$& $\chi^2/{\rm dof}$ & $p$  &  [$t_{\rm min}$, $t_{\rm max}$] & $aM_{B}$& $\chi^2/{\rm dof}$ & $p$ \\ \hline
      0.005  & [7:30] & 3.0638(13) &  0.92 &  57\% & [10:30] & 3.1020(12) &  0.46 &  98\% \\
      0.01   & [7:30] & 3.0727(13) &  0.74 &  80\% & [10:30] & 3.1028(13) &  1.13 &  31\% \\
      0.004  & [9:30] & 2.3203(14) &  1.40 &  11\% & [13:30] & 2.3509(11) &  1.53 &   7\% \\
      0.006  & [9:30] & 2.3241(10) &  0.78 &  74\% & [13:30] & 2.3520(09) &  0.57 &  92\% \\
      0.008  & [9:30] & 2.3274(12) &  0.85 &  66\% & [13:30] & 2.3533(12) &  0.95 &  51\% \\
    \end{tabular}
  \end{ruledtabular}    
\end{table*}

\begingroup
\begin{table*}[h]
  \begin{ruledtabular}
   \caption{
   Three-point correlator ratios $R^{B\pi}_{3,\mu}$ on all ensembles.  Results are shown for pion momenta $\vec{p}_\pi^2 = (2\pi\vec{n}/L)^2$ through $n^2=3$.}
    \label{tab:data_R_BtoPi}
    \begin{tabular}{cc|cccc|cccc} 
     $m_l$ & $n^2$
     &  [$t_{\rm min}$, $t_{\rm max}$]  & $M_{B_s}^{1/2}R^{B\pi}_{3,i}/p_\pi^i$ & $\chi^2/{\rm dof}$ & $p$
     &  [$t_{\rm min}$, $t_{\rm max}$]   & $M_{B_s}^{-1/2}R^{B\pi}_{3,0}$ & $\chi^2/{\rm dof}$ & $p$ 
     \\ \hline
     0.005 & 0 &        &           &      &      & [6:10] & 0.2523(36) & 1.49 & 20\% \\
           & 1 & [6:10] & 1.086(31) & 1.19 & 31\% & [6:10] & 0.2034(52) & 0.16 & 96\% \\
           & 2 & [6:10] & 0.827(46) & 1.71 & 15\% & [6:10] & 0.1819(93) & 0.38 & 82\% \\
           & 3 & [6:10] & 0.755(78) & 0.60 & 66\% & [6:10] & 0.154(17)  & 0.62 & 65\% \\ \hline
     0.01  & 0 &        &           &      &      & [6:10] & 0.2513(35) & 2.03 &  9\% \\
           & 1 & [6:10] & 1.049(23) & 1.07 & 37\% & [6:10] & 0.2060(43) & 1.17 & 32\% \\
           & 2 & [6:10] & 0.798(30) & 0.53 & 72\% & [6:10] & 0.1837(73) & 0.55 & 70\% \\
           & 3 & [6:10] & 0.706(47) & 0.88 & 47\% & [6:10] & 0.180(13)  & 0.04 &100\% \\ \hline
     0.004 & 0 &        &           &      &      & [8:13] & 0.3679(79) & 1.25 & 28\% \\
           & 1 & [8:13] & 1.546(71) & 0.88 & 50\% & [8:13] & 0.288(12)  & 1.58 & 16\% \\
           & 2 & [8:13] & 1.100(92) & 0.40 & 85\% & [8:13] & 0.256(18)  & 1.93 &  9\% \\
           & 3 & [8:13] & 0.96(20)  & 0.33 & 89\% & [8:13] & 0.186(32)  & 0.70 & 63\% \\ \hline
     0.006 & 0 &        &           &      &      & [8:13] & 0.3528(51) & 1.19 & 31\% \\
           & 1 & [8:13] & 1.529(42) & 1.01 & 41\% & [8:13] & 0.2739(63) & 0.30 & 91\% \\
           & 2 & [8:13] & 1.219(58) & 0.41 & 84\% & [8:13] & 0.240(11)  & 0.76 & 58\% \\
           & 3 & [8:13] & 0.921(91) & 0.46 & 80\% & [8:13] & 0.224(21)  & 1.14 & 34\% \\ \hline 
     0.008 & 0 &        &           &      &      & [8:13] & 0.3432(58) & 1.45 & 20\% \\
           & 1 & [8:13] & 1.483(38) & 1.43 & 21\% & [8:13] & 0.2829(67) & 1.02 & 40\% \\
           & 2 & [8:13] & 1.183(53) & 0.63 & 68\% & [8:13] & 0.2349(98) & 1.01 & 41\% \\
           & 3 & [8:13] & 0.888(89) & 0.52 & 76\% & [8:13] & 0.211(18)  & 0.20 & 96\% \\
    \end{tabular}
  \end{ruledtabular}
\end{table*}
\endgroup

\begingroup
\begin{table*}[H]
    \begin{ruledtabular}
     \caption{Three-point correlator ratios $R^{B_sK}_{3,\mu}$ on all ensembles.  Results are shown for kaon momenta $\vec{p}_K^2 = (2\pi\vec{n}/L)^2$ through $n^2=4$.}
    \label{tab:data_R_BstoK}
    \begin{tabular}{cc|cccc|cccc} 
     $m_l$ & $n^2$ 
     &  [$t_{\rm min}$, $t_{\rm max}$]  & $M_{B_s}^{1/2}R^{B_s K}_{3,i}/p_K^i$ & $\chi^2/{\rm dof}$ & $p$
     &  [$t_{\rm min}$, $t_{\rm max}$]   & $M_{B_s}^{-1/2}R^{B_s K}_{3,0}$ & $\chi^2/{\rm dof}$ & $p$ \\ \hline
     0.005 & 0 &        &           &      &      & [6:10] & 0.2394(27) & 0.99 & 41\% \\
           & 1 & [6:10] & 0.984(16) & 2.17 &  7\% & [6:10] & 0.2036(27) & 0.60 & 66\% \\
           & 2 & [6:10] & 0.763(16) & 2.53 &  4\% & [6:10] & 0.1791(36) & 0.46 & 77\% \\
           & 3 & [6:10] & 0.623(22) & 1.60 & 17\% & [6:10] & 0.1609(57) & 0.52 & 72\% \\
           & 4 & [6:10] & 0.543(31) & 0.98 & 42\% & [6:10] & 0.1568(88) & 0.68 & 61\% \\ \hline
     0.01  & 0 &        &           &      &      & [6:10] & 0.2421(28) & 2.35 &  5\% \\
           & 1 & [6:10] & 0.963(15) & 0.95 & 44\% & [6:10] & 0.2060(30) & 1.00 & 40\% \\
           & 2 & [6:10] & 0.749(16) & 0.13 & 97\% & [6:10] & 0.1847(40) & 0.17 & 95\% \\
           & 3 & [6:10] & 0.634(21) & 0.38 & 83\% & [6:10] & 0.1737(61) & 0.16 & 96\% \\
           & 4 & [6:10] & 0.573(30) & 0.30 & 88\% & [6:10] & 0.1664(81) & 0.12 & 98\% \\ \hline
     0.004 & 0 &        &           &      &      & [8:13] & 0.3264(49) & 1.14 & 34\% \\
           & 1 & [8:13] & 1.340(26) & 0.57 & 72\% & [8:13] & 0.2751(51) & 1.25 & 28\% \\
           & 2 & [8:13] & 1.039(26) & 1.14 & 33\% & [8:13] & 0.2444(64) & 1.84 & 10\% \\
           & 3 & [8:13] & 0.830(39) & 1.65 & 14\% & [8:13] & 0.2148(89) & 1.80 & 11\% \\
           & 4 & [8:13] & 0.690(50) & 0.88 & 49\% & [8:13] & 0.195(13)  & 1.18 & 32\% \\ \hline
     0.006 & 0 &        &           &      &      & [8:13] & 0.3312(35) & 1.12 & 34\% \\
           & 1 & [8:13] & 1.336(20) & 1.02 & 40\% & [8:13] & 0.2770(36) & 1.19 & 31\% \\
           & 2 & [8:13] & 1.068(20) & 0.71 & 61\% & [8:13] & 0.2419(46) & 0.60 & 70\% \\
           & 3 & [8:13] & 0.868(26) & 1.49 & 19\% & [8:13] & 0.2168(68) & 0.53 & 75\% \\
           & 4 & [8:13] & 0.737(37) & 0.57 & 72\% & [8:13] & 0.206(10)  & 0.37 & 87\% \\ \hline
     0.008 & 0 &        &           &      &      & [8:13] & 0.3323(47) & 1.28 & 27\% \\
           & 1 & [8:13] & 1.332(25) & 1.21 & 30\% & [8:13] & 0.2848(47) & 1.16 & 32\% \\
           & 2 & [8:13] & 1.056(25) & 0.86 & 51\% & [8:13] & 0.2476(58) & 1.23 & 29\% \\
           & 3 & [8:13] & 0.856(34) & 1.17 & 32\% & [8:13] & 0.2264(87) & 0.80 & 55\% \\
           & 4 & [8:13] & 0.736(45) & 0.83 & 52\% & [8:13] & 0.200(13)  & 0.61 & 69\% \\ 
    \end{tabular}
    \end{ruledtabular}
\end{table*}
\endgroup

\section{$z$-fit results}
\label{app:zfits}

Here we present the complete results for the $z$-fits described in Secs.~\ref{Sec:LatzFit} and~\ref{Sec:Vub}.  Tables~\ref{tab:BtoPiLatzFits} and~\ref{tab:BstoKLatzFits} show the results of fits to only our lattice form factors, while Tables~\ref{tab:BtoPiExpZFits} and~\ref{tab:VubZFits} show the result of joint fits to the lattice and experimental data.


\begin{sidewaystable}[p!]\caption{\label{tab:BtoPiLatzFits} Results for $z$-fits of the $B\to\pi\ell\nu$ synthetic form-factor data to the BCL parametrization, Eqs.~(\ref{eq:BCLConstraint}) and~(\ref{eq:f0BtoPiBCL}).  The top panels show fits without constraints on the BCL coefficients for (i) $f_+$ only (first panel), (ii) $f_0$ only (second panel), and (iii) $f_+$ and $f_0$ imposing the kinematic constraint $f_+(q^2=0)=f_0(q^2=0)$ (third panel).   The bottom panel shows combined fits of $f_+$ and $f_0$ imposing both the kinematic constraint 
and the heavy-quark estimate for the sum of the vector-form-factor coefficients $\sum B_{mn} b_m b_n$, Eq.~(\ref{eq:HQConstraint}).  We show results for different truncations $K=(1),2,3$.  For fits that include the heavy-quark constraint, we quote the augmented $\chi^2/{\rm dof}$ as defined in Eq.~(\ref{eq:chi2Aug}).   Note that we cannot quote a goodness-of-fit for cases where the fits have the same number of free parameters as data points. }
\begin{center}
\begin{ruledtabular}
\begin{tabular}{lccccl|lccccl|lcc}
\multicolumn{6}{c|}{$f_+^{B \pi}$} & \multicolumn{6}{c|}{$f_0^{B\pi}$} &  \\[1mm]
$K$ & $b^{(0)}$ & $b^{(1)}/b^{(0)}$ & $b^{(2)}/b^{(0)}$  & $b^{(3)}/b^{(0)}$ & $\sum B_{mn} b_m b_n$ 
& $K$ & $b^{(0)}$ & $b^{(1)}/b^{(0)}$ & $b^{(2)}/b^{(0)}$  & $b^{(3)}/b^{(0)}$ & $\sum B_{mn} b_m b_n$ & $f(q^2=0)$ & $\chi^2/{\rm dof}$ & $p$  \\ \hline
1& 0.447(36) &            &          &           &  0.00394(63)& &           &           &           &        &           &\ 0.447(36) & 4.02 &  2\% \\
2& 0.410(39) &  -1.30(52) &          &           &  0.0120(59) & &           &           &           &        &           &\ 0.241(83) & 0.30 & 58\% \\
3& 0.420(43) &  -1.46(59) & -4.7(7.2)&           &  0.15(42)   & &           &           &           &        &           &\ 0.07(32)  &     &      \\ \hline
 &           &            &          &           &             &1& 0.460(61) &           &           &        & 0.0225(60)&\  0.460(61)& 90.1 & 0\% \\
 &           &            &          &           &             &2& 0.516(61) &  -4.09(55)&           &        & 0.408(63) &  -0.074(73)& 0.03 & 87\% \\
 &           &            &          &           &             &3& 0.516(61) &  -3.94(97)& 0.7(3.8)  &        & 0.32(41)  &  -0.02(28) &     &     \\ \hline
2& 0.366(37) & -2.79(54) &           &           & 0.0337(85)  &2& 0.587(58) & -3.33(38) &           &        & 0.346(55) &\ 0.040(65) & 6.18 & 0\% \\
3& 0.427(40) & -1.62(46) & -7.7(1.5) &           & 0.38(15)    &2& 0.521(60) & -4.03(52) &           &        & 0.404(62) & -0.066(70) & 0.10 & 91\% \\
2& 0.410(39) & -1.24(51) &           &           & 0.0113(56)  &3& 0.520(60) & -3.12(42) & 4.5(1.3)  &        & 0.41(17)  &\ 0.248(82) & 0.58 & 56\% \\ 
3& 0.424(41) & -1.50(57) & -6.0(5.0) &           & 0.24(38)    &3& 0.519(60) & -3.81(81) & 1.2(3.4)  &        & 0.27(25)  &\ 0.01(24)  & 0.07 & 79\% \\ \hline
2& 0.368(37) & -2.70(51) &           &           & 0.0320(79)  &2& 0.592(57) & -3.27(36) &           &        & 0.338(53) &\ 0.051(63) & 4.78 & 0\%  \\
3& 0.384(38) & -2.22(49) & -3.18(74) &           & 0.066(21)   &2& 0.579(57) & -3.34(37) &           &        & 0.339(52) &\ 0.038(63) & 4.60 & 0\%  \\
2& 0.410(39) & -1.24(51) &           &           & 0.0112(54)  &3& 0.520(60) & -3.12(41) & 4.5(1.3)  &        & 0.41(17)  &\ 0.249(81) & 0.39 & 76\% \\ 
3& 0.412(39) & -1.24(50) & -1.3(1.5) &           & 0.019(24)   &3& 0.520(60) & -3.19(44) & 4.1(1.5)  &        & 0.36(18)  &\ 0.224(95) & 0.51 & 60\%\\
4& 0.412(39) & -1.25(50) & -1.2(1.5) & -0.8(2.7) & 0.019(25)   &3& 0.520(60) & -3.20(45) & 4.1(1.5)  &        & 0.35(18)  &\ 0.220(97) & 0.99 & 32\% \\ 
3& 0.411(39) & -1.29(51) & -0.9(2.0) &           & 0.015(25)   &4& 0.510(61) & -3.81(90) & 3.8(1.7)  & 10(12) & 4.6(8.6)  &\ 0.23(11)  & 0.29 & 59\% \\ 
4& 0.411(39) & -1.29(51) & -0.9(2.0) & -0.3(4.2) & 0.015(26)   &4& 0.510(61) & -3.81(90) & 3.7(1.8)  & 10(13) & 4.6(8.8)  &\ 0.23(12)  &     &     \\ 
\end{tabular}
\end{ruledtabular}
\end{center}
\end{sidewaystable}
\begin{sidewaystable}[p!]
\caption{\label{tab:BstoKLatzFits} Results for $z$-fits of the $B_s \to K \ell\nu$ synthetic form-factor data to the BCL parametrization. Entry meanings are the same as in Table~\ref{tab:BtoPiLatzFits}.}
\begin{center}
\begin{ruledtabular}
\begin{tabular}{lccccl|lccccl|lcc}
\multicolumn{6}{c|}{$f_+^{B_s  K}$} & \multicolumn{6}{c|}{$f_0^{B_s K}$} &  \\[1mm]
$K$ & $b^{(0)}$ & $b^{(1)}/b^{(0)}$ & $b^{(2)}/b^{(0)}$ & $b^{(3)}/b^{(0)}$ & $\sum B_{mn} b_m b_n$ 
& $K$ & $b^{(0)}$ & $b^{(1)}/b^{(0)}$ & $b^{(2)}/b^{(0)}$ & $b^{(3)}/b^{(0)}$ & $\sum B_{mn} b_m b_n$ & $f(q^2=0)$ & $\chi^2/{\rm dof}$ & $p$  \\ \hline
1& 0.393(23) &            &           &           & 0.00178(21)& &           &           &            &          &             & 0.393(23) & 48.8 & 0\%  \\
2& 0.337(24) &  -3.52(48) &           &           & 0.0249(44) & &           &           &            &          &             & 0.152(34) & 0.52 & 47\% \\
3& 0.324(30) &  -5.1(2.3) & -11(14)   &           & 0.21(47)   & &           &           &            &          &             & -0.00(21) &     &     \\ \hline
 &           &            &           &           &            &1& 0.209(14) &           &            &          & 0.00406(54) & 0.209(14) & 2.88 &  6\% \\
 &           &            &           &           &            &2& 0.222(15) & 0.49(23)  &            &          & 0.0064(18)  & 0.238(20) & 1.48 & 22\% \\
 &           &            &           &           &            &3& 0.208(19) & -1.1(1.4) & -7.3(6.5)  &          & 0.26(41)    & 0.142(82) &     &     \\ \hline
2& 0.376(20) & -2.33(19)  &           &           & 0.0152(21) &2& 0.225(15) &  0.44(22) &            &          & 0.0062(17) & 0.239(20) & 3.65 & 1\% \\
3& 0.347(22) & -2.63(24)  & 2.7(1.4)  &           & 0.039(19)  &2& 0.222(15) &  0.48(23) &            &          & 0.0062(17) & 0.237(20) & 0.98 & 38\% \\
2& 0.338(24) & -3.52(48)  &           &           & 0.0249(44) &3& 0.209(16) & -0.94(54) & -6.5(2.2)  &          & 0.21(12)   & 0.151(34) & 0.27 & 76\% \\
3& 0.337(24) & -3.61(89)  & -2.3(4.5) &           & 0.028(29)  &3& 0.208(19) & -1.1(1.4) & -7.2(6.5)  &          & 0.26(40)   & 0.143(82) & 0.53 & 47\% \\ \hline
2& 0.376(20) & -2.33(19)  &           &           & 0.0152(21) &2& 0.225(15) &  0.45(22) &            &          & 0.0062(17) & 0.239(20) & 2.75 & 3\%\\
3& 0.351(22) & -2.58(22)  & 2.2(1.2)  &           & 0.032(13)  &2& 0.222(15) &  0.48(23) &            &          & 0.0063(17) & 0.238(20) & 0.88 & 45\%\\
2& 0.338(24) & -3.49(47)  &           &           & 0.0246(43) &3& 0.209(16) & -0.91(54) & -6.4(2.2)  &          & 0.21(12)   & 0.153(33) & 0.26 & 85\%\\
3& 0.338(24) & -3.43(67)  & -1.4(3.0) &           & 0.023(10)  &3& 0.210(17) & -0.8(1.0) & -5.9(4.4)  &          & 0.18(23)   & 0.159(59) & 0.38 & 68\%\\
4& 0.338(24) & -3.47(70)  & -1.7(3.3) & -1.5(6.1) & 0.016(18)  &3& 0.209(18) & -0.9(1.1) & -6.4(4.9)  &          & 0.20(27)   & 0.153(63) & 0.63 & 43\%\\	
3& 0.338(24) & -3.46(73)  & -1.6(3.7) &           & 0.024(14)  &4& 0.211(18) & -0.8(1.0) & -6.3(6.7)  &  -3(35)  & 0.3(1.4)   & 0.156(68) & 0.76 & 38\%\\
4& 0.337(24) & -3.54(72)  & -2.2(3.6) & -1.9(5.8) & 0.018(22)  &4& 0.210(18) & -0.9(1.0) & -7.4(6.6)  &  -7(35)  & 0.5(2.6)   & 0.145(67) &     & \\
\end{tabular}
\end{ruledtabular}
\end{center}\end{sidewaystable}


\begin{table*}[tb]
\caption{\label{tab:BtoPiExpZFits} Results for $K=3$ $z$-fits of the $B\to\pi\ell\nu$ experimental data to the BCL parametrization, Eqs.~(\ref{eq:BCLConstraint}) and~(\ref{eq:f0BtoPiBCL}).  
}
\begin{ruledtabular}
\begin{tabular}{lcccccc}
 & $ |V_{ub}| b^{(0)}_+ \times 10^3$ & $b^{(1)}_+/b^{(0)}_+$ & $b^{(2)}_+/b^{(0)}_+$ & $|V_{ub}|^2  \sum B_{mn} b_m b_n \times 10^6 $ & $\chi^2/{\rm dof}$ & $p$  \\ \hline
All                                    & 1.517(32) & -1.03(19) &   -1.18(61) &  0.23(12) & 1.36 &  5\% \\ \hline
BaBar 2010~\cite{delAmoSanchez:2010af} & 1.361(75) & -0.91(47) & \  0.5(1.5) &  0.06(11) & 1.98 & 11\% \\
BaBar 2012~\cite{Lees:2012vv}          & 1.497(58) & -0.33(44) &   -3.5(1.3) &  1.19(77) & 0.45 & 91\% \\

Belle 2010~\cite{Ha:2010rf}            & 1.602(61) & -1.31(27) &   -0.8(0.9) &  0.21(12) & 1.19 & 29\% \\ 
Belle 2013~\cite{Sibidanov:2013rkk}    & 1.562(84) & -1.88(50) & \  1.3(1.6) &  0.40(53) & 1.24 & 22\% \\
\end{tabular}
\end{ruledtabular}
\end{table*}


\begin{table*}[tb]
\caption{\label{tab:VubZFits} Determinations of $|V_{ub}|$ from combined fits of $B\to\pi\ell\nu$ experimental and lattice form-factor data to the BCL parametrization, Eq.~(\ref{eq:BCLConstraint}), for different truncations $K$. }
\centering
\begin{ruledtabular}
\begin{tabular}{l|c|l|llllll|cc}
 & $K$ & $|V_{ub}| \times 10^3 $ & $b^{(0)}$ & $b^{(1)}_+/b^{(0)}_+$ & $b^{(2)}_+/b^{(0)}_+$ & $b^{(3)}_+/b^{(0)}_+$  & $b^{(4)}_+/b^{(0)}_+$  
& $\sum B_{mn} b_m b_n $ & $\chi^2/{\rm dof}$ & $p$  \\ \hline
All                                    & 3 & 3.61(32) & 0.422(35) & -1.07(17) &  -1.08(56) &           &        & 0.0162(86) & 1.32 & 7\%\\
                                       & 4 & 3.61(32) & 0.421(37) & -1.07(18) &  -1.0(1.6) & -0.6(4.1) &        & 0.015(19)  & 1.35 & 5\%\\
                                       & 5 & 3.61(32) & 0.421(37) & -1.10(27) &  -1.0(1.6) & -1(12)    & -5(26) & 0.2(2.4)   & 1.37 & 4\%\\ \hline
BaBar 2010~\cite{delAmoSanchez:2010af} & 3 & 3.31(36) & 0.414(37) & -1.03(31) & \ 0.9(1.1) &           &        & 0.009(11)  & 1.35 & 24\%\\ 
                                       & 4 & 3.29(35) & 0.428(41) & -1.04(30) &  -1.4(2.6) & -7.5(7.2) &        & 0.47(97)   & 1.48 & 20\%\\ 
                                       & 5 & 3.29(35) & 0.428(41) & -1.21(39) &  -1.0(2.7) & 17(15)    & -20(39)& 7(20)      & 1.82 & 14\%\\ \hline
BaBar 2012~\cite{Lees:2012vv}          & 3 & 3.52(35) & 0.436(38) & -0.79(28) &  -2.20(86) &           &        & 0.043(31)  & 0.61 & 82\%\\ 
                                       & 4 & 3.56(36) & 0.425(40) & -0.81(29) &  -0.5(2.4) & -6.2(6.5) &        & 0.24(54)   & 0.61 & 81\%\\ 
                                       & 5 & 3.54(36) & 0.428(40) & -1.04(36) &  -0.3(2.4) &  9(15)    & -41(35)& 15(27)     & 0.54 & 85\%\\ \hline
Belle 2010~\cite{Ha:2010rf}            & 3 & 3.90(37) & 0.411(35) & -1.29(23) &  -0.86(79) &           &        & 0.0141(82) & 1.01 & 43\%\\ 
                                       & 4 & 3.93(38) & 0.403(36) & -1.27(23) & \ 0.6(2.0) & -4.8(6.0) &        & 0.14(38)   & 1.05 & 40\%\\ 
                                       & 5 & 3.94(38) & 0.401(36) & -1.41(34) & \ 1.0(2.2) &  2(13)    & -24(36)& 4(13)      & 1.12 & 34\% \\ \hline
Belle 2013~\cite{Sibidanov:2013rkk}    & 3 & 3.88(44) & 0.397(37) & -1.52(36) & \ 0.2(1.2) &           &        & 0.0109(56) & 1.17 & 27\%\\ 
                                       & 4 & 3.84(42) & 0.425(41) & -1.59(32) &  -5.0(2.6) & -17.0(7.6)&        & 2.6(2.5)   & 1.01 & 45\%\\ 
                                       & 5 & 3.86(42) & 0.429(41) & -1.28(38) &  -6.2(2.7) & -2(15)    & 73(41) & 41(52)     & 0.95 & 51\%\\ 
\end{tabular}
\end{ruledtabular}
\end{table*}

\clearpage
\bibliography{B_meson}
\bibliographystyle{apsrev4-1} 
\end{document}